\documentclass[a4paper,11pt]{article}
\usepackage[english]{babel}
\usepackage{jheppub}
\usepackage{subfig}
\usepackage{float}
\usepackage{caption}

\usepackage{graphicx}
\usepackage{epsfig}
\usepackage{amsmath}
\usepackage{amssymb}
\usepackage{float}
\usepackage{placeins}
\usepackage{braket}
\usepackage{slashed}
\usepackage{mathdots}
\usepackage{lipsum}
\usepackage{bigints}
\usepackage{cleveref}

\usepackage{longtable}

\usepackage{array}

\allowdisplaybreaks[4]

\title{Two-loop form factors for $P$-wave quarkonium production and decay}

\author[a]{Melih~A.~Ozcelik}
\affiliation[a]{Universit\'e Paris-Saclay, CNRS, IJCLab, 91405 Orsay, France}
\emailAdd{melih.ozcelik@ijclab.in2p3.fr}

\abstract{We present the analytical results for the two-loop form factors needed for $\chi_{Q,J}$ production and decay. We consider the two-loop corrections to the process $\gamma \gamma \leftrightarrow {^3 P_J^{[1]}}$, that has been known only numerically before, and the processes $gg \leftrightarrow {^3 P_J^{[1]}}$, $\gamma g \leftrightarrow {^3 P_J^{[8]}}$ and $gg \leftrightarrow {^3 P_J^{[8]}}$, which have not been computed before. We observe that the NRQCD pole structure of the two-loop amplitude in the $gg$ channel is more involved for the spin-triplet $P$-wave case than for the pseudo-scalar $S$-wave case. It involves in addition to the standard Coulomb singularity also a new singularity whose cancellation requires the inclusion of the $gg \leftrightarrow {^3S_1^{[8]}}$ form factor. We give the high precision numerical results for the hard functions that can be used to compute $\chi_{Q,J}$ production and decay up to NNLO accuracy. 

}

\DeclareMathOperator{\Tr}{Tr}
\newcommand{\msbar}{{$\overline{\text{MS}}$}}

\DeclareMathOperator{\Cl}{Cl}
\DeclareMathOperator{\Li}{Li}
\DeclareMathOperator{\Ree}{Re}
\DeclareMathOperator{\Ime}{Im}
\DeclareMathOperator{\hpli}{HPLI}

\date{}

\begin{document}
\maketitle
\flushbottom

\section{Introduction}
\label{sec:introduction}

With the advent of the High-Luminosity phase at the LHC (HL-LHC) and the upcoming Electron-Ion Collider (EIC) and Future Circular Collider (FCC) experiments, it becomes necessary to improve the theory predictions for various observables to match their experimental precision. For instance, this will allow us to measure the fundamental quantities in the Standard Model, as the strong coupling constant, $\alpha_s$, and the mass of the top quark, $m_t$, to higher precision \cite{dEnterria:2022hzv, Proceedings:2015eho, Nason:2023asn, Benitez:2024nav, Azzi:2019yne, Frixione:2014ala, Juste:2013dsa, Cortiana:2015rca}. Equally, new measurements will allow us to understand the underlying structure of the colliding hadrons as the proton to higher accuracy.

On the theory side, this requires the inclusion of various kind of corrections as, for instance, perturbative QCD (pQCD) corrections, which can yield sizable contributions. However, the task of computing higher-order pQCD corrections is a highly non-trivial one because of the appearance of multi-loop Feynman integrals. Their evaluation can be quite challenging due to the presence of special mathematical functions and singularity structures. Therefore, a significant part of the effort is devoted in understanding and improving the evaluation of multi-loop amplitudes and integrals \cite{Travaglini:2022uwo, Abreu:2022mfk, Weinzierl:2022eaz}.

Within this precision program, the study of quarkonium physics in hadron-hadron and hadron-lepton collisions is quite interesting as it allows us to access the structure of the colliding hadrons at rather low scales on the order of the charm quark mass. Quarkonia are bound states consisting of a $Q\bar{Q}$ heavy quark pair, in the case of charm and bottom quarks, they form charmonium ($c\bar{c}$) and bottomonium ($b\bar{b}$) bound states respectively.\footnote{The hypothetical toponium bound state, consisting of a $t\bar{t}$ quark pair, has not been observed so far experimentally. The top quark decays via the electro-weak interaction before it can form a bound state.} As Parton Distribution Functions (PDFs) of protons are rather unconstrained at scales close to the charm quark mass, the study of charmonium production can set valuable constraints on PDFs \cite{Halzen:1984rq, Martin:1987vw, Martin:1987ww, Jung:1992uj, Lansberg:2020ejc}. Furthermore, as the strong coupling is not so small at these scales, it is interesting to study the interplay between the perturbative and non-perturbative regimes. For reviews and prospects in quarkonium physics, we guide the reader to refs.~\cite{Kramer:2001hh, Lansberg:2019adr, QuarkoniumWorkingGroup:2004kpm, Lansberg:2006dh, Andronic:2015wma, Chapon:2020heu}.

In refs.~\cite{Ozcelik:2019qze, Lansberg:2020ejc}, we have studied the inclusive hadro-production of pseudo-scalar quarkonium, $\eta_Q$, at Next-to-Leading Order (NLO) in the strong coupling and observed that the result is strongly dependent on the choice of the proton PDF. In particular, we observed that the NLO cross sections in collinear factorisation can turn negative for some scale choices which was noted already back in ref.~\cite{Schuler:1994hy}. Similarly issues were encountered in inclusive $J/\psi$ photo-production in ref.~\cite{Kramer:1995nb} and for $\chi_{c,0}$ and $\chi_{c,2}$ hadro-production in ref.~\cite{Mangano:1996kg} where the issue was most significant.

We have traced back the appearance of the negative cross sections to an over-subtraction of the initial-state collinear singularities into the proton PDF via the Altarelli-Parisi counterterm in the \msbar-scheme. In ref.~\cite{Lansberg:2020ejc}, we have then cured this issue at NLO by fixing the collinear factorisation scale choice based on the high-energy limit of the partonic cross section. We performed a similar treatment for $J/\psi$ photo-production in ref.~\cite{ColpaniSerri:2021bla}. The partonic high-energy limit on which the scale choice is based on has a connection to the $k_T$-factorisation approach which we have investigated in refs.~\cite{Lansberg:2021vie, Lansberg:2023kzf}.

The current state-of-the-art phenomenology of quarkonium hadro-production is based on NLO cross sections \cite{Kuhn:1992qw,Schuler:1994hy,Petrelli:1997ge}. However in view of the large scale uncertainties encountered at NLO \cite{Ozcelik:2019qze,Lansberg:2020ejc,Schuler:1994hy,Mangano:1996kg,Kramer:1995nb}, it will be essential and crucial to include the Next-to-Next-to-Leading Order (NNLO) contributions which may provide an alternative treatment of the aforementioned negative cross sections. In this regard, we have studied and computed in a companion paper \cite{Abreu:2022cco} the two-loop virtual amplitudes needed for $\eta_Q$ hadro-production and decay up to NNLO accuracy.

In this paper, we continue with the precision program and focus on the two-loop virtual amplitudes needed for $\chi_{Q,J}$ hadro-production and decay up to NNLO accuracy. The subscript $J$ indicates the total spin of the bound state and can take values $J=0$, $J=1$ and $J=2$. This involves the computation of form factors with the $Q\bar{Q}$ pair being in $P$-wave configuration which is more involved than the $S$-wave configuration as was the case for $\eta_Q$ form factors. Furthermore, we work in the framework of Non-Relativistic QCD (NRQCD) factorisation \cite{Bodwin:1994jh}, where the cross section can be factorised into non-perturbative Long-Distance Matrix Elements (LDMEs) and the perturbative short-distance coefficients.

We compute the two-loop form factors for the processes $\gamma \gamma \leftrightarrow {^3 P_J^{[1]}}$ and $gg \leftrightarrow {^3 P_J^{[1]}}$ with the heavy quark pair being in colour-singlet configuration and also the processes $\gamma g \leftrightarrow {^3 P_J^{[8]}}$ and $gg \leftrightarrow {^3 P_J^{[8]}}$ where the heavy quark pair is in colour-octet configuration. While the process $\gamma \gamma \leftrightarrow {^3 P_J^{[1]}}$, needed for the exclusive decay to di-photon\footnote{We note that the form factor $\gamma \gamma \leftrightarrow {^3 P_J^{[1]}}$ could also be used as ingredient for $\chi_{Q,J}$ production in Ultra-Peripheral Collisions (UPC) of heavy ions where photon-fusion processes are dominant \cite{Bertulani:2005ru, Baltz:2007kq, Contreras:2015dqa, Shao:2022cly}.}, has been studied before in numerical form only in ref.~\cite{Sang:2015uxg}, the other form factors have not been computed before and are thus new. The form factor $gg \leftrightarrow {^3 P_J^{[1]}}$ is part of the virtual NNLO contribution for $\chi_{Q,J}$ hadro-production.

Similarly as in the $S$-wave case in ref.~\cite{Abreu:2022cco}, the two-loop amplitudes can be decomposed into a linear combination of two-loop master integrals via virtue of Integration-By-Parts (IBP) identities \cite{Chetyrkin:1979bj,Chetyrkin:1981qh}. We have studied and computed these two-loop master integrals analytically and provided high-precision numerical results in our companion paper \cite{Abreu:2022vei}. These master integrals can be expressed analytically in terms of special functions as the multiple polylogarithms (MPLs) \cite{GoncharovMixedTate} and their generalisation, the elliptic multiple polylogarithms (eMPLs) \cite{brown2013multiple,Broedel:2014vla,Broedel:2017kkb,ManinModular,Brown:mmv}.

The structure of the paper is built up as follows: In Section~\ref{sec:computationalsetuphelicity} we provide the necessary background to compute the two-loop form factors. In Section~\ref{sec:barecoefficientsdescription}, we discuss the bare amplitude structure, before discussing the UV renormalisation in Section~\ref{sec:UVrenormalis}. In Section~\ref{sec:IRpolestruc}, we outline the infra-red pole structure which turns out to be more involved in the $P$-wave case than in the $S$-wave case. In Section~\ref{sec:formfactors}, we provide the finite remainders and hard functions of the different form factors. We conclude in Section~\ref{sec:conclusions}.

\section{Computational setup and helicity decomposition}
\label{sec:computationalsetuphelicity}

Within the framework of NRQCD factorisation \cite{Bodwin:1994jh}, we can express the partonic cross section of inclusive quarkonium production in the partonic channel $ab$ as
\begin{equation}
    {\rm d}\hat{\sigma}_{ab}{\left(\mathcal{Q}+\{k\}\right)}=\sum_n {\rm d}\hat{\sigma}_{ab}{\left(Q\bar{Q}[n]+\{k\}\right)}\langle \mathcal{O}^n_{\mathcal{Q}} \rangle,
\end{equation}
where $\mathcal{Q}$ is the quarkonium bound state and $\{k\}$ signifies any additional final-state partons. As mentioned in the introduction, the cross section can be decomposed into a perturbative short-distance part, where the heavy quark pair is produced onto a specific quantum configuration $n$, and the non-perturbative LDME $\langle \mathcal{O}^n_{\mathcal{Q}} \rangle$. The sum proceeds over all quantum configurations $n$, but only a few will be relevant for phenomenology.

We can label these quantum configurations via their spectroscopic notation $^{2S+1}L_{J}^{[1,8]}$, where $S$ is the intrinsic spin of the $Q\bar{Q}$ pair, $L$ is the orbital angular momentum and $J$ is the total spin of the bound state. For $S$-wave states, we have that $L=0$, and for $P$-wave states, we have $L=1$. In addition, the superscript $[1,8]$ indicates the colour configuration of the $Q\bar{Q}$ pair, either $[1]$ for colour-singlet or $[8]$ for colour-octet state.

NRQCD factorisation provides a clear separation of the different scales in quarkonium physics. While the perturbative part proceeds at a hard scale on the order of the heavy quark mass, $\mu\sim m_Q$, the non-perturbative LDMEs are guided at the much lower NRQCD scale $\mu_{\Lambda} \sim m_Q\, v$, where $v$ is the relative velocity of the $Q\bar{Q}$ pair in the rest frame of the bound state. For charmonia, we have that $v^2\sim 0.3$, and for bottomonia $v^2 \sim 0.1$. Apart from the expansion in the strong coupling, the short-distance coefficients also admit an expansion in the relative velocity $v^2$ which are called relativistic corrections. However, in this paper, we focus only on the leading term in $v^2$. 

For $\chi_{Q,J}$ production, the LDMEs for the quantum states $^3P_J^{[1]}$ and $^3S_1^{[8]}$ have the same power counting in $v^2$ and hence for phenomenology, it will be necessary to include both contributions together with their corresponding short-distance coefficients. For the process $\gamma \gamma \rightarrow \chi_{Q,J}$, the short-distance coefficient of the $\gamma \gamma \rightarrow {^3S_1^{[8]}}$ trivially vanishes via colour conservation, while for the gluon fusion case, $gg \rightarrow \chi_{Q,J}$, it turns out that the tree-level contribution to the process $gg \rightarrow {^3S_1^{[8]}}$ vanishes as well.\footnote{We compute the one-loop contribution of $gg \leftrightarrow {^3S_1^{[8]}}$ in Appendix~\ref{sec:3S18contribution}.} The leading order contributions to inclusive $\chi_{Q,J}$ hadro-production will then be the $^3P_J^{[1]}$ contribution in the $gg$ channel and the $^3S_1^{[8]}$ contribution in the $q\bar{q}$ channel.\footnote{The form factors $q\bar{q} \leftrightarrow {^3P_J^{[1]}}$ vanish for $J=0$ and are loop induced for the $J=1$ and $J=2$ cases, hence the one-loop amplitude starts to contribute at NNLO only.}

Similarly as was done in the $S$-wave case \cite{Abreu:2022cco}, in this paper, we focus on the bosonic channels $\gamma \gamma$, $\gamma g$ and $gg$. In particular, we focus on the two-loop corrections to the form factors $\gamma \gamma \leftrightarrow {^3 P_J^{[1]}}$, which can be used for the exclusive decay of $\chi_{Q,J}$ to di-photon, the channel $gg \leftrightarrow {^3 P_J^{[1]}}$ that can be used for $\chi_{Q,J}$ hadro-production and hadronic decay.  In addition to this, we also compute the colour-octet form factors, $\gamma g \leftrightarrow {^3 P_J^{[8]}}$ and $gg \leftrightarrow {^3 P_J^{[8]}}$ up to two-loop order. While $\gamma \gamma \leftrightarrow {^3 P_J^{[1]}}$ has been computed numerically before in ref.~\cite{Sang:2015uxg}, the other form factors were previously not known in the literature and are thus new.

In order to compute the short-distance coefficients for these $^3 P_J$ form factors, we consider the process
\begin{equation}
a(k_1) b(k_2) \rightarrow Q{(p+q)}\bar{Q}{(p-q)},
\label{eq:processinit}
\end{equation}
where $a$ and $b$ stand for the initial particles and $Q\bar{Q}$ is the heavy quark pair. In contrast to the $S$-wave case considered in ref.~\cite{Abreu:2022cco}, we have to take into account the relative momentum $q$ of the heavy quark pair. The kinematics for this process read
\begin{equation}
k_1^2=k_2^2=0,\qquad\qquad\qquad k_1 \cdot k_2 = 2 m_Q^2 - 2 q^2,\qquad\qquad\qquad \left(k_1 + k_2\right)\cdot q = 0,
\end{equation}
where we used the fact that $p=\frac{1}{2}\left(k_1 + k_2\right)$ and $m_Q$ is the mass of the heavy quark.

We generate the Feynman diagrams and amplitude for this process in eq.~\eqref{eq:processinit} up to two-loop order using the $\texttt{FeynArts}$ package \cite{Hahn:2000kx} and process its Lorentz and colour structure using the $\texttt{FeynCalc}$ package \cite{Shtabovenko:2016sxi, Shtabovenko:2020gxv}.

In a first step, we will need to project out the $Q\bar{Q}$ pair onto the spin-triplet configuration using the replacement of the heavy quark spinors as \cite{Petrelli:1997ge},\footnote{We can make use of the cyclicity of the trace and the fact that the amplitude is a scalar to write $\bar{u}\mathcal{T}v = \Tr{\left[v \bar{u}\mathcal{T}\right]}$ where $\mathcal{T}$ is the Dirac structure.}
\begin{equation}
v{(p-q)}\, \bar{u}{(p+q)} \rightarrow \frac{1}{\sqrt{8 m_Q^3}} \left(\slashed{p} - \slashed{q} - m_Q\right) \slashed{\varepsilon}_{\mathcal{Q}} \left(\slashed{p} + \slashed{q} + m_Q\right) C_{ij}^{[1,8]},
\end{equation}
where we have defined the colour-projection operator as
\begin{equation}
C_{ij}^{[1,8]}=\begin{cases} \delta_{ij}/\sqrt{N_c} & \text{colour-singlet [1],} \\ \sqrt{2} t_{ij}^{c} & \text{colour-octet [8],} \end{cases}
\end{equation}
to project the heavy quark pair either onto a colour-singlet or a colour-octet state.
In the equation above, $i$ and $j$ are the colour indices of the heavy quark and anti-quark respectively and $c$ is the colour index of the quarkonium bound state in its colour-octet configuration. We have that $N_c=3$.

In a second step, we now need to remove the polarisation vector of the spin-triplet quarkonium state, $\varepsilon_{\mathcal{Q}}^{\mu}$, and expand the amplitude in the relative momentum $q$ to retain its first subleading term. This is achieved by taking the vectorial derivative of the amplitude with respect to $q$ as
\begin{equation}
\mathcal{A}=\varepsilon_{\mu \nu} \mathcal{A}^{\mu \nu}= \varepsilon_{\mu \nu} \left.\frac{\partial}{\partial q_{\nu}} \mathcal{A}^{\mu} \right\vert_{q=0},
\label{eq:amplitudederivative}
\end{equation}
where $\varepsilon_{\mu \nu}$ is the polarisation tensor of the $^3 P_J$ state. We have that, in the case of $J=0$ and $J=2$, $\varepsilon_{\mu \nu}$ is symmetric under exchange of the indices $\mu$ and $\nu$, while in the $J=1$ case, the tensor is anti-symmetric. We have further that $\varepsilon_{\mu \nu}$ is traceless in the $J=2$ case.

After setting $q=0$, the amplitude depends only on a single independent scale, the quark mass $m_Q$, and the kinematics can now be simplified to
\begin{equation}
k_1^2=k_2^2=0,\qquad\qquad\qquad k_1 \cdot k_2 = 2 m_Q^2.
\label{eq:newkinematics}
\end{equation}
In addition to the two independent external four-momenta, $k_1$ and $k_2$, we have two polarisation vectors, $\varepsilon_1$ and $\varepsilon_2$, associated to the initial-state particles $a{\left(k_1\right)}$ and $b{\left(k_2\right)}$ respectively, and the polarisation rank-two tensor, $\varepsilon_{\mu \nu}$, of the $^3 P_J$ state.

Using the axial gauge condition for the vectors $\varepsilon_{i, \mu}$ and the Lorentz gauge condition for the tensor $\varepsilon_{\mu \nu}$, we can put following constraints
\begin{equation}
\begin{split}
\varepsilon_1 \cdot k_1=&\varepsilon_2 \cdot k_2=0,
\\
\varepsilon_1 \cdot k_2=&\varepsilon_2 \cdot k_1=0,
\\
\varepsilon_{\mu \nu} \left(k_1 + k_2\right)^\mu=&\varepsilon_{\mu \nu} \left(k_1 + k_2\right)^\nu=0.
\end{split}
\label{eq:polarisationconstraints}
\end{equation}
We have collected in Appendix~\ref{sec:appendixpolarisation} identities related to the polarisation tensor $\varepsilon_{\mu \nu}$ of the $^3 P_J$ state which are needed in the calculation, as for instance the identity $\sum_{\text{pol}} \varepsilon_{\mu \nu} \varepsilon^*_{\mu' \nu'}$, which depends on the total spin $J$ of the bound state.

Using Lorentz invariance, Bose symmetry, parity conservation and the constraints given in eq.~\eqref{eq:polarisationconstraints}, we can then express the amplitude as a linear combination of three independent coefficients $c_i$
\begin{equation}
\mathcal{A}= \sum_i^3 c_i\,T_i,
\label{eq:helicityamplitude}
\end{equation}
with the following Lorentz structures\footnote{For the process $gg \leftrightarrow {^3P_1^{[8]}}$, the amplitude can in principle also exhibit the anti-symmetric Lorentz structure, $T_{\rm a-s}= \varepsilon_{\mu \nu} \left(\varepsilon_1^\mu \varepsilon_2^\nu - \varepsilon_2^\mu \varepsilon_1^\nu \right)$, where Bose symmetry of the two gluons can be restored via the anti-symmetric colour structure $f^{abc}$. It could only exist for the $J=1$ state which has an anti-symmetric polarisation tensor $\varepsilon_{\mu \nu}$ under exchange of $\mu$ and $\nu$. However, an explicit calculation shows that this form factor vanishes throughout up and including the two-loop order. \label{footnote:check3P18}}
\begin{align}
T_1=&g^{\mu \nu}\; \varepsilon_{\mu \nu}\; \varepsilon_1 \cdot \varepsilon_2,
\\
T_2=&\frac{1}{m_Q^2}\left(k_1^\mu k_1^\nu + k_2^\mu k_2^\nu\right)\; \varepsilon_{\mu \nu}\; \varepsilon_1 \cdot \varepsilon_2,
\\
T_3=&\varepsilon_{\mu \nu}\; \left(\varepsilon_1^\mu \varepsilon_2^\nu + \varepsilon_2^\mu \varepsilon_1^\nu \right).
\end{align}

In order to project out the individual coefficients $c_i$, we can follow the strategy described in refs.~\cite{Peraro:2019cjj, Peraro:2020sfm} and construct the matrix\footnote{The 5 non-vanishing elements of the matrix $M_{ij}$ read
\begin{equation*}
\begin{split}
M_{11} = (d-2) (d-1), \qquad M_{22} = 4 (d-2), \qquad M_{33} = 2 (d-2) (d-1),
\\
M_{31} = M_{13}= 2 (d-2), \qquad M_{12} = M_{21} = -2 (d-2), \qquad M_{23} = M_{32} = 0.
\end{split}
\end{equation*}}
\begin{equation}
M_{ij}=\sum_{\rm pol} T_i^\dagger T_j,
\end{equation}
where the subscript in the sum indicates that we have summed over all polarisation configurations and the different $J$ states.

The projection operator $P_i$ is constructed as follows
\begin{equation}
P_i=\sum_j \left(M\right)^{-1}_{ij} \, T_j^\dagger,
\end{equation}
where $\left(M\right)^{-1}_{ij}$ is the inverse matrix and trivially satisfies
\begin{equation}
\sum_j \left(M\right)^{-1}_{kj} M_{ji}=\delta_{ki}.
\end{equation}
It is then easy to see that applying the projection operator to the amplitude will extract the corresponding coefficient
\begin{equation}
\sum_{\rm pol} P_k\; \mathcal{A} = \sum_i \sum_j c_i\, \left(M\right)^{-1}_{kj} \, \sum_{\rm pol}  T_j^\dagger T_i = \sum_i c_i\, \delta_{ki}=c_k.
\end{equation}

In order to facilitate this step above in the calculation, we can isolate the polarisations from the amplitude and absorb them into the projection operator,
\begin{equation}
\sum_{\rm pol} P_k\; \mathcal{A}= \sum_{\rm pol} \varepsilon^{\mu \nu}\,\varepsilon_{1}^{\rho}\,\varepsilon_{2}^{\sigma}\, P_k\, \mathcal{A}_{\mu \nu \rho \sigma} = P_k^{\mu \nu \rho \sigma}\, \mathcal{A}_{\mu \nu \rho \sigma} = c_k,
\end{equation}
where the projection tensors can be written as
\begin{equation}
    P_i^{\mu \nu \rho \sigma} = \sum_{\text{pol}} \varepsilon^{\mu \nu}\,\varepsilon_{1}^{\rho}\,\varepsilon_{2}^{\sigma}\, P_i = \sum_j \left(M\right)^{-1}_{ij} \, \tilde{T}_j^{\mu \nu \rho \sigma},
\end{equation}
using the three $\tilde{T}_j$ tensors defined as
\begin{equation}
\begin{split}
\tilde{T}_1^{\mu \nu \rho \sigma} = & \left(g^{\mu \nu} - \frac{1}{4 m_Q^2} \left(k_1^\mu+k_2^\mu\right) \left(k_1^\nu+k_2^\nu\right)\right) \left(g^{\rho \sigma} - \frac{1}{2 m_Q^2} k_2^\rho k_1^\sigma - \frac{1}{2 m_Q^2} k_1^\rho k_2^\sigma \right),
\\
\tilde{T}_2^{\mu \nu \rho \sigma} = & \frac{1}{2m_Q^2}\left(k_1^\mu - k_2^\mu \right)\left(k_1^\nu-k_2^\nu\right) \left(g^{\rho \sigma} - \frac{1}{2 m_Q^2} k_2^\rho k_1^\sigma - \frac{1}{2 m_Q^2} k_1^\rho k_2^\sigma \right),
\\
\tilde{T}_3^{\mu \nu \rho \sigma} = & \left[\left( g^{\mu \rho} - \frac{1}{2 m_Q^2} k_1^\mu k_2^\rho - \frac{1}{2 m_Q^2} k_2^\mu k_1^\rho \right) \left( g^{\nu \sigma} - \frac{1}{2 m_Q^2} k_2^\nu k_1^\sigma - \frac{1}{2 m_Q^2} k_1^\nu k_2^\sigma \right) \right.
\\
& \left. + \left( g^{\mu \sigma} - \frac{1}{2 m_Q^2} k_1^\mu k_2^\sigma - \frac{1}{2 m_Q^2} k_2^\mu k_1^\sigma \right) \left( g^{\nu \rho} - \frac{1}{2m_Q^2} k_1^\nu k_2^\rho - \frac{1}{2m_Q^2} k_2^\nu k_1^\rho \right) \right].
\end{split}
\end{equation}

From these three coefficients $c_i$ in eq.~\eqref{eq:helicityamplitude} we can then construct the helicity amplitudes for the different polarisation states of the initial and final-state particles. We make use of the explicit representations of the polarisation vectors and tensors in Appendix~\ref{sec:appendixpolarisation}. We will denote the helicity configurations as $\mathcal{A}^{(J)}_{J_z,\lambda_1,\lambda_2}$. The non-vanishing configurations read\footnote{As expected from the Landau-Yang theorem \cite{Landau:1948kw,Yang:1950rg}, the amplitude for the $^3P_1$ decay to two massless particles vanishes. See also footnote~\ref{footnote:check3P18}.}
\begin{align}
\mathcal{A}^{(2)}_{+2,-,+} =\; \mathcal{A}^{(2)}_{-2,+,-} = \mathcal{A}_1 =&\; 2 c_3,
\label{eq:helicity1}
\\
\mathcal{A}^{(2)}_{0,+,+} =\; \mathcal{A}^{(2)}_{0,-,-} = \mathcal{A}_2 =&\; \frac{1}{\sqrt{\left(d-2\right)\left(d-1\right)}}\big(2\left(d-2\right) c_2 + 2 c_3 \big),
\label{eq:helicity2}
\\
\mathcal{A}^{(0)}_{0,+,+} =\; \mathcal{A}^{(0)}_{0,-,-} = \mathcal{A}_3 =&\; \frac{1}{\sqrt{d-1}}\big(\left(1-d\right) c_1 + 2 c_2 -2 c_3 \big).
\label{eq:helicity3}
\end{align}

Having described the setup of the amplitude generation, we can compute each coefficient $c_i$ up to two-loop order. Due to the kinematics described in eq.~\eqref{eq:newkinematics}, some of the Feynman integrals will exhibit linearly dependent propagators as was already encountered in the $^1S_0$-state case \cite{Abreu:2022cco}. We eliminate these dependences by performing partial-fraction decomposition using the package $\texttt{Apart}$ \cite{Feng:2012iq}. We can then make use of the \textit{Integration-By-Parts (IBP)} reduction technique \cite{Chetyrkin:1979bj,Chetyrkin:1981qh} by using packages as $\texttt{FIRE}$ \cite{Smirnov:2008iw,Smirnov:2014hma,Smirnov:2019qkx,Smirnov:2023yhb}, \texttt{LiteRed} \cite{Lee:2012cn,Lee:2013mka} and $\texttt{Kira}$ \cite{Maierhofer:2017gsa} to reduce the Feynman integrals to the master integrals.

The two-loop bare amplitudes for the $^3P_J$ form factors can be written as a linear combination of the same 76 master integrals that already appeared in the $^1S_0$ case \cite{Abreu:2022cco}.\footnote{Due to the derivative in the relative momentum $q$ in eq.~\eqref{eq:amplitudederivative}, the two-loop integrals before IBP appear with raised propagator powers and numerator terms. This creates, in principle, the possibility of emergence of new master integrals beyond the ones considered in the $^1S_0$ state case \cite{Abreu:2022cco, Abreu:2022vei}. However, we confirm that the set of integrals that appear is the same for the $^1S_0$ and $^3P_J$ states. This may apply presumably also to $C$-even quarkonia with higher $L$ states such as $D$ waves ($L=2$).} These master integrals have been computed both analytically and numerically to high precision in our companion paper \cite{Abreu:2022vei}. The analytical structure of the integrals exhibit in addition to the functions in the class of \textit{multiple polylogarithms} \cite{GoncharovMixedTate}, also elliptic functions in the class of \textit{elliptic multiple polylogarithms} \cite{brown2013multiple,Broedel:2014vla,Broedel:2017kkb} and \textit{iterated Eisenstein integrals} \cite{ManinModular,Brown:mmv}.\footnote{As the amplitudes can be expressed in terms of the same set of master integrals, we make the same observation already done in the $^1S_0$ case, that the amplitude exhibits polylogarithms up to weight $w=4$ and up to length $l=4$ for the elliptic functions.}

\section{Bare amplitude structure}
\label{sec:barecoefficientsdescription}

In the following, we will define the structure of the bare coefficients $c_i$ as an expansion in the bare coupling $\alpha_s^B$ up to two-loop order as
\begin{equation}
    \mathcal{F}_{p,c_i}=c_i/\mathcal{N}_{p, c_i}=\left(\frac{\alpha_s^B}{\pi}\right)^q \left[\mathcal{F}_{p,c_i}^{(0)}+\left(\frac{\alpha_s^B}{\pi}\right) \mathcal{F}_{p,c_i}^{(1)}+\left(\frac{\alpha_s^B}{\pi}\right)^2 \mathcal{F}_{p,c_i}^{(2)} + \mathcal{O}{\left({\alpha_s^B}\right)^3}\right],
\end{equation}
where the subscripts $p$ and $c_i$ correspond to the process and the coefficient under consideration. The exponent $q$ in the equation above indicates the leading power in the strong coupling. We have that $q=0$ for the $\gamma \gamma$ channel, $q=1/2$ for the $\gamma g$ channel and $q=1$ for the $gg$ channel. The factor $\mathcal{N}_{p, c_i}$ is a normalisation factor chosen to normalise $\mathcal{F}_{p,c_i}^{(0)}=1$. However, in the case of the $c_1$ coefficient, we find that the tree-level contribution vanishes,
\begin{equation}
    \mathcal{F}_{p,c_1}^{(0)} = 0, \qquad\qquad\qquad \mathcal{F}_{p,c_2}^{(0)} = 1, \qquad\qquad\qquad \mathcal{F}_{p,c_3}^{(0)} = 1.
\end{equation}
Hence, for this case, we choose the same normalisation as for the $c_2$ coefficient, $\mathcal{N}_{p, c_1}=\mathcal{N}_{p, c_2}$.

We can then write this normalisation factor as
\begin{equation}
    \mathcal{N}_{p, c_i}=-\frac{4\sqrt{2}\,\pi^2}{m_Q^{3/2}}\, \mathcal{C}_{c_i}^{\text{coef}} \, \mathcal{C}_{p}^{\text{col.em.}},
    \label{eq:gloablfactorcoefficient1}
\end{equation}
where
\begin{equation}
    \mathcal{C}_{c_i}^{\text{coef}}=\begin{cases} 1 & i=1, \\ 1 & i=2, \\ -2 & i=3, \end{cases} \qquad\qquad  \mathcal{C}_{p}^{\text{col.em.}}=\begin{cases} \sqrt{N_c}\, e_Q^2\, \alpha_{em}/\pi & \gamma \gamma \leftrightarrow {^3P_J^{[1]}}, \\ T_F\, \delta^{ab}/\sqrt{N_c} & gg \leftrightarrow {^3P_J^{[1]}}, \\ \sqrt{2}\,T_F\, \delta^{bc}\, e_Q \sqrt{\alpha_{em}/\pi} & \gamma g \leftrightarrow {^3P_J^{[8]}}, \\ \sqrt{2}\,T_F\, d^{abc}/2 & gg \leftrightarrow {^3P_J^{[8]}}.  \end{cases}
    \label{eq:gloablfactorcoefficient2}
\end{equation}
In the factor $\mathcal{C}_{p}^{\text{col.em.}}$ above, $\alpha_{em}$ is the electromagnetic fine-structure constant and $e_Q$ is the fractional charge of the quark. We have that $e_c=2/3$ and $e_b=-1/3$. As for definition of the colour factors, we have that
\begin{equation}
    \Tr{\left[t^a t^b\right]} = T_F \delta^{ab}, \qquad\qquad \Tr{\left[t^a t^b t^c\right]} = T_F \frac{1}{2} \left(d^{abc}+i f^{abc}\right),
\end{equation}
where $d^{abc}$ and $f^{abc}$ are the symmetric and anti-symmetric tensors under interchange of the colour indices respectively, and we have that $T_F=1/2$.

At one-loop and two-loop level, the amplitude will exhibit divergences in the dimensional regulator $\epsilon$. We can write the coefficients at $n$-loop order as
\begin{equation}
    \mathcal{F}_{p,c_i}^{(n)} = S_{\epsilon}^{n} \left(m_Q^2\right)^{-n\epsilon} \sum_{k \geq -2n} \epsilon^k\, \mathcal{F}_{p,c_i}^{(n,k)},
\end{equation}
where $S_{\epsilon}=\left(4\pi\right)^{4\epsilon}e^{-\epsilon \gamma_E}$.
At one-loop level, we can further structure the form factor $\mathcal{F}_{p,c_i}^{(1)}$ by their colour composition as
\begin{equation}
    \mathcal{F}_{p,c_i}^{(1,k)} = C_F\, \mathcal{F}_{p,c_i; F}^{(1,k)} + C_A\, \mathcal{F}_{p,c_i; A}^{(1,k)},
\end{equation}
where $C_F$ and $C_A$ are the Casimir invariants of $\text{SU}{(N_c)}$,
\begin{equation}
    t_{ik}^a t_{kj}^b = C_F\, \delta_{ij}=\frac{N_c^2-1}{2N_c}\,\delta_{ij} \quad\text{and}\quad f^{acd} f^{bcd}=C_A\, \delta^{ab}=N_c\, \delta^{ab}.
\end{equation}

\begin{figure}
  \begin{center}
    \subfloat[]{\includegraphics[width=4.5cm]{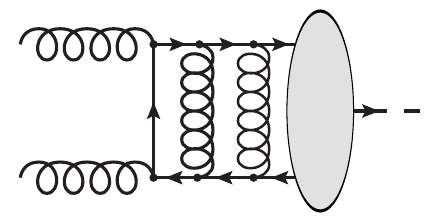}
    \label{fig:ggabelian}}\, 
    \subfloat[]{\includegraphics[width=4.9cm]{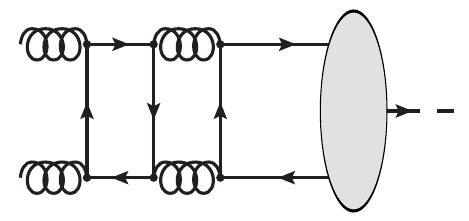} 
    \label{fig:gglbl}}\, 
    \subfloat[]{\includegraphics[width=4.5cm]{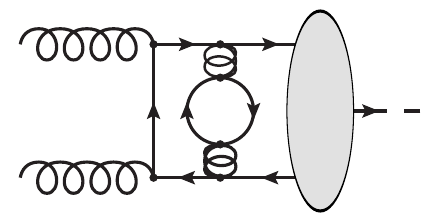} 
    \label{fig:ggvac}}
  \end{center}
  \caption{Two-loop diagrams for the form factor 
  $g g \leftrightarrow {^3P_J^{[1,8]}}$ with 
  (a) regular contributions, (b) light-by-light contributions and 
  (c) vacuum polarisation contributions.}
  \label{fig:typesdiagrams}
\end{figure}
At two-loop level, we can express the amplitude into different type of contributions that are each \textit{gauge invariant}. We have \textit{regular} contributions, $\mathcal{F}_{p,c_i;\text{reg}}$, representing contributions without any fermion loops as shown in Fig.~\ref{fig:ggabelian}, then \textit{light-by-light} contributions, $\mathcal{F}_{p,c_i;\text{lbl}}$, where the closed fermion loop is connected to both external bosons as in Fig.~\ref{fig:gglbl}, and \textit{vacuum polarisation} contributions, $\mathcal{F}_{p,c_i;\text{vac}}$, where the fermion loop is inserted either in the gluon propagator or the triple gluon vertex as in Fig.~\ref{fig:ggvac}. The amplitude then reads
\begin{equation}
    \mathcal{F}_{p,c_i}^{(2,k)} = \mathcal{F}_{p,c_i;\text{reg}}^{(2,k)} + \mathcal{F}_{p,c_i;\text{lbl}}^{(2,k)} + \mathcal{F}_{p,c_i;\text{vac}}^{(2,k)},
\end{equation}
with
\begin{align}
    \mathcal{F}_{p,c_i;\text{reg}}^{(2,k)} =& C_F^2\, \mathcal{F}_{p,c_i;FF}^{(2,k)} + C_F C_A\, \mathcal{F}_{p,c_i;FA}^{(2,k)} + C_A^2\, \mathcal{F}_{p,c_i;AA}^{(2,k)},
    \\[5pt]
    \begin{split}
    \mathcal{F}_{p,c_i;\text{lbl}}^{(2,k)} =& C_F T_F n_h\, \mathcal{F}_{p,c_i;Fh;\text{lbl}}^{(2,k)} + C_F T_F \tilde{n}_l\, \mathcal{F}_{p,c_i;Fl;\text{lbl}}^{(2,k)}
    \\
    & + C_A T_F n_h\, \mathcal{F}_{p,c_i;Ah;\text{lbl}}^{(2,k)} + C_A T_F \tilde{n}_l\, \mathcal{F}_{p,c_i;Al;\text{lbl}}^{(2,k)},
    \end{split}
    \\[5pt]
    \begin{split}
    \mathcal{F}_{p,c_i;\text{vac}}^{(2,k)} =& C_F T_F n_h\, \mathcal{F}_{p,c_i;Fh;\text{vac}}^{(2,k)} + C_F T_F n_l\, \mathcal{F}_{p,c_i;Fl;\text{vac}}^{(2,k)}
    \\
    & + C_A T_F n_h\, \mathcal{F}_{p,c_i;Ah;\text{vac}}^{(2,k)} + C_A T_F n_l\, \mathcal{F}_{p,c_i;Al;\text{vac}}^{(2,k)}.
    \end{split}
\end{align}
In the expression above, $n_h$ indicates the contributions with a massive fermion loop of same flavour as the heavy quark forming the bound state, and $n_l$ indicates the contribution with massless fermion loops. We have also introduced the factor $\tilde{n}_l$ in the light-by-light contribution to account for the different electric charge of the massless quark flavours,
\begin{equation}
    \tilde{n}_l = \begin{cases} \sum_{i}^{n_l} e_i^2/e_Q^2 & \gamma \gamma \quad \text{channel}, \\ \sum_{i}^{n_l} e_i/e_Q & \gamma g \quad \text{channel}, \\ n_l & gg \quad\text{channel}. \end{cases}
    \label{eq:tildenldefinition}
\end{equation}

We make the same observations already done in the $^1S_0$ case \cite{Abreu:2022cco} that, for a given form factor $c_i$, all abelian type corrections, $\mathcal{F}^{(1)}_{F}$ at one-loop order and $\mathcal{F}^{(2)}_{FF}$, $\mathcal{F}^{(2)}_{Fh;\text{vac}}$, $\mathcal{F}^{(2)}_{Fl;\text{vac}}$ at two-loop order are identical for all channels considered. As for the abelian coefficients in the light-by-light contributions, $\mathcal{F}^{(2)}_{Fh;\text{lbl}}$ and $\mathcal{F}^{(2)}_{Fl;\text{lbl}}$, they are the same for channels, where the quarkonium is in the same colour-configuration, colour-singlet $[1]$ and colour-octet $[8]$, but differ between the $[1]$ and $[8]$ states by a factor of two as
\begin{equation}
    \mathcal{F}^{(2), [8]}_{Fh;\text{lbl}}=2 \times \mathcal{F}^{(2), [1]}_{Fh;\text{lbl}}, \qquad\qquad \mathcal{F}^{(2), [8]}_{Fl;\text{lbl}}=2 \times \mathcal{F}^{(2), [1]}_{Fl;\text{lbl}}.
\end{equation}

We have collected in Appendix~\ref{sec:barepolestructure}, all colour-structure coefficients for the different form factors $c_i$ and the different processes, $\gamma \gamma \leftrightarrow {^3 P_J^{[1]}}$, $gg \leftrightarrow {^3 P_J^{[1]}}$ , $\gamma g \leftrightarrow {^3 P_J^{[8]}}$ and $gg \leftrightarrow {^3 P_J^{[8]}}$, at one-loop level up to $\mathcal{O}{\left(\epsilon^2\right)}$ and at two-loop level up to $\mathcal{O}{\left(\epsilon^0\right)}$.

\section{UV renormalisation}
\label{sec:UVrenormalis}

In this section, we discuss the UV renormalisation of the form factors and follow closely the procedure discussed in ref.~\cite{Abreu:2022cco}. We employ the on-shell renormalisation scheme for the gluon wavefunction, the heavy-quark wavefunction and the heavy-quark mass with the renormalisation factors $Z_g$, $Z_Q$ and $Z_m$ respectively, while for the strong coupling constant we employ the \msbar-scheme with $Z_{\alpha_s}$ \cite{Barnreuther:2013qvf, Broadhurst:1993mw, Melnikov:2000qh, Mitov:2006xs}.

The $Z_{\kappa}$ factors can be expanded in the renormalised strong coupling constant as follows
\begin{equation}
    Z_{\kappa}=1+\left(\frac{\alpha_s^{\left(n_f\right)}}{\pi}\right) Z_{\kappa}^{(1)} + \left(\frac{\alpha_s^{\left(n_f\right)}}{\pi}\right)^2 Z_{\kappa}^{(2)} + \mathcal{O}{\left({\alpha_s^{(n_l)}}\right)^3},
    \label{eq:Zexpansion}
\end{equation}
where the running of $\alpha_s$ involves $n_f=n_h+n_l$ quark flavours. The relation between the bare and the renormalised coupling in the \msbar-scheme reads
\begin{equation}
    \alpha_s^B = S_{\epsilon}^{-1} \mu_R^{2\epsilon} Z_{\alpha_s} \alpha_s^{\left(n_f\right)}.
\end{equation}
In order to keep only the number of light quark flavours, $n_l$, in the running of the coupling, we can use the following decoupling relation \cite{Bernreuther:1981sg}
\begin{equation}
    \alpha_s^{\left(n_l+n_h\right)} = \zeta_{\alpha_s} \alpha_s^{\left(n_l\right)}.
\end{equation}
Similarly as in eq.~\eqref{eq:Zexpansion}, the decoupling factor $\zeta_{\alpha_s}$ can be expanded in the renormalised strong coupling with $n_l$ flavours in the evolution. The coefficients for the renormalisation factors, $Z_i$, and the decoupling factor, $\zeta_{\alpha_s}$, are given in Appendix C of our companion paper in ref.~\cite{Abreu:2022cco}.

We can expand the renormalised form factor in terms of the renormalised coupling as follows
\begin{equation}
\begin{split}
    \overline{\mathcal{F}}_{p,c_i}=& \overline{c}_i/\mathcal{N}_{p, c_i} 
    \\
    =&\left(\frac{\alpha_s^{(n_l)}}{\pi}\right)^q \left[\overline{\mathcal{F}}_{p,c_i}^{(0)}+\left(\frac{\alpha_s^{(n_l)}}{\pi}\right) \overline{\mathcal{F}}_{p,c_i}^{(1)}+\left(\frac{\alpha_s^{(n_l)}}{\pi}\right)^2 \overline{\mathcal{F}}_{p,c_i}^{(2)} + \mathcal{O}{\left({\alpha_s^{(n_l)}}\right)^3}\right],
    \end{split}
    \label{eq:UVrenormalisationcifactor}
\end{equation}
where we have that at $n$-loop order
\begin{equation}
    \overline{\mathcal{F}}_{p,c_i}^{(n)}=\mu_R^{2n\epsilon}S_{\epsilon}^{-n}\mathcal{F}_{p,c_i}^{(n)}+\mathcal{F}_{p,c_i}^{(n,\textrm{CT})}+\mathcal{F}_{p,c_i}^{(n,\textrm{decoupling})}.
\label{eq:UVrenormalisationtermatordern}
\end{equation}
The terms $\mathcal{F}_{p,c_i}^{(n,\textrm{CT})}$ are the counterterm expressions expanded in $\alpha_s^{(n_f)}$ with $n_f=n_l+n_h$ flavours in the running of the coupling. In order to convert to the scheme where we keep only $n_l$ flavours in the coupling, $\alpha_s^{(n_l)}$, we need to add the decoupling contribution $\mathcal{F}_{p,c_i}^{(n,\textrm{decoupling})}$ which can be expressed as
\begin{align}
    \mathcal{F}_{p,c_i}^{(1,\textrm{decoupling})} = & q \zeta_{\alpha_s}^{(1)} \mathcal{F}_{p,c_i}^{(0)},
    \\[5pt]
    \begin{split}
    \mathcal{F}_{p,c_i}^{(2,\textrm{decoupling})} = & \left( q \zeta_{\alpha_s}^{(2)} + \frac{1}{2}q\left(q-1\right)\left(\zeta_{\alpha_s}^{(1)}\right)^2 \right) \mathcal{F}_{p,c_i}^{(0)}
    \\
    &+ \left(q+1\right)\zeta_{\alpha_s}^{(1)}\left(\overline{\mathcal{F}}_{p,c_i}^{(1)}-\mathcal{F}_{p,c_i}^{(1,\textrm{decoupling})}\right).
    \end{split}
\end{align}

We can write the one-loop and two-loop counterterms as follows
\begin{align}
    \mathcal{F}_{p,c_i}^{(1, \textrm{CT})}=& \left(q \left(Z_g^{(1)}+Z_{\alpha_s}^{(1)}\right)+Z_Q^{(1)}\right) \mathcal{F}_{p,c_i}^{(0)} +Z_m^{(1)} \mathcal{F}_{p,c_i}^{(0, 1 \textrm{ mass CT})},
    \\[5pt]
    \label{eq:renormalisationCT1expression}
    \begin{split}
    \mathcal{F}_{p,c_i}^{(2, \textrm{CT})}=& S_{\epsilon}^{-1} \mu_R^{2\epsilon} \mathcal{F}_{p,c_i}^{(1)} \left[q Z_g^{(1)}+\left(1+q\right)Z_{\alpha_s}^{(1)}
    +Z_Q^{(1)}\right]+\mathcal{F}_{p,c_i}^{(0)} \bigg[ q Z_{\alpha_s}^{(1)}\left(q Z_g^{(1)}+Z_Q^{(1)}\right)
    \\
    &+ q Z_{\alpha_s}^{(2)}+q Z_g^{(2)}+Z_Q^{(2)}+q Z_g^{(1)} Z_Q^{(1)} + \frac{1}{2}q\left(q-1\right)\left[\left(Z_{\alpha_s}^{(1)}\right)^2
    +\left(Z_{g}^{(1)}\right)^2\right] \bigg]
    \\
    & + \left[ \left(q Z_{\alpha_s}^{(1)} + q Z_g^{(1)} + Z_Q^{(1)}\right) Z_m^{(1)} + Z_m^{(2)} \right] \mathcal{F}_{p,c_i}^{(0, 1 \text{ mass CT})}
    \\
    &+\left(Z_m^{(1)}\right)^2 \mathcal{F}_{p,c_i}^{(0, 2 \text{ mass CT})} + Z_m^{(1)}\mathcal{F}_{p,c_i}^{(1, 1 \text{ mass CT})},
    \end{split}
\end{align}
where $\mathcal{F}_{p,c_i}^{(n, k \text{ mass CT})}$ refers to the $n$-loop contribution with $k$ mass counterterm insertions. In general, we find that these mass counterterm expressions depend on the tensor structure $c_i$. For instance, for the $k=1$ and $k=2$ mass counterterm coefficients at tree-level, $n=0$, we find that\footnote{In the pseudo-scalar case, $^1S_0$, discussed in ref.~\cite{Abreu:2022cco}, we wrote out the mass counterterm coefficients explicitly as we have only a single Lorentz structure there. We identify $\mathcal{F}_{^1S_0}^{(0, 1 \text{ mass CT})}=-1$ and $\mathcal{F}_{^1S_0}^{(0, 2 \text{ mass CT})}=\frac{1}{2}$.}
\begin{equation}
    \mathcal{F}_{p,c_i}^{(0, 1 \text{ mass CT})} = \begin{cases} -2, & i=1, \\ -2, & i=2, \\ -1, & i=3,\end{cases} \qquad\qquad \mathcal{F}_{p,c_i}^{(0, 2 \text{ mass CT})} = \begin{cases} 2, & i=1, \\ 2, & i=2, \\ \frac{1}{2}, & i=3.\end{cases}
\end{equation}

We can expand the renormalised form factors in the dimensional regulator $\epsilon$ at one-loop and two-loop level as
\begin{equation}
    \overline{\mathcal{F}}_{p,c_i}^{(1)}=\sum_{k\geq -2} \epsilon^k \overline{\mathcal{F}}_{p,c_i}^{(1,k)},\qquad\qquad    \overline{\mathcal{F}}_{p,c_i}^{(2)}=\sum_{k\geq -4} \epsilon^k \overline{\mathcal{F}}_{p,c_i}^{(2,k)}.
\label{eq:expandamplitudeepsorder}
\end{equation}
We can further organise the form factors via their colour structure. At one-loop level, the structure now reads
\begin{equation}
    \overline{\mathcal{F}}_{p,c_i}^{(1,k)} = C_A\, \overline{\mathcal{F}}_{p,c_i;A}^{(1,k)} + C_F\, \overline{\mathcal{F}}_{p,c_i;F}^{(1,k)} + T_F n_l\, \overline{\mathcal{F}}_{p,c_i;l}^{(1,k)}.
\end{equation}

At two-loop level, we can divide the colour structure as follows
\begin{equation}
    \overline{\mathcal{F}}_{p,c_i}^{(2,k)} = \overline{\mathcal{F}}_{p,c_i;\text{reg}}^{(2,k)} + \overline{\mathcal{F}}_{p,c_i;\text{lbl}}^{(2,k)} + \overline{\mathcal{F}}_{p,c_i;\text{vac}}^{(2,k)},
\end{equation}
with\footnote{We note that, for the bare amplitudes in the $gg$ channel, the light-by-light contributions proportional to $C_A$ contain a simple pole in $\epsilon$ for $c_1$ and $c_3$ (see Appendix~\ref{sec:barepolestructure}). This can be traced back to the fact that the light-by-light contributions are effectively a loop correction to the amplitude with the four-gluon vertex. For the $c_2$ case, the amplitude with the four-gluon vertex vanishes, hence the absence of singularities there. In order to facilitate the renormalisation step, for the renormalised amplitudes, we keep the finite piece of the light-by-light contributions and group all other renormalisation contributions into the vacuum polarisation part.}
\begin{align}
    \overline{\mathcal{F}}_{p,c_i;\text{reg}}^{(2,k)} =& C_F^2\, \overline{\mathcal{F}}_{p,c_i;FF}^{(2,k)} + C_F C_A\, \overline{\mathcal{F}}_{p,c_i;FA}^{(2,k)} + C_A^2\, \overline{\mathcal{F}}_{p,c_i;AA}^{(2,k)},
    \\[5pt]
    \begin{split}
    \overline{\mathcal{F}}_{p,c_i;\text{lbl}}^{(2,k)} =& C_F T_F n_h\, \overline{\mathcal{F}}_{p,c_i;Fh;\text{lbl}}^{(2,k)} + C_F T_F \tilde{n}_l\, \overline{\mathcal{F}}_{p,c_i;Fl;\text{lbl}}^{(2,k)}
    \\
    & + C_A T_F n_h\, \overline{\mathcal{F}}_{p,c_i;Ah;\text{lbl}}^{(2,k)} + C_A T_F \tilde{n}_l\, \overline{\mathcal{F}}_{p,c_i;Al;\text{lbl}}^{(2,k)},
    \end{split}
    \\[5pt]
    \begin{split}
    \overline{\mathcal{F}}_{p,c_i;\text{vac}}^{(2,k)} =& C_F T_F n_h\, \overline{\mathcal{F}}_{p,c_i;Fh;\text{vac}}^{(2,k)} + C_F T_F n_l\, \overline{\mathcal{F}}_{p,c_i;Fl;\text{vac}}^{(2,k)}
    \\
    & + C_A T_F n_h\, \overline{\mathcal{F}}_{p,c_i;Ah;\text{vac}}^{(2,k)} + C_A T_F n_l\, \overline{\mathcal{F}}_{p,c_i;Al;\text{vac}}^{(2,k)} + T_F^2 n_l^2\, \overline{\mathcal{F}}_{p,c_i;ll}^{(2,k)}.
    \end{split}
\end{align}

Similarly as in the case of the bare amplitude in the previous section, we have collected the coefficients for the renormalised form factors in Appendix~\ref{sec:renpole}. We observe that after UV renormalisation, in the $\gamma\gamma$ channel, we are left with only a simple pole in $\epsilon$ at two-loop level. In the case of the form factors in the $gg$ channel, the pole structure of the two-loop amplitude starts at the quadruple order in $\epsilon$. We make here the striking observation that, at the simple pole, there is a term proportional to the number of heavy flavours $n_h$. As this gauge structure appears for the first time at two-loop level, this feature cannot be explained a priori by standard QCD. The meaning and origin of this term will become clear in the next section.

\section{Infrared singularities}
\label{sec:IRpolestruc}

Within the framework of NRQCD, the pole structure involves in addition to standard QCD singularities also NRQCD singularities that have to be absorbed into the bound state wavefunction. As such, we can no longer focus on the individual form factor coefficients $c_i$, but need to discuss the helicity form factors $\mathcal{A}_i$ discussed in eqs.~\eqref{eq:helicity1}, \eqref{eq:helicity2} and \eqref{eq:helicity3} because these directly correspond to the different $^3P_J{\left(J_z\right)}$ states. Using the UV-renormalised form factors $\overline{c}_i$ defined in eq.~\eqref{eq:UVrenormalisationcifactor}, we can then write
\begin{equation}
\begin{split}
\overline{\mathcal{A}}_1 =&\; 2 \overline{c}_3,
\\
\overline{\mathcal{A}}_2 =&\; \frac{1}{\sqrt{\left(d-2\right)\left(d-1\right)}}\big(2\left(d-2\right) \overline{c}_2 + 2 \overline{c}_3 \big),
\\
\overline{\mathcal{A}}_3 =&\; \frac{1}{\sqrt{d-1}}\big(\left(1-d\right) \overline{c}_1 + 2 \overline{c}_2 -2 \overline{c}_3 \big).
\end{split}
\end{equation}

In order to keep the notation and discussion as close as possible to the previous sections, we denote the normalised helicity form factors as
\begin{equation}
\begin{split}
    \overline{\mathcal{F}}_{p,\mathcal{A}_i}=&\overline{\mathcal{A}}_i/\mathcal{N}_{p,\mathcal{A}_i}
    \\
    =&\left(\frac{\alpha_s^{(n_l)}}{\pi}\right)^q \left[\overline{\mathcal{F}}_{p,\mathcal{A}_i}^{(0)}+\left(\frac{\alpha_s^{(n_l)}}{\pi}\right) \overline{\mathcal{F}}_{p,\mathcal{A}_i}^{(1)}+\left(\frac{\alpha_s^{(n_l)}}{\pi}\right)^2 \overline{\mathcal{F}}_{p,\mathcal{A}_i}^{(2)} + \mathcal{O}{\left({\alpha_s^{(n_l)}}\right)^3}\right],
\end{split}
\label{eq:definitionfinitexpanAi}
\end{equation}
with the following normalisation
\begin{equation}
    \mathcal{N}_{p, \mathcal{A}_i}=-\frac{4\sqrt{2}\,\pi^2}{m_Q^{3/2}}\, \mathcal{C}_{\mathcal{A}_i}^{\text{coef}} \, \mathcal{C}_{p}^{\text{col.em.}}, \qquad\qquad
    \mathcal{C}_{\mathcal{A}_i}^{\text{coef}}=\begin{cases} -4 & i=1, \\ 12/\sqrt{6} & i=2, \\ 6/\sqrt{3} & i=3, \end{cases}
    \label{eq:gloablfactorcoefficientAi}
\end{equation}
where $\mathcal{C}_{p}^{\text{col.em.}}$ was already given in eq.~\eqref{eq:gloablfactorcoefficient2}. In this decomposition, we have that the tree-level contribution to $\mathcal{A}_2$ vanishes at $\epsilon=0$,
\begin{equation}
    \left.\overline{\mathcal{F}}_{p,\mathcal{A}_1}^{(0)}\right\vert_{\epsilon=0}=1, \qquad\qquad\qquad \left.\overline{\mathcal{F}}_{p,\mathcal{A}_2}^{(0)}\right\vert_{\epsilon=0}=0, \qquad\qquad\qquad \left.\overline{\mathcal{F}}_{p,\mathcal{A}_3}^{(0)}\right\vert_{\epsilon=0}=1.
\end{equation}

We can then construct the finite remainder of the form factors as follows
\begin{equation}
    \mathcal{F}^{\text{fin}}_{p,\mathcal{A}_i}=\mathbf{Z}^{-1}_{\text{NRQCD}}\mathbf{Z}^{-1}_{\text{IR}}\, \overline{\mathcal{F}}_{p,\mathcal{A}_i},
\end{equation}
where $\mathbf{Z}_{\text{IR}}$ is the standard QCD subtraction factor and $\mathbf{Z}_{\text{NRQCD}}$ subtracts the remaining NRQCD singularities into the bound state. Both factors, $\mathbf{Z}_{\rm IR}$ and $\mathbf{Z}_{\rm NRQCD}$, can be expanded in the strong coupling constant $\alpha_s^{(n_l)}$ similarly as the renormalisation factors $Z_{\kappa}$.\footnote{However, the expansion proceeds at different scales in the coupling. While the renormalised form factors are expanded at the renormalisation scale $\mu_R$, the expansion proceeds for $\mathbf{Z}_{\text{IR}}$ factor at scale $\mu_F$ and for the $\mathbf{Z}_{\text{NRQCD}}$ factor at the NRQCD scale $\mu_{\Lambda}$. We employ the renormalisation group equation for the strong coupling to evolve its scale to the renormalisation scale $\mu_R$. \label{footnote:scaledependeevol}} For a detailed discussion on the construction of the $\mathbf{Z}_{\text{IR}}$ factor, we guide the reader to our companion paper \cite{Abreu:2022cco}.

In contrast to the $^1S_0$ form factor case, the cancellation of the NRQCD singularities is more involved in the $^3P_J$ case. After subtracting the standard IR singularities using the $\mathbf{Z}_{\text{IR}}$ factor we are left with a simple pole at two-loop order $\mathcal{O}{\left(\alpha_s^{q+2}\right)}$,
\begin{equation}
    \mathbf{Z}^{-1}_{\text{IR}}\, \overline{\mathcal{F}}_{p,\mathcal{A}_i} = \left(\frac{\alpha_s^{(n_l)}}{\pi}\right)^{q+2} \frac{1}{\epsilon} \gamma_{{\rm NRQCD}}^{p,\mathcal{A}_i} + \mathcal{O}{\left(\epsilon^0, \alpha_s^{q+3}\right)}.
\end{equation}

In the $\gamma \gamma$ channel, where we have trivially $\mathbf{Z}_{\text{IR}}=1$, and in the $\gamma g$ channel, where the final state is in colour-octet state, we find that the remaining singularity is indeed the Coulomb singularity that has to be absorbed into the bound state wave function, similarly as was done in ref.~\cite{Abreu:2022cco},
\begin{equation}
    \begin{split}
    \gamma_{{\rm NRQCD}}^{\gamma \gamma,\mathcal{A}_i}=& \frac{1}{4}\,\gamma_{\rm Coulomb}^{^3P_J^{[1]}}\, \overline{\mathcal{F}}_{\gamma \gamma, \mathcal{A}_i}^{(0)},
    \\
    \gamma_{{\rm NRQCD}}^{\gamma g,\mathcal{A}_i}=& \frac{1}{4}\,\gamma_{\rm Coulomb}^{^3P_J^{[8]}}\, \overline{\mathcal{F}}_{\gamma g, \mathcal{A}_i}^{(0)}.
    \end{split}
\end{equation}

We observe that the anomalous dimensions for the colour-singlet states takes the form
\begin{align}
    \gamma_{\rm Coulomb}^{^3P_0^{[1]}} =& -\pi^2 \left(\frac{2}{3}C_F^2 + \frac{1}{6}C_F C_A\right),
    \\
    \gamma_{\rm Coulomb}^{^3P_2^{[1]}} =& -\pi^2 \left(\frac{13}{60}C_F^2 + \frac{1}{6}C_F C_A\right),
\end{align}
which is in agreement with the literature \cite{Hoang:2006ty, Sang:2015uxg}. For the colour-octet states, we obtain the anomalous dimensions here for the first time\footnote{We note the following observation that the anomalous dimension results for the colour-octet states can be deduced from the colour-singlet ones by performing the replacement $C_F \rightarrow C_F - \frac{1}{2}C_A$, which is consistent with the behaviour observed in the pseudo-scalar case, $^1S_0$, in ref.~\cite{Abreu:2022cco}.}
\begin{align}
    \gamma_{\rm Coulomb}^{^3P_0^{[8]}} =& -\pi^2 \left(\frac{2}{3}C_F^2-\frac{1}{2}C_F C_A+\frac{1}{12}C_A^2\right),
    \\
    \gamma_{\rm Coulomb}^{^3P_2^{[8]}} =& -\pi^2 \left(\frac{13}{60}C_F^2-\frac{1}{20}C_F C_A-\frac{7}{240}C_A^2\right).
\end{align}

In contrast to this, we find that, for the two form factors with gluons in the initial state, $gg \leftrightarrow {^3P_J^{[1]}}$ and $gg \leftrightarrow {^3P_J^{[8]}}$, the NRQCD pole structure involves in addition to the Coulomb singularity also another singularity component and we can write
\begin{equation}
\begin{split}
    \gamma_{{\rm NRQCD}}^{gg,[1],\mathcal{A}_i} =& \frac{1}{4}\,\gamma_{\rm Coulomb}^{^3P_J^{[1]}}\, \overline{\mathcal{F}}_{gg, [1], \mathcal{A}_i}^{(0)} + \tilde{\gamma}_{{\rm NRQCD}}^{gg,[1],\mathcal{A}_i},
    \\
    \gamma_{{\rm NRQCD}}^{gg,[8],\mathcal{A}_i} =& \frac{1}{4}\,\gamma_{\rm Coulomb}^{^3P_J^{[8]}} \, \overline{\mathcal{F}}_{gg, [8], \mathcal{A}_i}^{(0)} + \tilde{\gamma}_{{\rm NRQCD}}^{gg,[8],\mathcal{A}_i}.
\end{split}
\label{eq:additionalsingularitystruc}
\end{equation}
In particular, we observe that this additional term only exists for the helicity configuration where both gluons have $\lambda_1=\lambda_2=\pm 1$ and where the quarkonium is longitudinal polarised with $J_z=0$, which correspond to the helicity amplitudes $\mathcal{A}_2$ and $\mathcal{A}_3$. This additional term turns out to be independent of the total spin $J$.

\begin{figure}
  \begin{center}
    \subfloat[]{\includegraphics[width=4.5cm]{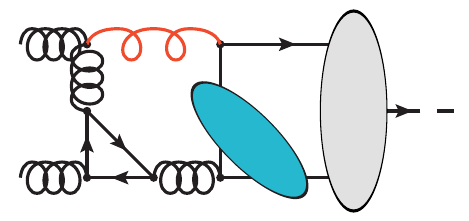}
    \label{fig:gglblultrasoft}}\;\;\;\;
    \subfloat[]{\includegraphics[width=4.5cm]{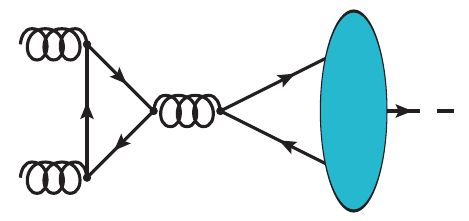} 
    \label{fig:gglbl3S18}}
  \end{center}
  \caption{(a) Two-loop diagram for $g g \leftrightarrow {^3P_J^{[1,8]}}$ where one of the intermediate gluons, depicted in red colour, is ultra-soft, and the second blob indicates the ${^3S_1^{[8]}}$ state, (b) One-loop diagram for the process $g g \leftrightarrow {^3S_1^{[8]}}$.}
  \label{fig:ultrasoftcase}
\end{figure}

It is well known that one encounters infra-red singularities beyond standard QCD in inclusive decays of $P$-wave quarkonia \cite{Bodwin:1992ye,Huang:1996cs}. For instance, the decay width in the channel $^3P_J \rightarrow g q\bar{q}$ exhibits an infra-red singularity that results from the phase-space integration of the final-state gluon in the soft limit.
Similar issues are also encountered in inclusive $P$-wave quarkonium production after performing phase-space integrations \cite{Bodwin:1994jh, Petrelli:1997ge, AH:2024ueu}.

Within the framework of NRQCD factorisation \cite{Bodwin:1994jh}, this singularity is then absorbed into the LDME of the $^3S_1^{[8]}$ state. As such, one has to combine it with the contribution coming from the $^3S_1^{[8]} \rightarrow q\bar{q}$ decay. The inclusive decay width of $\chi_{Q,J}$ into light hadrons ($lh$) can then be written as contribution from two parts
\begin{equation}
    \Gamma_{\chi_{Q,J}\rightarrow lh}=\langle \mathcal{O}^{^3P_J^{[1]}}_{\chi_{Q,J}}\rangle\; \hat{\Gamma}_{^3P_{J}^{[1]} \rightarrow lh} + \langle \mathcal{O}^{^3S_1^{[8]}}_{\chi_{Q,J}}\rangle\; \hat{\Gamma}_{^3S_{1}^{[8]} \rightarrow lh},
\end{equation}
where $\hat{\Gamma}$ is the corresponding short-distance decay contribution and $\langle \mathcal{O}^{n}_{\chi_{Q,J}}\rangle$ is the non-perturbative LDME.
However, our case is fundamentally different, as the singularity originates from the loop integration rather than the phase-space integration, and initial and final states do not factorise easily.

It was first observed already in ref.~\cite{Kuhn:1979bb} that, in the case of the leptonic decay mode $\chi_{Q,J} \rightarrow e^+e^-$, which is loop induced at leading order, the amplitude contains logarithmic singularities in the binding energy. This was attributed to the E1 transition from the $^3P_J$ state to the $^3S_1+\gamma$ state. In refs.~\cite{Yang:2012gk, Kivel:2015iea}, this singularity was then re-expressed as a pole in the more accustomed dimensional regulator $\epsilon$. It was shown that, in order to cancel this singularity, one has to include also the contribution coming from the higher Fock state $\vert ^3S_1^{[1]}\gamma \rangle$ and consider the amplitude with the virtual photon in the ultra-soft mode. More recently, we note ref.~\cite{Jia:2024dzm}, where similar conclusions were drawn and connections to the Lamb shift were shown.

Returning to our process, it will be instructive to look at the Feynman diagram in Fig.~\ref{fig:gglblultrasoft}. The gluon coloured in red, that is connected to the heavy quark line of the bound state, is understood to be ultra-soft. The $Q\bar{Q}$ pair then changes its configuration from the $^3P_J^{[1,8]}$ state to the ${^3S_1^{[8]}}$ state which is depicted by the second blob. Therefore, in order to cancel this singularity, we will need to take into account also contributions coming from the higher Fock state $\vert ^3S_1^{[8]}g \rangle$ and integrate out the ultra-soft virtual gluon. This singularity is indeed different than that from the aforementioned situation encountered in $P$-wave inclusive decays \cite{Bodwin:1994jh,Huang:1996cs}, where the additional singularities can be simply absorbed into the LDME of the $^3S_1^{[8]}$ state.

To the best of our knowledge, this is the first time, that such a new singularity involving the transition $^3P_J^{[1,8]} \rightarrow {^3S_1^{[8]}} + g$ appears and is discussed at two-loop level. In the following, we will construct the $\mathbf{Z}_{\rm NRQCD}$ factor and subtract this singularity using the \msbar-scheme. Hence, it will be necessary to include the form factor for the process $gg \leftrightarrow {^3S_1^{[8]}}$ up to one-loop level as shown in Fig.~\ref{fig:gglbl3S18}. We have given details on the calculation and the result of this new form factor in Appendix~\ref{sec:3S18contribution}.

According to the Landau-Yang theorem \cite{Landau:1948kw,Yang:1950rg}, the form factors $\gamma \gamma \leftrightarrow {^3S_1^{[1]}}$, $gg \leftrightarrow {^3S_1^{[1]}}$ and $\gamma g \leftrightarrow {^3S_1^{[8]}}$ are forbidden. This explains the absence of additional singularities for the $\gamma \gamma$ and $\gamma g$ form factors. As was shown in ref.~\cite{Cacciari:2015ela}, the Landau-Yang theorem does not necessarily extend to cases that exhibit an anti-symmetric colour structure as is the case for the $gg \leftrightarrow {^3S_1^{[8]}}$ form factor.

It turns out that, while the tree-level amplitude for the $gg \leftrightarrow {^3S_1^{[8]}}$ form factor vanishes, the one-loop amplitude does not and is finite.\footnote{The fact that this new contribution starts at $\mathcal{O}{\left(\alpha_s^2\right)}$ and taking into account the additional coupling factor of $\alpha_s$ coming from the ultra-soft gluon, they explain the appearance of this new singularity at $\mathcal{O}{\left(\alpha_s^3\right)}$ for the $gg \leftrightarrow {^3P_J^{[1,8]}}$ form factors. In particular, this also clarifies the presence of the aforementioned $n_h$ term at the simple pole.}
The form factor $gg \leftrightarrow {^3S_1^{[8]}}$ has only a single allowed helicity configuration which we denote as $\mathcal{A}_4$, namely when $\lambda_1=\lambda_2=\pm 1$ and the $^3S_1^{[8]}$ state has longitudinal polarisation with $J_z=0$. This configuration is identical to the cases of $\mathcal{A}_2$ and $\mathcal{A}_3$, where we observed this additional singularity.

We can then write
\begin{align}
    \tilde{\gamma}_{{\rm NRQCD}}^{gg,[1],\mathcal{A}_2} = \tilde{\gamma}_{{\rm NRQCD}}^{gg,[1],\mathcal{A}_3} =& \frac{1}{2}\, \gamma_{{^3S_1^{[8]}}}^{{^3P_J^{[1]}}} \;\overline{\mathcal{F}}_{{^3S_1^{[8]}},\mathcal{A}_4}^{(1)},
    \\
    \tilde{\gamma}_{{\rm NRQCD}}^{gg,[8],\mathcal{A}_2} = \tilde{\gamma}_{{\rm NRQCD}}^{gg,[8],\mathcal{A}_3} =& \frac{1}{2}\, \gamma_{{^3S_1^{[8]}}}^{{^3P_J^{[8]}}} \;\overline{\mathcal{F}}_{{^3S_1^{[8]}},\mathcal{A}_4}^{(1)},
\end{align}
where $\overline{\mathcal{F}}_{{^3S_1^{[8]}},\mathcal{A}_4}^{(1)}$ is the one-loop amplitude of $\mathcal{A}_4$ given in Appendix~\ref{sec:3S18contribution} and where the cross anomalous dimensions read\footnote{We note that this depends on the choice of normalisations $\mathcal{N}$ in the $^3P_J$ and the $^3S_1^{[8]}$ form factors.}
\begin{equation}
    \gamma_{{^3S_1^{[8]}}}^{{^3P_J^{[1]}}} = -\frac{1}{3}C_A, \qquad\qquad\qquad \gamma_{{^3S_1^{[8]}}}^{{^3P_J^{[8]}}} = -\frac{1}{6}C_A.
\end{equation}

Having discussed the origin of this new singularity, we are now in a position to construct the $\mathbf{Z}_{\rm NRQCD}$ factor which consists of two parts
\begin{equation}
    \mathbf{Z}_{\rm NRQCD} = \mathbf{Z}_{\rm Coul.}^{^3P_J^{[1,8]}} + \mathbf{Z}_{{^3S_1^{[8]}}}^{{^3P_J^{[1,8]}}},
\end{equation}
where the first term subtracts the Coulomb singularity and the second term takes into account the contribution from the $^3S_1^{[8]}$ state. Similarly as done in our companion paper \cite{Abreu:2022cco}, we can express the first term as 
\begin{equation}
    \mathbf{Z}_{\rm Coul.}^{^3P_J^{[1,8]}} = 1 + \frac{1}{4\epsilon}\left(\frac{\alpha_s^{(n_l)}}{\pi}\right)^2 \gamma_{\rm Coulomb}^{{^3P_J^{[1,8]}}} + \mathcal{O}{\left(\alpha_s^3\right)}.
\end{equation}
However, the second term $\mathbf{Z}_{{^3S_1^{[8]}}}^{{^3P_J^{[1,8]}}}$ is a matrix in helicity space which is applied to by the vector
\begin{equation}
    \vec{v}=\left( \overline{\mathcal{F}}_{p,\mathcal{A}_1}, \overline{\mathcal{F}}_{p,\mathcal{A}_2}, \overline{\mathcal{F}}_{p,\mathcal{A}_3}, \overline{\mathcal{F}}_{p,\mathcal{A}_4} \right)^{\rm T},
\end{equation}
and where its only two non-zero components are
\begin{equation}
    \left(\mathbf{Z}_{{^3S_1^{[8]}}}^{{^3P_J^{[1,8]}}}\right)_{2,4} = \left(\mathbf{Z}_{{^3S_1^{[8]}}}^{{^3P_J^{[1,8]}}}\right)_{3,4} = \frac{1}{2\epsilon}\left(\frac{\alpha_s^{(n_l)}}{\pi}\right) \gamma_{{^3S_1^{[8]}}}^{{^3P_J^{[1,8]}}} + \mathcal{O}{\left(\alpha_s^2\right)}.
\end{equation}

We can then express the finite remainder as expansion of $\alpha_s{(\mu_R)}$ as
\begin{equation}
\begin{split}
    \mathcal{F}^{\text{fin}}_{p,\mathcal{A}_i}= &\left(\frac{\alpha_s^{(n_l)}}{\pi}\right)^q \Bigg[\mathcal{F}^{\text{fin},(0)}_{p,\mathcal{A}_i}+\left(\frac{\alpha_s^{(n_l)}}{\pi}\right) \mathcal{F}^{\text{fin},(1)}_{p,\mathcal{A}_i} +\left(\frac{\alpha_s^{(n_l)}}{\pi}\right)^2 \mathcal{F}^{\text{fin},(2)}_{p,\mathcal{A}_i}\Bigg]+ \mathcal{O}{\left({\alpha_s^{q+3}}\right)},
\end{split}
\label{eq:deffiniteremainder}
\end{equation}
with
\begin{align}
    \mathcal{F}^{\text{fin},(0)}_{p,\mathcal{A}_i} =& \overline{\mathcal{F}}_{p,\mathcal{A}_i}^{(0)},
    \\
    \mathcal{F}^{\text{fin},(1)}_{p,\mathcal{A}_i} =& \left[ \overline{\mathcal{F}}_{p,\mathcal{A}_i}^{(1)} - \left(\frac{\mu_R^2}{\mu_F^2}\right)^{\epsilon} \mathbf{Z}^{(1)}_{\text{IR}}\, \overline{\mathcal{F}}_{p,\mathcal{A}_i}^{(0)} - \left(\frac{\mu_R^2}{\mu_{\Lambda}^2}\right)^{\epsilon} \mathbf{Z}^{(1)}_{\text{NRQCD}}\, \overline{\mathcal{F}}_{p,\mathcal{A}_i}^{(0)} \right],
    \\
    \begin{split}
    \mathcal{F}^{\text{fin},(2)}_{p,\mathcal{A}_i} =& \bigg[\overline{\mathcal{F}}_{p,\mathcal{A}_i}^{(2)}-\left(\frac{\mu_R^2}{\mu_F^2}\right)^{\epsilon}\mathbf{Z}^{(1)}_{\text{IR}} \left(\overline{\mathcal{F}}_{p,\mathcal{A}_i}^{(1)}+\frac{\beta_0}{4\epsilon}\left(\left(\frac{\mu_R^2}{\mu_F^2}\right)^{\epsilon}-1\right)\overline{\mathcal{F}}_{p,\mathcal{A}_i}^{(0)}\right)
    \\
    &-\left(\frac{\mu_R^2}{\mu_F^2}\right)^{2\epsilon}\left(\mathbf{Z}^{(2)}_{\text{IR}}-\left(\mathbf{Z}^{(1)}_{\text{IR}}\right)^2\right)\overline{\mathcal{F}}_{p,\mathcal{A}_i}^{(0)} +\left(\frac{\mu_R^2}{\mu_{F}^2}\right)^{\epsilon} \left(\frac{\mu_R^2}{\mu_{\Lambda}^2}\right)^{\epsilon} \mathbf{Z}^{(1)}_{\text{IR}}\, \mathbf{Z}^{(1)}_{\text{NRQCD}} \overline{\mathcal{F}}_{p,\mathcal{A}_i}^{(0)}
    \\
    &-\left(\frac{\mu_R^2}{\mu_{\Lambda}^2}\right)^{\epsilon}\mathbf{Z}^{(1)}_{\text{NRQCD}} \left(\overline{\mathcal{F}}_{p,\mathcal{A}_i}^{(1)}+\frac{\beta_0}{4\epsilon}\left(\left(\frac{\mu_R^2}{\mu_{\Lambda}^2}\right)^{\epsilon}-1\right)\overline{\mathcal{F}}_{p,\mathcal{A}_i}^{(0)}\right)
    \\
    &-\left(\frac{\mu_R^2}{\mu_{\Lambda}^2}\right)^{2\epsilon}\left(\mathbf{Z}^{(2)}_{\text{NRQCD}}-\left(\mathbf{Z}^{(1)}_{\text{NRQCD}}\right)^2\right)\overline{\mathcal{F}}_{p,\mathcal{A}_i}^{(0)} \bigg],
    \end{split}
\end{align}
where the scales $\mu_F$ and $\mu_{\Lambda}$ are associated to the $\mathbf{Z}_{\text{IR}}$ and $\mathbf{Z}_{\text{NRQCD}}$ subtraction factors respectively (see also footnote~\ref{footnote:scaledependeevol}).

\section{Form factors}
\label{sec:formfactors}

After having subtracted the IR singularities, the finite remainders at one-loop and two-loop level in eq.~\eqref{eq:deffiniteremainder} have dependencies on the scales discussed in the previous section, $\mu_R$, $\mu_F$ and $\mu_{\Lambda}$.

We can express the finite remainders into scale-dependent and scale-independent terms as follows,
\begin{align}
    \mathcal{F}_{p,\mathcal{A}_i}^{{\rm fin}, (0)}= & \mathcal{F}_{p,\mathcal{A}_i; \rm reg}^{{\rm fin}, (0)},
    \\[5pt]
    \mathcal{F}_{p,\mathcal{A}_i}^{{\rm fin}, (1)}= & \mathcal{F}_{p,\mathcal{A}_i; \rm reg}^{{\rm fin}, (1)} + \mathcal{F}_{p,\mathcal{A}_i; \rm reg}^{{\rm fin}, (0)} \left(\mathcal{C}_{\mu_R}^{(1)} + \mathcal{C}_{\mu_F}^{(1)}\right),
    \\[5pt]
    \begin{split}
    \mathcal{F}_{p,\mathcal{A}_i}^{{\rm fin}, (2)}= & \mathcal{F}_{p,\mathcal{A}_i; \rm reg}^{{\rm fin}, (2)} + \mathcal{F}_{p,\mathcal{A}_i; \rm lbl}^{{\rm fin}, (2)} + \mathcal{F}_{p,\mathcal{A}_i; \rm vac}^{{\rm fin}, (2)} + \mathcal{F}_{p,\mathcal{A}_i; \rm reg}^{{\rm fin}, (1)} \left(\mathcal{D}_{\mu_R}^{(1)} + \mathcal{C}_{\mu_F}^{(1)}\right)
    \\
    & + \mathcal{F}_{p,\mathcal{A}_i; \rm reg}^{{\rm fin}, (0)} \left(\mathcal{C}_{\mu_R}^{(2)} + \mathcal{C}_{\mu_F}^{(2)} + \mathcal{C}_{\mu_{\Lambda}}^{(2)} + \mathcal{D}_{\mu_R}^{(1)} \mathcal{C}_{\mu_F}^{(1)}\right) + \mathcal{G}^{{\rm fin}, (1)}_{p,\mathcal{A}_i}\,\overline{\mathcal{D}}^{(1)}_{\mu_{\Lambda}},
    \end{split}
\end{align}
where in the last line we had introduced the quantity
\begin{equation}
    \mathcal{G}^{{\rm fin}, (1)}_{p,\mathcal{A}_i} = \begin{cases} \mathcal{F}^{{\rm fin}, (1)}_{^3S_1^{[8]}, \mathcal{A}_4}, & \text{if in $gg$ channel and, $i=2$ or $i=3$}, \\ 0, & \text{otherwise}, \end{cases}
\end{equation}
and where the result for $\mathcal{F}^{{\rm fin}, (1)}_{^3S_1^{[8]}, \mathcal{A}_4}$ has been collected in Appendix~\ref{sec:3S18contribution}.

The scale dependence, $\mu$, of the finite remainders is entirely encoded in the corresponding scale-dependent coefficients $\mathcal{C}_{\mu}^{(i)}$, $\mathcal{D}_{\mu}^{(i)}$ and $\overline{\mathcal{D}}_{\mu}^{(i)}$. While the coefficients $\mathcal{C}_{\mu}^{(i)}$ and $\mathcal{D}_{\mu}^{(i)}$ were derived in our previous paper \cite{Abreu:2022cco}, the coefficient $\overline{\mathcal{D}}_{\mu}^{(i)}$ is new and appears here due to the inclusion of the $gg \leftrightarrow {^3S_1^{[8]}}$ form factor. It is defined as
\begin{equation}
    \overline{\mathcal{D}}_{\mu_{\Lambda}}^{(1)} = \frac{1}{2}\, \gamma_{{^3S_1^{[8]}}}^{{^3P_J^{[1,8]}}}\, l_{\mu_{\Lambda}}\,.
\end{equation}
For convenience, we have collected the remaining coefficients in Appendix~\ref{sec:appendixscaledependence}. In the case, where the scale is set equal to the mass of the heavy quark, $\mu=m_Q$, the corresponding coefficients $\mathcal{C}_{\mu}^{(i)}=\mathcal{D}_{\mu}^{(i)}=\overline{\mathcal{D}}^{(i)}_{\mu}=0$ vanish.

The form factors at one-loop can be further structured according to the colour factors
\begin{align}
    \mathcal{F}_{p,\mathcal{A}_i; \rm reg}^{{\rm fin}, (1)} = C_F \, a^{(1)}_{p,\mathcal{A}_i; F} + C_A \, a^{(1)}_{p,\mathcal{A}_i; A}.
\end{align}
At two-loop level, we can structure the form factors into three different class of contributions, regular contributions, $\mathcal{F}_{p,\mathcal{A}_i; \rm reg}^{{\rm fin}, (2)}$, without any fermion loops, and fermion-loop contributions of type light-by-light, $\mathcal{F}_{p,\mathcal{A}_i; \rm lbl}^{{\rm fin}, (2)}$, and type vacuum polarisation, $\mathcal{F}_{p,\mathcal{A}_i; \rm vac}^{{\rm fin}, (2)}$. These contributions read\footnote{In the $gg$ channel case, the finite contributions originating from the $\mathcal{O}{\left(\epsilon\right)}$ terms in the $^3S_1^{[8]}$ form factor are included in the coefficients $a^{(2)}_{p,\mathcal{A}_i; FA}$, $a^{(2)}_{p,\mathcal{A}_i; AA}$, $c^{(2)}_{p,\mathcal{A}_i; Ah}$ and $c^{(2)}_{p,\mathcal{A}_i; Al}$.}
\begin{align}
    \mathcal{F}_{p,\mathcal{A}_i; \rm reg}^{{\rm fin}, (2)} =& \, C_F^2 \, a^{(2)}_{p,\mathcal{A}_i; FF} + C_F C_A \, a^{(2)}_{p,\mathcal{A}_i; FA} + C_A^2 \, a^{(2)}_{p,\mathcal{A}_i; AA},
    \\[5pt]
    \begin{split}
    \mathcal{F}_{p,\mathcal{A}_i; \rm lbl}^{{\rm fin}, (2)} =& \, C_F T_F n_h \, b^{(2)}_{p,\mathcal{A}_i; Fh} + C_F T_F \tilde{n}_l\, b^{(2)}_{p,\mathcal{A}_i; Fl}
    \\
    &+ C_A T_F n_h \, b^{(2)}_{p,\mathcal{A}_i; Ah} + C_A T_F \tilde{n}_l\, b^{(2)}_{p,\mathcal{A}_i; Al},
    \end{split}
    \\[5pt]
    \begin{split}
    \mathcal{F}_{p,\mathcal{A}_i; \rm vac}^{{\rm fin}, (2)} =& \, C_F T_F n_h \, c^{(2)}_{p,\mathcal{A}_i; Fh} + C_F T_F n_l \, c^{(2)}_{p,\mathcal{A}_i; Fl}
    \\
    &+ C_A T_F n_h \, c^{(2)}_{p,\mathcal{A}_i; Ah} + C_A T_F n_l \, c^{(2)}_{p,\mathcal{A}_i; Al},
    \end{split}
\end{align}
where $\tilde{n}_l$ was defined in eq.~\eqref{eq:tildenldefinition}.

Similarly as was done in ref.~\cite{Abreu:2022cco}, we can construct the hard function as follows
\begin{equation}
\begin{split}
    \mathcal{H}_{p,\mathcal{A}_i}=&\, \left\vert \sum_{k=0} \left(\frac{\alpha_s^{(n_l)}}{\pi}\right)^k \mathcal{F}_{p,\mathcal{A}_i}^{{\rm fin}, (k)} \right\vert^2
    \\
    =&\, \sum_{k=0} \left(\frac{\alpha_s^{(n_l)}}{\pi}\right)^k \mathcal{H}_{p,\mathcal{A}_i}^{(k)},
    \end{split}
    \label{eq:definitionhardfunctionA}
\end{equation}
and we will denote the scale dependence of the hard function using the shorthand notation
\begin{equation}
    l_{\mu}=\log{\frac{\mu^2}{m_Q^2}}.
    \label{eq:scaledepnotation}
\end{equation}

\subsection{Form factor coefficients}
\label{sec:FFcoefficient}

We can now construct a linear bases of independent form factor coefficients for each of the contributions above. At loop order $k$, we denote them with $a_i^{(k)}$ for the regular part, $b_i^{(k)}$ for the light-by-light part and $c_i^{(k)}$ for the vacuum polarisation contribution.

As the analytical expressions at two-loop level are rather lengthy, in this section, we only display the numerical evaluation of these two-loop bases up to 20 digits accuracy. As for the analytical expressions, we have collected them in Appendix~\ref{sec:analyticsFF} and provide them also in electronic form in ref.~\cite{formFactorPwaveGit}.

At one-loop level, we can express the regular part in terms of the following bases
\begin{align}
    a_1^{(1)}=&\; -2,
    \\[5pt]
    \begin{split}
    a_2^{(1)}=&\; \frac{\pi ^2}{8}-2 \log{2}+\frac{1}{3}=\; 0.18073952234961254185 \,,
    \end{split}
    \\[5pt]
    \begin{split}
    a_3^{(1)}=&\; \frac{\pi ^2}{8}-\frac{7}{6}= \; 0.06703388346950316069\,,
    \end{split}
    \\[5pt]
    \begin{split}
    a_4^{(1)}=&\; \frac{\pi ^2}{24}-\log^2{2}+\frac{5}{6}\log{2}+\frac{1}{6}+i \pi  \left[\log{2}-\frac{11}{12}\right]
    \\
    =& \; 0.67506981992714294230 - i\, 0.70220717548704167142\,,
    \end{split}
    \\[5pt]
    \begin{split}
    a_5^{(1)}=&\; \frac{\pi ^2}{6}-\log^2{2}+\frac{1}{6} + i \pi \log{2}
    \\
    =& \; 1.33114771959669167847 + i\, 2.17758609030360213050\,,
    \end{split}
    \\[5pt]
    \begin{split}
    a_6^{(1)}=&\; \frac{7 \pi ^2}{48}-\frac{1}{2} \log^2{2} + \frac{3}{4}\log{2}+\frac{3}{8}=\; 2.0939511869530564016\,.
    \end{split}
\end{align}
At two-loop level, the regular part can be expressed in terms of the following bases
\begin{align}
    a_1^{(2)}=a^{(2)}_{\gamma \gamma, [1], \mathcal{A}_1; FF}=&\; -5.93023532518981563672\,,
    \\
    a_2^{(2)}=a^{(2)}_{\gamma \gamma, [1], \mathcal{A}_2; FF}=&\; -0.5300774288390386517\,,
    \\
    a_3^{(2)}=a^{(2)}_{\gamma \gamma, [1], \mathcal{A}_3; FF}=&\; -9.1475107737851465282\,,
    \\
    a_4^{(2)}=a^{(2)}_{\gamma \gamma, [1], \mathcal{A}_1; FA}=&\; -5.7820492249031375348\,,
    \\
    a_5^{(2)}=a^{(2)}_{\gamma \gamma, [1], \mathcal{A}_2; FA}=&\; 0.71091563462233753591\,,
    \\
    a_6^{(2)}=a^{(2)}_{\gamma \gamma, [1], \mathcal{A}_3; FA}=&\; -1.69821087211618022213\,,
    \\
    a_7^{(2)}=a^{(2)}_{gg, [1], \mathcal{A}_1; FA}=&\; -5.89023646266125024983 + i\, 1.40441435097408334285\,,
    \\
    a_8^{(2)}=a^{(2)}_{gg, [1], \mathcal{A}_2; FA}=&\; 1.12950192508940774226 - i\, 0.45813906360367321855\,,
    \\
    a_9^{(2)}=a^{(2)}_{gg, [1], \mathcal{A}_3; FA}=&\; 0.85311075698117693513 + i\, 0.14597205222222265176\,,
    \\
    a_{10}^{(2)}=a^{(2)}_{\gamma g, [8], \mathcal{A}_1; FA}=&\; -3.7619315957838279695\,,
    \\
    a_{11}^{(2)}=a^{(2)}_{\gamma g, [8], \mathcal{A}_2; FA}=&\; 1.6096378399719816087\,,
    \\
    a_{12}^{(2)}=a^{(2)}_{\gamma g, [8], \mathcal{A}_3; FA}=&\; 5.62927894430618740156\,,
    \\
    a_{13}^{(2)}=a^{(2)}_{gg, [1], \mathcal{A}_1; AA}=&\; -1.27056838663114079024 - i\, 2.25089811312768810948\,,
    \\
    a_{14}^{(2)}=a^{(2)}_{gg, [1], \mathcal{A}_2; AA}=&\; -0.07457847030222826630 + i\, 0.55780467363317568241\,,
    \\
    a_{15}^{(2)}=a^{(2)}_{gg, [1], \mathcal{A}_3; AA}=&\; -5.24902027471099440945 + i\, 11.88155649574916378744\,,
    \\
    a_{16}^{(2)}=a^{(2)}_{\gamma g, [8], \mathcal{A}_1; AA}=&\; 7.36074295578544341944\,,
    \\
    a_{17}^{(2)}=a^{(2)}_{\gamma g, [8], \mathcal{A}_2; AA}=&\; -0.92202892301657378772\,,
    \\
    a_{18}^{(2)}=a^{(2)}_{\gamma g, [8], \mathcal{A}_3; AA}=&\; 1.82287137980929676742\,,
    \\
    a_{19}^{(2)}=a^{(2)}_{gg, [8], \mathcal{A}_1; AA}=&\; 7.94217462983074650361 - i\, 2.50457943196915473464\,,
    \\
    a_{20}^{(2)}=a^{(2)}_{gg, [8], \mathcal{A}_2; AA}=&\; -1.36125919482522218959 + i\, 0.10572368842822388739\,,
    \\
    a_{21}^{(2)}=a^{(2)}_{gg, [8], \mathcal{A}_3; AA}=&\; -0.17203625568493402772 + i\, 2.90457889382683733345\,.
\end{align}
For the light-by-light contribution, we define the bases
\begin{align}
    b_1^{(2)}=b^{(2)}_{\gamma \gamma, [1], \mathcal{A}_1; Fh}=&\; -0.08823049707449080918 + i\, 0.18723934739940723983\,,
    \\
    b_2^{(2)}=b^{(2)}_{\gamma \gamma, [1], \mathcal{A}_2; Fh}=&\; -0.00659026544329967262 + i\, 0.00373145946604155785\,,
    \\
    b_3^{(2)}=b^{(2)}_{\gamma \gamma, [1], \mathcal{A}_3; Fh}=&\; -0.12032418591273448000 + i\, 0.39854668592833680473\,,
    \\
    b_4^{(2)}=b^{(2)}_{\gamma \gamma, [1], \mathcal{A}_1; Fl}=&\; -0.66987349811640208651 - i\, 1.25660453776445208421\,,
    \\
    b_5^{(2)}=b^{(2)}_{\gamma \gamma, [1], \mathcal{A}_2; Fl}=&\; 0.11995003589488380828 + i\, 0.01522464802238932080\,,
    \\
    b_6^{(2)}=b^{(2)}_{\gamma \gamma, [1], \mathcal{A}_3; Fl}=&\; 0.95374124621806418235 + i\, 0.52359877559829887308\,,
    \\
    b_7^{(2)}=b^{(2)}_{gg, [1], \mathcal{A}_1; Ah}=&\; -0.96161008954308369807 + i\, 0.12598978258968944149\,,
    \\
    b_8^{(2)}=b^{(2)}_{gg, [1], \mathcal{A}_2; Ah}=&\; 0.22927334842246826528 - i\, 0.02205490782376391947\,,
    \\
    b_9^{(2)}=b^{(2)}_{gg, [1], \mathcal{A}_3; Ah}=&\; 1.45468597025539068464 + i\, 0.44068058094632827221\,,
    \\
    b_{10}^{(2)}=b^{(2)}_{gg, [1], \mathcal{A}_1; Al}=&\; -0.75619363865595256192 - i\, 0.44057558420087164750\,,
    \\
    b_{11}^{(2)}=b^{(2)}_{gg, [1], \mathcal{A}_2; Al}=&\; 0.32077373084306341351 + i\, 0.01581924566167270828\,,
    \\
    b_{12}^{(2)}=b^{(2)}_{gg, [1], \mathcal{A}_3; Al}=&\; 1.56292540681705650580 + i\, 0.22206429929488992551\,.
\end{align}
For the vacuum polarisation contribution, the bases are defined as
\begin{align}
    c_1^{(2)}=c^{(2)}_{\gamma \gamma, [1], \mathcal{A}_1; Fh}=&\; 0.0217165020132035420\,,
    \\
    c_2^{(2)}=c^{(2)}_{\gamma \gamma, [1], \mathcal{A}_2; Fh}=&\; 0.0053139547426124988\,,
    \\
    c_3^{(2)}=c^{(2)}_{\gamma \gamma, [1], \mathcal{A}_3; Fh}=&\; 0.0929247940360905018\,,
    \\
    c_4^{(2)}=c^{(2)}_{\gamma \gamma, [1], \mathcal{A}_1; Fl}=&\; 1.81129471951252935812\,,
    \\
    c_5^{(2)}=c^{(2)}_{\gamma \gamma, [1], \mathcal{A}_2; Fl}=&\; -0.2406757452914943874\,,
    \\
    c_6^{(2)}=c^{(2)}_{\gamma \gamma, [1], \mathcal{A}_3; Fl}=&\; -0.81386890705354170789\,,
    \\
    c_7^{(2)}=c^{(2)}_{gg, [1], \mathcal{A}_1; Ah}=&\; 0.97521077435968101669 - i\, 0.14608775747964754712\,,
    \\
    c_8^{(2)}=c^{(2)}_{gg, [1], \mathcal{A}_2; Ah}=&\; -0.17245425273105110968 + i\, 0.03920549405008300023\,,
    \\
    c_9^{(2)}=c^{(2)}_{gg, [1], \mathcal{A}_3; Ah}=&\; -1.35855412031220396834 - i\, 0.50818381920840848875\,,
    \\
    c_{10}^{(2)}=c^{(2)}_{\gamma g, [8], \mathcal{A}_1; Ah}=&\; -0.0067785557422028523\,,
    \\
    c_{11}^{(2)}=c^{(2)}_{\gamma g, [8], \mathcal{A}_2; Ah}=&\; -0.0010853877090717154\,,
    \\
    c_{12}^{(2)}=c^{(2)}_{\gamma g, [8], \mathcal{A}_3; Ah}=&\; -0.0516806494932764653\,,
    \\
    c_{13}^{(2)}=c^{(2)}_{gg, [1], \mathcal{A}_1; Al}=&\; 0.82736241752488066639 + i\, 2.58248215993987668378\,,
    \\
    c_{14}^{(2)}=c^{(2)}_{gg, [1], \mathcal{A}_2; Al}=&\; -0.27391879587309758029 - i\, 0.22448965882761501674\,,
    \\
    c_{15}^{(2)}=c^{(2)}_{gg, [1], \mathcal{A}_3; Al}=&\; -1.08300713307448943692 - i\, 2.53492077304042705450\,,
    \\
    c_{16}^{(2)}=c^{(2)}_{\gamma g, [8], \mathcal{A}_1; Al}=&\; -2.7676682083605183278\,,
    \\
    c_{17}^{(2)}=c^{(2)}_{\gamma g, [8], \mathcal{A}_2; Al}=&\; 0.21833235568409157812\,,
    \\
    c_{18}^{(2)}=c^{(2)}_{\gamma g, [8], \mathcal{A}_3; Al}=&\; -0.7989670822746665226\,.
\end{align}

\subsection{$\gamma \gamma \leftrightarrow {^3P_J^{[1]}}$}

In the following, we express the finite remainders for the helicity amplitudes in the $\gamma \gamma$ channel at one-loop and two-loop level in terms of the bases defined in the previous subsection and give also the hard functions needed for phenomenological applications.\footnote{We note that the results presented here could also be used to compute the pure two-loop QED corrections to the di-photon decay of $\chi_{Q,J}$ and also of $P$-wave leptonium states  \cite{Berko:1980gg,Pwavepostronium:1953,Pwavepostronium:1958}, which are QED bound states of a lepton pair. One has to take the appropriate replacements of colour factors and flavour numbers (see the discussion around eq.~(5.15) in ref.~\cite{Abreu:2022cco}).}

For the finite remainder at one-loop level, the helicity amplitudes $\mathcal{A}_1$, $\mathcal{A}_2$ and $\mathcal{A}_3$ can be expressed in terms of the aforementioned one-loop bases as 
\begin{align}
    a_{\gamma \gamma, [1], \mathcal{A}_i; F}^{(1)}=&\begin{cases}
        a_1^{(1)}, & i=1,\\
        a_2^{(1)}, & i=2,\\
        a_3^{(1)}, & i=3,
    \end{cases}
    \qquad\quad
    a_{\gamma \gamma, [1], \mathcal{A}_i; A}^{(1)}=\begin{cases}
        0, & i=1,\\
        0, & i=2,\\
        0, & i=3.
    \end{cases}
\end{align}
At two-loop level, the finite remainders for the helicity amplitudes read
\begin{align}
    a_{\gamma \gamma, [1], \mathcal{A}_i; FF}^{(2)}=&\begin{cases}
        a_1^{(2)}, & i=1,\\
        a_2^{(2)}, & i=2,\\
        a_3^{(2)}, & i=3,
    \end{cases}
    \qquad\quad  a_{\gamma \gamma, [1], \mathcal{A}_i; FA}^{(2)}=\begin{cases}
        a_4^{(2)}, & i=1,\\
        a_5^{(2)}, & i=2,\\
        a_6^{(2)}, & i=3,
    \end{cases}
    \nonumber \\
    a_{\gamma \gamma, [1], \mathcal{A}_i; AA}^{(2)}=&\begin{cases}
        0, & i=1,\\
        0, & i=2,\\
        0, & i=3,
    \end{cases}
    \nonumber \\
    b_{\gamma \gamma, [1], \mathcal{A}_i; Fh}^{(2)}=&\begin{cases}
        b_1^{(2)}, & i=1,\\
        b_2^{(2)}, & i=2,\\
        b_3^{(2)}, & i=3,
    \end{cases}
    \qquad\quad b_{\gamma \gamma, [1], \mathcal{A}_i; Fl}^{(2)}=\begin{cases}
        b_4^{(2)}, & i=1,\\
        b_5^{(2)}, & i=2,\\
        b_6^{(2)}, & i=3,
    \end{cases}
    \nonumber \\
    b_{\gamma \gamma, [1], \mathcal{A}_i; Ah}^{(2)}=&\begin{cases}
        0, & i=1,\\
        0, & i=2,\\
        0, & i=3,
    \end{cases}
    \qquad\quad b_{\gamma \gamma, [1], \mathcal{A}_i; Al}^{(2)}=\begin{cases}
        0, & i=1,\\
        0, & i=2,\\
        0, & i=3,
    \end{cases}
    \nonumber \\
    c_{\gamma \gamma, [1], \mathcal{A}_i; Fh}^{(2)}=&\begin{cases}
        c_1^{(2)}, & i=1,\\
        c_2^{(2)}, & i=2,\\
        c_3^{(2)}, & i=3,
    \end{cases}
    \qquad\quad c_{\gamma \gamma, [1], \mathcal{A}_i; Fl}^{(2)}=\begin{cases}
        c_4^{(2)}, & i=1,\\
        c_5^{(2)}, & i=2,\\
        c_6^{(2)}, & i=3,
    \end{cases}
    \nonumber \\
    c_{\gamma \gamma, [1], \mathcal{A}_i; Ah}^{(2)}=&\begin{cases}
        0, & i=1,\\
        0, & i=2,\\
        0, & i=3,
    \end{cases}
    \qquad\quad c_{\gamma \gamma, [1], \mathcal{A}_i; Al}^{(2)}=\begin{cases}
        0, & i=1,\\
        0, & i=2,\\
        0, & i=3.
    \end{cases}
\end{align}

Comparing our result with the ones from ref.~\cite{Sang:2015uxg}, where the two-loop form factors have been computed numerically, we find overall agreement for all three helicity amplitudes both at one-loop and two-loop level. This serves as a cross-check of our calculation. We note that our result is much more precise having at disposal the complete analytical result and the high precision numerical evaluation up to 1000 digits. In particular, we observe that the numerical result for the light-by-light contribution proportional to $n_h$ in ref.~\cite{Sang:2015uxg} has uncertainties in the last two digits.

Furthermore, we remark that in ref.~\cite{Sang:2015uxg}, the helicity amplitude decomposition for $\mathcal{A}_2$ in terms of the $c_2$ and $c_3$ form factors has been taken to be strictly four-dimensional rather than in $d$ dimensions as in eq.~\eqref{eq:helicity2}. We observe no difference between the two schemes for the $\gamma \gamma$ case. As the $c_2$ form factor is non-vanishing at the Born order, the $\epsilon$ pre-factor of $c_2$ can be considered as a global factor when subtracting the Coulomb singularity with the $\mathbf{Z}_{\rm Coul.}$ factor.

In contrast to this, in the case of the $\mathcal{A}_3$ helicity amplitude, it turns out to be crucial to take the $d$-dimensional version in eq.~\eqref{eq:helicity3} as was also done in ref.~\cite{Sang:2015uxg}. The key point is that the $c_1$ form factor vanishes at tree-level but has a simple pole at two-loop level (see Appendix~\ref{sec:renpole}). Hence this pole cannot be removed by applying the $\mathbf{Z}_{\rm Coul.}$ factor to $c_1$ alone, but one has to include the tree-level contributions of $c_2$ and $c_3$ in eq.~\eqref{eq:helicity3} as well. Therefore, any finite piece originating from the $\epsilon$ pre-factor in front of $c_1$ does not get removed and survives.

The hard function for the helicity amplitude $\mathcal{A}_1$ reads
{\small
\begin{align}
    \mathcal{H}_{p,\mathcal{A}_1}^{(0)} =& 1\,,
    \nonumber \\
    \mathcal{H}_{p,\mathcal{A}_1}^{(1)} =& -5.33333333333333333333\,,
    \nonumber \\
    \mathcal{H}_{p,\mathcal{A}_1}^{(2)} =& -60.3192491710928277876 -14.6666666666666666667\; l_{\mu_R}
    \nonumber \\
    & -10.3813616663310290657\; l_{\mu_{\Lambda}} +2.4150596260167058108\; n_{l}
    \nonumber \\
    & + 0.8888888888888888889 \; n_{l} \, l_{\mu_R} -0.8931646641552027820\; \tilde{n}_{l}\,.
\end{align}
}

\noindent
For $\mathcal{A}_2$, we find that up to third order
{\small
\begin{align}
    \mathcal{H}_{p,\mathcal{A}_2}^{(0)} =& 0\,,
    \nonumber \\
    \mathcal{H}_{p,\mathcal{A}_2}^{(1)} =& 0\,,
    \nonumber \\
    \mathcal{H}_{p,\mathcal{A}_2}^{(2)} =& 0.0580742665585174960\,,
    \nonumber \\
    \mathcal{H}_{p,\mathcal{A}_2}^{(3)} =& 0.9159646639718927032+0.3194084660718462282\; l_{\mu_R}
    \nonumber \\
    &-0.0773326564357719206\; n_{l}-0.0193580888528391653\; n_{l}\, l_{\mu_R}
    \nonumber \\
    &+0.0385417105661514288\; \tilde{n}_{l}\,,
\end{align}
}

\noindent
and for $\mathcal{A}_3$, we find
{\small
\begin{align}
    \mathcal{H}_{p,\mathcal{A}_3}^{(0)} =& 1\,,
    \nonumber \\
    \mathcal{H}_{p,\mathcal{A}_3}^{(1)} =& 0.17875702258534176183\,,
    \nonumber \\
    \mathcal{H}_{p,\mathcal{A}_3}^{(2)} =& -46.1387137323868161964 +0.4915818121096898450 \; l_{\mu_R}
    \nonumber \\
    &-18.2770451872025159608\; l_{\mu_{\Lambda}}-1.0851585427380556105 \; n_{l}
    \nonumber \\
    &-0.0297928370975569603\; n_{l}\, l_{\mu_R}+1.2716549949574189098 \; \tilde{n}_{l}\,.
\end{align}
}

\noindent
In order to distinguish the charmonium and bottomonium cases, we will need to fix the number of light flavours as follows
\begin{equation}
    n_l = \begin{cases}
        3, & c\bar{c},
        \\
        4, & b\bar{b},
    \end{cases} \qquad\qquad\qquad
    \tilde{n}_l = \begin{cases}
        3/2, & c\bar{c},
        \\
        10, & b\bar{b}.
    \end{cases}
\end{equation}

\subsection{$gg \leftrightarrow {^3P_J^{[1]}}$}

We give the result for the three helicity amplitudes for the process $gg \leftrightarrow {^3P_J^{[1]}}$. This result is new and can be used for the hadro-production of $\chi_{Q,J}$ in collinear factorisation and also for the inclusive hadronic decay width up to NNLO accuracy.

For the finite remainder at one-loop level, we can write for the helicity amplitudes $\mathcal{A}_1$, $\mathcal{A}_2$ and $\mathcal{A}_3$, 
\begin{align}
    a_{gg, [1], \mathcal{A}_i; F}^{(1)}=&\begin{cases}
        a_1^{(1)}, & i=1,\\
        a_2^{(1)}, & i=2,\\
        a_3^{(1)}, & i=3,
    \end{cases}
    \qquad\quad
    a_{gg, [1], \mathcal{A}_i; A}^{(1)}=\begin{cases}
        a_4^{(1)}, & i=1,\\
        \frac{1}{48}a_1^{(1)}-\frac{1}{4}a_2^{(1)}-\frac{1}{4}a_3^{(1)}, & i=2,\\
        a_5^{(1)}, & i=3.
    \end{cases}
\end{align}
At the two-loop level, the finite remainder reads for the helicity amplitudes
\begin{align}
    a_{gg, [1], \mathcal{A}_i; FF}^{(2)}=&\begin{cases}
        a_1^{(2)}, & i=1,\\
        a_2^{(2)}, & i=2,\\
        a_3^{(2)}, & i=3,
    \end{cases}
    \qquad\quad  a_{gg, [1], \mathcal{A}_i; FA}^{(2)}=\begin{cases}
        a_7^{(2)}, & i=1,\\
        a_8^{(2)}, & i=2,\\
        a_9^{(2)}, & i=3,
    \end{cases}
    \nonumber \\    
    a_{gg, [1], \mathcal{A}_i; AA}^{(2)}=&\begin{cases}
        a_{13}^{(2)}, & i=1,\\
        a_{14}^{(2)}, & i=2,\\
        a_{15}^{(2)}, & i=3,
    \end{cases}
    \nonumber \\
    b_{gg, [1], \mathcal{A}_i; Fh}^{(2)}=&\begin{cases}
        b_1^{(2)}, & i=1,\\
        b_2^{(2)}, & i=2,\\
        b_3^{(2)}, & i=3,
    \end{cases}
    \qquad\quad b_{gg, [1], \mathcal{A}_i; Fl}^{(2)}=\begin{cases}
        b_4^{(2)}, & i=1,\\
        b_5^{(2)}, & i=2,\\
        b_6^{(2)}, & i=3,
    \end{cases}
    \nonumber \\
    b_{gg, [1], \mathcal{A}_i; Ah}^{(2)}=&\begin{cases}
        b_7^{(2)}, & i=1,\\
        b_8^{(2)}, & i=2,\\
        b_9^{(2)}, & i=3,
    \end{cases}
    \qquad\quad b_{gg, [1], \mathcal{A}_i; Al}^{(2)}=\begin{cases}
        b_{10}^{(2)}, & i=1,\\
        b_{11}^{(2)}, & i=2,\\
        b_{12}^{(2)}, & i=3,
    \end{cases}
    \nonumber \\
    c_{gg, [1], \mathcal{A}_i; Fh}^{(2)}=&\begin{cases}
        c_1^{(2)}, & i=1,\\
        c_2^{(2)}, & i=2,\\
        c_3^{(2)}, & i=3,
    \end{cases}
    \qquad\quad c_{gg, [1], \mathcal{A}_i; Fl}^{(2)}=\begin{cases}
        c_4^{(2)}, & i=1,\\
        c_5^{(2)}, & i=2,\\
        c_6^{(2)}, & i=3,
    \end{cases}
    \nonumber \\
    c_{gg, [1], \mathcal{A}_i; Ah}^{(2)}=&\begin{cases}
        c_7^{(2)}, & i=1,\\
        c_8^{(2)}, & i=2,\\
        c_9^{(2)}, & i=3,
    \end{cases}
    \qquad\quad c_{gg, [1], \mathcal{A}_i; Al}^{(2)}=\begin{cases}
        c_{13}^{(2)}, & i=1,\\
        c_{14}^{(2)}, & i=2,\\
        c_{15}^{(2)}, & i=3.
    \end{cases}
\end{align}
As mentioned in the $\gamma \gamma$ case, it is indeed crucial that one takes the $d$-dimensional helicity decomposition in terms of the $c_i$ form factors as done in eqs.~\eqref{eq:helicity1}, \eqref{eq:helicity2} and \eqref{eq:helicity3} due to the more involved NRQCD singularity cancellation by taking into account the $^3S_1^{[8]}$ contribution.

In the following, we give the hard functions for these three helicity amplitudes. Each of the hard functions represents the contribution due to a different polarisation configuration of the gluons and the bound state. In the case of unpolarised production or decay of $\chi_{Q,J=2}$, one will have to combine the necessary hard functions for $\mathcal{A}_1$ and $\mathcal{A}_2$ by accounting also for global factors as in eqs.~\eqref{eq:definitionfinitexpanAi} and \eqref{eq:gloablfactorcoefficientAi}. In the case of $\chi_{Q,J=0}$ production and decay, we only need the hard function for the helicity amplitude $\mathcal{A}_3$. As mentioned before, the hard functions exhibit dependencies on the three scales, $\mu_R$, $\mu_F$ and $\mu_{\Lambda}$.

The hard functions read for $\mathcal{A}_1$
{\small
\begin{align}
    \mathcal{H}_{p,\mathcal{A}_1}^{(0)} =& 1\,,
    \nonumber \\
    \mathcal{H}_{p,\mathcal{A}_1}^{(1)} =& -1.28291441377047567954 +5.5000000000000000000 \; l_{\mu_R}
    \nonumber \\
    &-1.34111691664032814350\; l_{\mu_F}-1.5000000000000000000 \; l_{\mu_F}^2 
    \nonumber \\
    &-0.3333333333333333333\; n_{l}\, \left(l_{\mu_R} - l_{\mu_F}\right)\,,
    \nonumber \\
    \mathcal{H}_{p,\mathcal{A}_1}^{(2)} =& -86.27596548516517162901 +2.16595608639357564379\; l_{\mu_R}
    \nonumber \\
    &+22.68750000000000000000\; l_{\mu_R}^2 -4.66401619959352461250 \; l_{\mu_F}
    \nonumber \\
    &-0.00619367650589536808 \; l_{\mu_F}^2 +4.76167537496049221524 \; l_{\mu_F}^3
    \nonumber \\
    &+1.1250000000000000000 \; l_{\mu_F}^4 -11.06421456228270718385 \; l_{\mu_R}\, l_{\mu_F}
    \nonumber \\
    &-12.37500000000000000000\; l_{\mu_R}\, l_{\mu_F}^2-10.3813616663310290657\; l_{\mu_{\Lambda}}
    \nonumber \\
    &+1.73540129846828734224\; n_{l}-0.94187612644809549356 \;n_{l} \, l_{\mu_R} 
    \nonumber \\
    &-2.75000000000000000000 \; n_{l} \, l_{\mu_R}^2 +0.11699415554232042645 \; n_{l} \, l_{\mu_F}
    \nonumber \\
    &-0.60046538193347005979 \; n_{l} \, l_{\mu_F}^2 -0.6666666666666666667 \; n_{l} \, l_{\mu_F}^3
    \nonumber \\
    &+3.42055845832016407175 \; n_{l} \, l_{\mu_R}\, l_{\mu_F} +0.7500000000000000000 \; n_{l} \, l_{\mu_R}\, l_{\mu_F}^2
    \nonumber \\
    &+ 0.0833333333333333333 \; n_{l}^2 \, \left(l_{\mu_R}-l_{\mu_F}\right)^2 \,.
\end{align}
}

\noindent
In the case of $\mathcal{A}_2$, we can write
{\small
\begin{align}
    \mathcal{H}_{p,\mathcal{A}_2}^{(0)} =& 0\,,
    \nonumber \\
    \mathcal{H}_{p,\mathcal{A}_2}^{(1)} =& 0\,,
    \nonumber \\
    \mathcal{H}_{p,\mathcal{A}_2}^{(2)} =& 0.0048781877674158435\,,
    \nonumber \\
    \mathcal{H}_{p,\mathcal{A}_2}^{(3)} =& -0.41750234614185436402+0.05366006544157427809\; l_{\mu_R}
    \nonumber \\
    &-0.00654222013742930219\; l_{\mu_F}-0.00731728165112376519\; l_{\mu_F}^2
    \nonumber \\
    &-0.0936034022118530043 \; l_{\mu_{\Lambda}}+0.00142501419652758927 \; n_{l}
    \nonumber \\
    &-0.00325212517827722898 \; n_{l} \, l_{\mu_R}+0.00162606258913861449 \; n_{l} \, l_{\mu_F}
    \nonumber \\
    &+0.0116406707608088980 \; n_{l} \, l_{\mu_{\Lambda}}\,,
\end{align}
}

\noindent
and for $\mathcal{A}_3$, we have that
{\small
\begin{align}
    \mathcal{H}_{p,\mathcal{A}_3}^{(0)} =& 1\,,
    \nonumber \\
    \mathcal{H}_{p,\mathcal{A}_3}^{(1)} =& 8.16564334016549183267+5.5000000000000000000 \; l_{\mu_R}
    \nonumber \\
    &-1.34111691664032814350\; l_{\mu_F}-1.5000000000000000000 \; l_{\mu_F}^2
    \nonumber \\
    &-0.3333333333333333333 \; n_{l} \, \left(l_{\mu_R} - l_{\mu_F}\right)\,,
    \nonumber \\
    \mathcal{H}_{p,\mathcal{A}_3}^{(2)} =& -60.58373519700638084127+80.11655755636530761949\; l_{\mu_R}
    \nonumber \\
    &+22.68750000000000000000\; l_{\mu_R}^2 -17.33563684125019366924 \;l_{\mu_F}
    \nonumber \\
    &-14.17903030740984663639 \; l_{\mu_F}^2 +4.76167537496049221524 \;l_{\mu_F}^3
    \nonumber \\
    &+1.1250000000000000000 \; l_{\mu_F}^4-11.06421456228270718385\; l_{\mu_R}\, l_{\mu_F}
    \nonumber \\
    &-12.37500000000000000000\; l_{\mu_R}\, l_{\mu_F}^2 -16.9368675156690271932 \; l_{\mu_{\Lambda}}
    \nonumber \\
    &+1.62625127344706450595\; n_{l}-5.66615500341607924967\; n_{l} \, l_{\mu_R}
    \nonumber \\
    &-2.75000000000000000000\; n_{l} \, l_{\mu_R}^2 +3.26651340685430959719\; n_{l} \, l_{\mu_F}
    \nonumber \\
    &-0.60046538193347005979 \; n_{l} \, l_{\mu_F}^2-0.6666666666666666667 \; n_{l} \, l_{\mu_F}^3 
    \nonumber \\
    &+3.42055845832016407175 \; n_{l} \, l_{\mu_R} \, l_{\mu_F} +0.7500000000000000000 \; n_{l} \, l_{\mu_R}\, l_{\mu_F}^2 
    \nonumber \\
    &-0.1666666666666666667 \; n_{l} \, l_{\mu_{\Lambda}}+ 0.0833333333333333333 \; n_{l}^2 \left( l_{\mu_R} - l_{\mu_F}\right)^2 \,.
\end{align}
}

\noindent
We remark that due to the involvement of the $^3S_1^{[8]}$ form factor for the $\mathcal{A}_2$ and $\mathcal{A}_3$ helicity amplitudes, the NRQCD scale dependence $\mu_{\Lambda}$ is more involved and differs from the one in the $\gamma \gamma$ case. In contrast to this, in the case of the $\mathcal{A}_1$ amplitude, the $\mu_{\Lambda}$ scale dependence agrees between the $gg$ and $\gamma \gamma$ channel due to the absence of the $^3S_1^{[8]}$ form factor. For charmonium bound state, we use $n_l=3$, while in the bottomonium case, we set $n_l=4$.

\subsection{$\gamma g \leftrightarrow {^3P_J^{[8]}}$}

We discuss the finite remainders and the hard functions for the process $\gamma g \leftrightarrow {^3P_J^{[8]}}$ which could be used for quarkonium photo-production.

At one-loop level, we can express the helicity amplitudes as 
\begin{align}
    a_{\gamma g, [8], \mathcal{A}_i; F}^{(1)}=&\begin{cases}
        a_1^{(1)}, & i=1,\\
        a_2^{(1)}, & i=2,\\
        a_3^{(1)}, & i=3,
    \end{cases}
    \nonumber \\
    a_{\gamma g, [8], \mathcal{A}_i; A}^{(1)}=&\begin{cases}
        a_6^{(1)}, & i=1,\\
        \frac{5}{96}a_1^{(1)}-\frac{1}{8}a_2^{(1)}-\frac{1}{8}a_3^{(1)}, & i=2,\\
        \frac{71}{96}a_1^{(1)}+\frac{1}{8}a_2^{(1)}-\frac{9}{8}a_3^{(1)}+a_6^{(1)}, & i=3,
    \end{cases}
\end{align}
while at the two-loop level, the expressions read
\begin{align}
    a_{\gamma g, [8], \mathcal{A}_i; FF}^{(2)}=&\begin{cases}
        a_1^{(2)}, & i=1,\\
        a_2^{(2)}, & i=2,\\
        a_3^{(2)}, & i=3,
    \end{cases}
    \qquad\quad  a_{\gamma g, [8], \mathcal{A}_i; FA}^{(2)}=\begin{cases}
        a_{10}^{(2)}, & i=1,\\
        a_{11}^{(2)}, & i=2,\\
        a_{12}^{(2)}, & i=3,
    \end{cases}
    \nonumber \\
    a_{\gamma g, [8], \mathcal{A}_i; AA}^{(2)}=&\begin{cases}
        a_{16}^{(2)}, & i=1,\\
        a_{17}^{(2)}, & i=2,\\
        a_{18}^{(2)}, & i=3,
    \end{cases}
    \nonumber \\
    b_{\gamma g, [8], \mathcal{A}_i; Fh}^{(2)}=&\begin{cases}
        2b_1^{(2)}, & i=1,\\
        2b_2^{(2)}, & i=2,\\
        2b_3^{(2)}, & i=3,
    \end{cases}
    \qquad\quad b_{\gamma g, [8], \mathcal{A}_i; Fl}^{(2)}=\begin{cases}
        2b_4^{(2)}, & i=1,\\
        2b_5^{(2)}, & i=2,\\
        2b_6^{(2)}, & i=3,
    \end{cases}
    \nonumber \\
    b_{\gamma g, [8], \mathcal{A}_i; Ah}^{(2)}=&\begin{cases}
        -\frac{3}{4}b_{1}^{(2)}, & i=1,\\
        -\frac{3}{4}b_{2}^{(2)}, & i=2,\\
        -\frac{3}{4}b_{3}^{(2)}, & i=3,
    \end{cases}
    \qquad\quad b_{\gamma g, [8], \mathcal{A}_i; Al}^{(2)}=\begin{cases}
        -\frac{3}{4}b_{4}^{(2)}, & i=1,\\
        -\frac{3}{4}b_{5}^{(2)}, & i=2,\\
        -\frac{3}{4}b_{6}^{(2)}, & i=3,
    \end{cases}
    \nonumber \\
    c_{\gamma g, [8], \mathcal{A}_i; Fh}^{(2)}=&\begin{cases}
        c_1^{(2)}, & i=1,\\
        c_2^{(2)}, & i=2,\\
        c_3^{(2)}, & i=3,
    \end{cases}
    \qquad\quad c_{\gamma g, [8], \mathcal{A}_i; Fl}^{(2)}=\begin{cases}
        c_4^{(2)}, & i=1,\\
        c_5^{(2)}, & i=2,\\
        c_6^{(2)}, & i=3,
    \end{cases}
    \nonumber \\
    c_{\gamma g, [8], \mathcal{A}_i; Ah}^{(2)}=&\begin{cases}
        c_{10}^{(2)}, & i=1,\\
        c_{11}^{(2)}, & i=2,\\
        c_{12}^{(2)}, & i=3,
    \end{cases}
    \qquad\quad c_{\gamma g, [8], \mathcal{A}_i; Al}^{(2)}=\begin{cases}
        c_{16}^{(2)}, & i=1,\\
        c_{17}^{(2)}, & i=2,\\
        c_{18}^{(2)}, & i=3.
    \end{cases}
\end{align}

We can write the hard functions for $\mathcal{A}_1$ as
{\small
\begin{align}
    \mathcal{H}_{p,\mathcal{A}_1}^{(0)} =& 1\,,
    \nonumber \\
    \mathcal{H}_{p,\mathcal{A}_1}^{(1)} =& 7.2303737883850050765+2.7500000000000000000\; l_{\mu_R}
    \nonumber \\
    &-2.1705584583201640717\; l_{\mu_F}-0.7500000000000000000\; l_{\mu_F}^2
    \nonumber \\
    &-0.1666666666666666667 \;n_{l} \, \left( l_{\mu_R} - l_{\mu_F}\right)\,,
    \nonumber \\
    \mathcal{H}_{p,\mathcal{A}_1}^{(2)} =& 94.35407252325475973490+46.1420558361175279209 \;l_{\mu_R}
    \nonumber \\
    &+7.5625000000000000000 \;l_{\mu_R}^2-24.0147525761474680479\; l_{\mu_F}
    \nonumber \\
    &-2.4195496254016697518 \; l_{\mu_F}^2+3.0029188437401230538 \;l_{\mu_F}^3
    \nonumber \\
    &+0.2812500000000000000 \; l_{\mu_F}^4-11.9380715207609023946 \;l_{\mu_R}\, l_{\mu_F}
    \nonumber \\
    &-4.1250000000000000000 \; l_{\mu_R} \, l_{\mu_F}^2+0.7630666365657050414\; l_{\mu_{\Lambda}}
    \nonumber \\
    &-5.8879449990648491726 \;n_{l}-3.2017912627950016922\;  n_{l}\, l_{\mu_R}
    \nonumber \\
    &-0.9166666666666666667 \; n_{l}\, l_{\mu_R}^2+1.8940451114637403392 \; n_{l}\, l_{\mu_F}
    \nonumber \\
    &-0.5634729479133743513 \; n_{l}\,l_{\mu_F}^2-0.2083333333333333333 \; n_{l}\, l_{\mu_F}^3
    \nonumber \\
    &+1.6401861527733880239 \; n_{l}\, l_{\mu_R}\, l_{\mu_F}+0.2500000000000000000 \; n_{l} \, l_{\mu_R}\, l_{\mu_F}^2
    \nonumber \\
    &+0.0277777777777777778 \; n_{l}^2\, \left(l_{\mu_R}-l_{\mu_F}\right)^2-0.27911395754850086938 \; \tilde{n}_{l}\,,
\end{align}
}

\noindent
while for $\mathcal{A}_2$, we have that
{\small
\begin{align}
    \mathcal{H}_{p,\mathcal{A}_2}^{(0)} =& 0\,,
    \nonumber \\
    \mathcal{H}_{p,\mathcal{A}_2}^{(1)} =& 0\,,
    \nonumber \\
    \mathcal{H}_{p,\mathcal{A}_2}^{(2)} =& 0.0270368951802750304\,,
    \nonumber \\
    \mathcal{H}_{p,\mathcal{A}_2}^{(3)} =& 0.92130462969977375740 +0.2230543852372690004 \; l_{\mu_R} 
    \nonumber \\
    &-0.0586851615202616444 \; l_{\mu_F}-0.0202776713852062728\; l_{\mu_F}^2
    \nonumber \\
    &-0.0549350823671903104 \; n_{l}-0.0135184475901375152\; n_{l}\, l_{\mu_R}
    \nonumber \\
    &+0.0045061491967125051\; n_{l}\,l_{\mu_F}-0.00821802672425534226\; \tilde{n}_{l}\,,
\end{align}
}

\noindent
and for $\mathcal{A}_3$, the hard functions read
{\small
\begin{align}
    \mathcal{H}_{p,\mathcal{A}_3}^{(0)} =& 1\,,
    \nonumber \\
    \mathcal{H}_{p,\mathcal{A}_3}^{(1)} =& 3.5505400726467432434+2.7500000000000000000 \;l_{\mu_R}
    \nonumber \\
    &-2.1705584583201640717\; l_{\mu_F}-0.7500000000000000000\; l_{\mu_F}^2
    \nonumber \\
    &-0.1666666666666666667\; n_{l}\, \left(l_{\mu_R}-l_{\mu_F}\right)\,,
    \nonumber \\
    \mathcal{H}_{p,\mathcal{A}_3}^{(2)} =& 48.39174004108633491851+25.9029703995570878389 \; l_{\mu_R}
    \nonumber \\
    &+7.5625000000000000000 \; l_{\mu_R}^2-16.0274583792400655661 \;l_{\mu_F}
    \nonumber \\
    &+0.3403256614020266230 \;l_{\mu_F}^2+3.0029188437401230538 \;l_{\mu_F}^3
    \nonumber \\
    &+0.2812500000000000000\; l_{\mu_F}^4-11.9380715207609023946\; l_{\mu_R}\, l_{\mu_F}
    \nonumber \\
    &-4.1250000000000000000 \;l_{\mu_R}\, l_{\mu_F}^2+0.6396965815520880586\; l_{\mu_{\Lambda}}
    \nonumber \\
    &-3.4820597895620551783\; n_{l}-1.9751800242155810811\; n_{l}\,l_{\mu_R}
    \nonumber \\
    &-0.9166666666666666667 \; n_{l}\,l_{\mu_R}^2+1.2807394921740300337 \;n_{l}\,l_{\mu_F} 
    \nonumber \\
    &-0.5634729479133743513 \; n_{l}\,l_{\mu_F}^2-0.2083333333333333333 \; n_{l}\, l_{\mu_F}^3
    \nonumber \\
    &+1.6401861527733880239 \; n_{l}\,l_{\mu_R}\, l_{\mu_F}+0.2500000000000000000 \; n_{l}\,l_{\mu_R} \,l_{\mu_F}^2
    \nonumber \\
    &+0.0277777777777777778 \; n_{l}^2\, \left(l_{\mu_R}-l_{\mu_F}\right)^2+0.39739218592419340931\; \tilde{n}_{l}\,.
\end{align}
}

\noindent
We note that the NRQCD scale dependence, $\mu_{\Lambda}$, is significantly smaller in the colour-octet configuration than in the $\gamma \gamma$ case due to the different Coulomb singularity. Analogously as in the $\gamma\gamma$ channel, we will need to specify the number of light flavours for $\tilde{n}_l$. We make use of eq.~\eqref{eq:tildenldefinition} and have that
\begin{equation}
    n_l = \begin{cases}
        3, & c\bar{c},
        \\
        4, & b\bar{b},
    \end{cases} \qquad\qquad\qquad
    \tilde{n}_l = \begin{cases}
        0, & c\bar{c},
        \\
        -2, & b\bar{b},
    \end{cases}
\end{equation}
where the light-by-light contribution with light flavours in the loops is absent in the charmonium case.

\subsection{$gg \leftrightarrow {^3P_J^{[8]}}$}

In the following, we give the result for the finite remainders and the hard functions for the process $gg \leftrightarrow {^3P_J^{[8]}}$.

At one-loop level, the finite remainders can be expressed as 
\begin{align}
    a_{gg, [8], \mathcal{A}_i; F}^{(1)}=&\begin{cases}
        a_1^{(1)}, & i=1,\\
        a_2^{(1)}, & i=2,\\
        a_3^{(1)}, & i=3,
    \end{cases}
    \nonumber \\
    a_{gg, [8], \mathcal{A}_i; A}^{(1)}=&\begin{cases}
        \frac{1}{2}a_4^{(1)}+a_6^{(1)}, & i=1,\\
        \frac{1}{16}a_1^{(1)}-\frac{1}{4}a_2^{(1)}-\frac{1}{4}a_3^{(1)}, & i=2,\\
        \frac{71}{96}a_1^{(1)}+\frac{1}{8}a_2^{(1)}-\frac{9}{8}a_3^{(1)}+\frac{1}{2}a_5^{(1)}+a_6^{(1)}, & i=3,
    \end{cases}
\end{align}
while at the two-loop level, the finite remainder reads
\begin{align}
    a_{gg, [8], \mathcal{A}_i; FF}^{(2)}=&\begin{cases}
        a_1^{(2)}, & i=1,\\
        a_2^{(2)}, & i=2,\\
        a_3^{(2)}, & i=3,
    \end{cases}
    \qquad\quad  a_{gg, [8], \mathcal{A}_i; FA}^{(2)}=\begin{cases}
        a_{10}^{(2)}-\frac{1}{2}a_{4}^{(2)}+\frac{1}{2}a_{7}^{(2)}, & i=1,\\
        a_{11}^{(2)}-\frac{1}{2}a_{5}^{(2)}+\frac{1}{2}a_{8}^{(2)}, & i=2,\\
        a_{12}^{(2)}-\frac{1}{2}a_{6}^{(2)}+\frac{1}{2}a_{9}^{(2)}, & i=3,
    \end{cases}
    \nonumber \\
    a_{gg, [8], \mathcal{A}_i; AA}^{(2)}=&\begin{cases}
        a_{19}^{(2)}, & i=1,\\
        a_{20}^{(2)}, & i=2,\\
        a_{21}^{(2)}, & i=3,
    \end{cases}
    \nonumber \\
    b_{gg, [8], \mathcal{A}_i; Fh}^{(2)}=&\begin{cases}
        2b_1^{(2)}, & i=1,\\
        2b_2^{(2)}, & i=2,\\
        2b_3^{(2)}, & i=3,
    \end{cases}
    \qquad\quad b_{gg, [8], \mathcal{A}_i; Fl}^{(2)}=\begin{cases}
        2b_4^{(2)}, & i=1,\\
        2b_5^{(2)}, & i=2,\\
        2b_6^{(2)}, & i=3,
    \end{cases}
    \nonumber \\
    b_{gg, [8], \mathcal{A}_i; Ah}^{(2)}=&\begin{cases}
        -\frac{3}{4}b_{1}^{(2)}+\frac{1}{2}b_{7}^{(2)}, & i=1,\\
        -\frac{3}{4}b_{2}^{(2)}+\frac{1}{2}b_{8}^{(2)}, & i=2,\\
        -\frac{3}{4}b_{3}^{(2)}+\frac{1}{2}b_{9}^{(2)}, & i=3,
    \end{cases}
    \qquad\quad b_{gg, [8], \mathcal{A}_i; Al}^{(2)}=\begin{cases}
        -\frac{3}{4}b_{4}^{(2)}+\frac{1}{2}b_{10}^{(2)}, & i=1,\\
        -\frac{3}{4}b_{5}^{(2)}+\frac{1}{2}b_{11}^{(2)}, & i=2,\\
        -\frac{3}{4}b_{6}^{(2)}+\frac{1}{2}b_{12}^{(2)}, & i=3,
    \end{cases}
    \nonumber \\
    c_{gg, [8], \mathcal{A}_i; Fh}^{(2)}=&\begin{cases}
        c_1^{(2)}, & i=1,\\
        c_2^{(2)}, & i=2,\\
        c_3^{(2)}, & i=3,
    \end{cases}
    \qquad\quad c_{gg, [8], \mathcal{A}_i; Fl}^{(2)}=\begin{cases}
        c_4^{(2)}, & i=1,\\
        c_5^{(2)}, & i=2,\\
        c_6^{(2)}, & i=3,
    \end{cases}
    \nonumber \\
    c_{gg, [8], \mathcal{A}_i; Ah}^{(2)}=&\begin{cases}
        \frac{1}{2}c_{7}^{(2)}+c_{10}^{(2)}, & i=1,\\
        \frac{1}{2}c_{8}^{(2)}+c_{11}^{(2)}, & i=2,\\
        \frac{1}{2}c_{9}^{(2)}+c_{12}^{(2)}, & i=3,
    \end{cases}
    \qquad\quad c_{gg, [8], \mathcal{A}_i; Al}^{(2)}=\begin{cases}
        \frac{1}{2}c_{13}^{(2)}+c_{16}^{(2)}, & i=1,\\
        \frac{1}{2}c_{14}^{(2)}+c_{17}^{(2)}, & i=2,\\
        \frac{1}{2}c_{15}^{(2)}+c_{18}^{(2)}, & i=3.
    \end{cases}
\end{align}

For the helicity amplitude $\mathcal{A}_1$, we can write the hard functions as
{\small
\begin{align}
    \mathcal{H}_{p,\mathcal{A}_1}^{(0)} =& 1\,,
    \nonumber \\
    \mathcal{H}_{p,\mathcal{A}_1}^{(1)} =& 9.25558324816643390342+5.5000000000000000000\; l_{\mu_R}
    \nonumber \\
    &-2.84111691664032814350\; l_{\mu_F}-1.5000000000000000000\; l_{\mu_F}^2
    \nonumber \\
    &-0.3333333333333333333 \;n_{l}\, \left(l_{\mu_R}- l_{\mu_F}\right)\,,
    \nonumber \\
    \mathcal{H}_{p,\mathcal{A}_1}^{(2)} =&113.86383733219355673497 +89.10856179737307970321 \; l_{\mu_R}
    \nonumber \\
    &+22.68750000000000000000 \; l_{\mu_R}^2-37.80927494394188853729 \; l_{\mu_F}
    \nonumber \\
    &-10.61476479445076752727 \; l_{\mu_F}^2+7.01167537496049221524 \; l_{\mu_F}^3
    \nonumber \\
    &+1.1250000000000000000 \; l_{\mu_F}^4-23.43921456228270718385 \; l_{\mu_R} \, l_{\mu_F}
    \nonumber \\
    &-12.37500000000000000000 \; l_{\mu_R}\, l_{\mu_F}^2+0.7630666365657050414 \; l_{\mu_{\Lambda}}
    \nonumber \\
    &-6.06030578830995788525 \; n_{l}-6.21112495741655028504 \;  n_{l} \, l_{\mu_R}
    \nonumber \\
    &-2.75000000000000000000 \; n_{l} \,l_{\mu_R}^2 +4.04649337618795695410 \; n_{l} \,l_{\mu_F}
    \nonumber \\
    &-1.22546538193347005979 \; n_{l} \, l_{\mu_F}^2-0.6666666666666666667 \;n_{l} \,l_{\mu_F}^3
    \nonumber \\
    &+4.17055845832016407175 \; n_{l} \, l_{\mu_R}\, l_{\mu_F} +0.7500000000000000000 \; n_{l} \,l_{\mu_R}\, l_{\mu_F}^2
    \nonumber \\
    &+0.0833333333333333333\; n_{l}^2 \, \left(l_{\mu_R}-l_{\mu_F}\right)^2\,,
\end{align}
}

\noindent
while for $\mathcal{A}_2$, it reads
{\small
\begin{align}
    \mathcal{H}_{p,\mathcal{A}_2}^{(0)} =& 0\,,
    \nonumber \\
    \mathcal{H}_{p,\mathcal{A}_2}^{(1)} =& 0\,,
    \nonumber \\
    \mathcal{H}_{p,\mathcal{A}_2}^{(2)} =& 0.1023002000498425373\,,
    \nonumber \\
    \mathcal{H}_{p,\mathcal{A}_2}^{(3)} =& 3.75804752711292757650+1.12530220054826791081 \; l_{\mu_R}
    \nonumber \\
    &-0.29064682893729737317 \; l_{\mu_F} -0.15345030007476380602 \;l_{\mu_F}^2
    \nonumber \\
    &-0.2143239100476125981\; l_{\mu_{\Lambda}} -0.14532358173726532713 \;n_{l}
    \nonumber \\
    &-0.06820013336656169156 \; n_{l}\,l_{\mu_R}  + 0.03410006668328084578 \; n_{l}\,l_{\mu_F} 
    \nonumber \\
    &+0.0266536687137377823\; n_{l} \, l_{\mu_{\Lambda}}\,,
\end{align}
}

\noindent
and for $\mathcal{A}_3$, we have that
{\small
\begin{align}
    \mathcal{H}_{p,\mathcal{A}_3}^{(0)} =& 1\,,
    \nonumber \\
    \mathcal{H}_{p,\mathcal{A}_3}^{(1)} =& 7.54398323143681827885+5.5000000000000000000\; l_{\mu_R}
    \nonumber \\
    &-2.84111691664032814350 \; l_{\mu_F}-1.5000000000000000000\; l_{\mu_F}^2
    \nonumber \\
    &-0.3333333333333333333 \; n_{l} \, \left(l_{\mu_R}-l_{\mu_F}\right)\,,
    \nonumber \\
    \mathcal{H}_{p,\mathcal{A}_3}^{(2)} =& 44.57845659721516605638+74.98786165935375080055 \;l_{\mu_R}
    \nonumber \\
    &+22.68750000000000000000 \;l_{\mu_R}^2-32.94641918188950892704 \;l_{\mu_F}
    \nonumber \\
    &-8.04736476935634409043 \;l_{\mu_F}^2+7.01167537496049221524 \;l_{\mu_F}^3
    \nonumber \\
    &+1.1250000000000000000\; l_{\mu_F}^4-23.43921456228270718385 \;l_{\mu_R} \, l_{\mu_F}
    \nonumber \\
    &-12.37500000000000000000 \; l_{\mu_R}\, l_{\mu_F}^2+1.3097854173188324424\; l_{\mu_{\Lambda}}
    \nonumber \\
    &-2.36479019302401116571\; n_{l}-5.35532494905174247276 \;n_{l} \, l_{\mu_R}
    \nonumber \\
    &-2.75000000000000000000 \; n_{l} \, l_{\mu_R}^2 +3.47596003727808507925 \; n_{l} \, l_{\mu_F}
    \nonumber \\
    &-1.22546538193347005979 \; n_{l} \, l_{\mu_F}^2-0.6666666666666666667 \; n_{l} \, l_{\mu_F}^3
    \nonumber \\
    &+4.17055845832016407175 \; n_{l} \, l_{\mu_R} \, l_{\mu_F} +0.7500000000000000000 \; n_{l} \, l_{\mu_R} \, l_{\mu_F}^2
    \nonumber \\
    &-0.0833333333333333333 \; n_{l} \,l_{\mu_{\Lambda}}+ 0.0833333333333333333\; n_{l}^2\, \left(l_{\mu_R}- l_{\mu_F}\right)^2\,.
\end{align}
}

\noindent
As already observed in the colour-singlet case, due to the inclusion of the $^3S_1^{[8]}$ form factor in the $gg$ channel, the NRQCD scale dependence is more involved as in the $\gamma g$ case. As before, we use $n_l=3$ for the charmonium and $n_l=4$ for the bottomonium case.

\section{Conclusions}
\label{sec:conclusions}

In this paper, we have computed analytically the two-loop form factors and helicity amplitudes for the processes $\gamma \gamma \leftrightarrow {^3P_J^{[1]}}$, $gg \leftrightarrow {^3P_J^{[1]}}$, $\gamma g \leftrightarrow {^3P_J^{[8]}}$, $gg \leftrightarrow {^3P_J^{[8]}}$ and we have also available the high-precision numerical results up to 1000 digits which can be used for phenomenology.

For the process $\gamma \gamma \leftrightarrow {^3P_J^{[1]}}$, which is needed for the exclusive decay of $\chi_{Q,J}$ to di-photon, we find overall agreement between our result and the numerical result in ref.~\cite{Sang:2015uxg} which serves as cross-check of our calculation. The results presented here for the form factor $gg \leftrightarrow {^3P_J^{[1]}}$ can be used as ingredient for the virtual NNLO contribution for $\chi_{Q,J}$ hadro-production and decay. In addition to the $^3P_J^{[1]}$ state, the production and decay of $\chi_{Q,J}$ also receives contributions from the $^3S_1^{[8]}$ state. As the tree-level contribution to the form factor $gg \leftrightarrow {^3S_1^{[8]}}$ vanishes, it is the square of its one-loop amplitude which contributes at NNLO as well. We have computed and presented the one-loop result in Appendix~\ref{sec:3S18contribution}. The results for the form factors with ${^3P_J^{[8]}}$ state in colour-octet configuration contribute to the higher terms in the $v$-expansion of the LDME for $J/\psi$ and $\Upsilon$ production and decay.

We find that our two-loop amplitudes can be expressed in terms of the same 76 master integrals that we already encountered in the $S$-wave case in ref.~\cite{Abreu:2022cco}. We had computed these integrals analytically and also provided high-precision numerics in ref.~\cite{Abreu:2022vei}. However, compared to the $S$-wave case in our companion paper \cite{Abreu:2022cco}, the cancellation of the infra-red pole structure in the $P$-wave case is significantly more involved and subtle.

After performing the UV renormalisation for the form factor in the $\gamma \gamma$ channel, we are able to reproduce the known Coulomb singularities for the colour-singlet $^3P_J^{[1]}$ states \cite{Hoang:2006ty, Sang:2015uxg}. In the case of the $\gamma g$ channel, after subtracting the standard QCD infra-red singularities, we obtain here for the first time the Coulomb singularities for the $^3P_J^{[8]}$ states.

However in the $gg$-channel case, we find that the NRQCD pole structure at two-loop level involves, in addition to the Coulomb singularity, also a new singularity component that is directly related to the helicity configuration of the gluons. This entanglement between initial and final-state particles naively suggests a breakdown of NRQCD factorisation. However, we are able to trace back the origin of this new pole structure to the missing contribution from the higher Fock state $\vert {^3S_1^{[8]}}g \rangle$ and restore factorisation by including the form factor $gg \leftrightarrow {^3S_1^{[8]}}$.

We have presented the finite remainders of the two-loop helicity amplitudes for all processes in analytic form in Section~\ref{sec:formfactors}. The finite remainders are expressed in terms of a bases given in Appendix~\ref{sec:analyticsFF} that we have also made available in electronic form in ref.~\cite{formFactorPwaveGit}. We have also presented the hard functions for all helicity configurations including the dependence on the scales $\mu_R$, $\mu_F$ and $\mu_{\Lambda}$. The results presented here can be used to compute the NNLO corrections for $\chi_{Q,J}$ hadro-production and decay. We leave this for future work.

\acknowledgments{We acknowledge valuable discussions with Claude Duhr, Hua-Sheng Shao, Lukas Simon and Lorenzo Tancredi. This research was supported by the ANR PIA funding under grant ANR-20-IDEES-0002.}

\appendix

\section{Polarisation vectors and tensors}
\label{sec:appendixpolarisation}

In this Appendix, we give details regarding different polarisation identities for the vectors $\varepsilon_{\mu}$ and the rank-$2$ tensor $\varepsilon_{\mu \nu}$ needed in the calculation.

We can parameterise the polarisation sum for the massless initial-state particles as follows
\begin{equation}
\begin{split}
\sum_{\rm pol} \varepsilon_1^{\rho} \varepsilon^{*,\rho'}_1 =& -g^{\rho \rho'} + \frac{1}{k_1\cdot k_2}\left(k_1^\rho k_2^{\rho'}+k_2^\rho k_1^{\rho'}\right),
\\
\sum_{\rm pol} \varepsilon_2^{\sigma} \varepsilon^{*,\sigma'}_2 =& -g^{\sigma \sigma'} + \frac{1}{k_1\cdot k_2}\left(k_1^\sigma k_2^{\sigma'}+k_2^\sigma k_1^{\sigma'}\right).
\end{split}
\end{equation}
For a massive spin-1 particle, we can write the polarisation sum as
\begin{equation}
\begin{split}
\sum_{\rm pol} \varepsilon^{\mu} \varepsilon^{*,\nu} = \Pi^{\mu \nu}=-g^{\mu \nu} + \frac{1}{2 k_1 \cdot k_2}\left(k_1^\mu + k_2^\mu\right)\left(k_1^\nu + k_2^\nu\right).
\end{split}
\label{eq:massivespinsum}
\end{equation}

For the polarisation tensor of the $P$-wave bound state, the polarisation sums for the different $J$ states read \cite{Petrelli:1997ge}
\begin{equation}
\begin{split}
	\varepsilon_{\mu \nu}^{(0)} \varepsilon_{\mu' \nu'}^{(0),*}=&\frac{1}{d-1}\Pi_{\mu \nu} \Pi_{\mu' \nu'},
	\\
	\sum_{\rm pol} \varepsilon_{\mu \nu}^{(1)} \varepsilon_{\mu' \nu'}^{(1),*}=&\frac{1}{2}\left(\Pi_{\mu \mu'} \Pi_{\nu \nu'}-\Pi_{\mu \nu'} \Pi_{\mu' \nu}\right),
	\\
	\sum_{\rm pol} \varepsilon_{\mu \nu}^{(2)} \varepsilon_{\mu' \nu'}^{(2),*}=&\frac{1}{2}\left(\Pi_{\mu \mu'} \Pi_{\nu \nu'}+\Pi_{\mu \nu'} \Pi_{\mu' \nu}\right)-\frac{1}{d-1}\Pi_{\mu \nu} \Pi_{\mu' \nu'}.
\end{split}
\label{eq:polarisationPwave}
\end{equation}

In the following, we will give explicit representations for the different polarisation states for the vectors $\epsilon_{\mu}$ and the tensor $\varepsilon_{\mu \nu}$. For this, we first define the polarisation states for the massive particle as
\begin{equation}
\begin{split}
    \varepsilon_{\mu}{\left(+1\right)}=&\frac{1}{\sqrt{2}}\left(0,-1,-i,0\right),
    \\
    \varepsilon_{\mu}{\left(-1\right)}=&\frac{1}{\sqrt{2}}\left(0,1,-i,0\right),
    \\
    \varepsilon_{\mu}{\left(0\right)}=&\left(0,0,0,1\right).
\end{split}
\end{equation}
The polarisation states of the two massless particles in the initial state can then be expressed in terms of these vectors as
\begin{equation}
    \begin{split}
    \varepsilon_{1, \mu}{\left(+1\right)}=\varepsilon_{2, \mu}{\left(-1\right)}=\varepsilon_{\mu}{\left(+1\right)}, \qquad\qquad    \varepsilon_{1, \mu}{\left(-1\right)}=\varepsilon_{2, \mu}{\left(+1\right)}=\varepsilon_{\mu}{\left(-1\right)}.
    \end{split}
\end{equation}

Using the spin decomposition for $J=2$ particles, we can write \cite{Poppe:1986dq},
\begin{equation}
    \begin{split}
        \varepsilon^{(2)}_{\mu \nu}{\left(J_z=\pm 2\right)} =& \varepsilon_{\mu}{\left(\pm 1\right)} \varepsilon_{\nu}{\left(\pm 1\right)},
        \\
        \varepsilon^{(2)}_{\mu \nu}{\left(J_z=\pm 1\right)} =& \frac{1}{\sqrt{2}}\big(\varepsilon_{\mu}{\left(\pm 1\right)}\varepsilon_{\nu}{\left(0\right)} + \varepsilon_{\mu}{\left(0\right)}\varepsilon_{\nu}{\left(\pm 1\right)}\big),
        \\
        \varepsilon^{(2)}_{\mu \nu}{\left(J_z=0\right)} =& \frac{1}{\sqrt{6}}\big(2\varepsilon_{\mu}{\left(0\right)}\varepsilon_{\nu}{\left(0\right)} + \varepsilon_{\mu}{\left(1\right)}\varepsilon_{\nu}{\left(-1\right)} + \varepsilon_{\mu}{\left(-1\right)}\varepsilon_{\nu}{\left(1\right)}\big)
        \\
        =& \frac{1}{\sqrt{\left(d-2\right)\left(d-1\right)}}\big(\left(d-1\right)\varepsilon_{\mu}{\left(0\right)}\varepsilon_{\nu}{\left(0\right)} - \Pi_{\mu \nu}\big).
    \end{split}
\end{equation}
The last expression has been extended to $d$ dimensions by using the fact that its trace vanishes, $\varepsilon^{(2), \mu}_\mu = 0$, and using the properties $\varepsilon{(\pm 1)}=-\varepsilon^*{(\mp 1)}$, $\varepsilon{(0)}=\varepsilon^*{(0)}$ and the polarisation sum identity in eq.~\eqref{eq:massivespinsum}.

For the $J=1$ states, we can express the tensor as
\begin{equation}
    \begin{split}
        \varepsilon^{(1)}_{\mu \nu}{\left(J_z=\pm 1\right)} =& \frac{1}{\sqrt{2}}\big(\varepsilon_{\mu}{\left(\pm 1\right)}\varepsilon_{\nu}{\left(0\right)} - \varepsilon_{\mu}{\left(0\right)}\varepsilon_{\nu}{\left(\pm 1\right)}\big),
        \\
        \varepsilon^{(1)}_{\mu \nu}{\left(J_z=0\right)} =& \frac{1}{\sqrt{2}}\big(\varepsilon_{\mu}{\left(1\right)}\varepsilon_{\nu}{\left(-1\right)} - \varepsilon_{\mu}{\left(-1\right)}\varepsilon_{\nu}{\left(1\right)}\big),
    \end{split}
\end{equation}
and for the $J=0$ case, we have that
\begin{equation}
    \begin{split}
        \varepsilon^{(0)}_{\mu \nu}{\left(J_z=0\right)} =& \frac{1}{\sqrt{3}}\big(\varepsilon_{\mu}{\left(0\right)}\varepsilon_{\nu}{\left(0\right)} - \varepsilon_{\mu}{\left(1\right)}\varepsilon_{\nu}{\left(-1\right)} - \varepsilon_{\mu}{\left(-1\right)}\varepsilon_{\nu}{\left(1\right)}\big)
        \\
        =& \frac{1}{\sqrt{d-1}}\Pi_{\mu \nu}.
    \end{split}
\end{equation}

\section{Bare amplitude structure}
\label{sec:barepolestructure}

In this Appendix, we present the bare amplitude structure for the three different coefficients $c_1$, $c_2$ and $c_3$, as given in eq.~\eqref{eq:helicityamplitude}, for the different channels $\gamma \gamma \leftrightarrow {^3P_J^{[1]}}$, $gg \leftrightarrow {^3P_J^{[1]}}$, $\gamma g \leftrightarrow {^3P_J^{[8]}}$ and $gg \leftrightarrow {^3P_J^{[8]}}$. As outlined in Section~\ref{sec:barecoefficientsdescription}, the tree-level amplitude for the coefficient $c_1$ vanishes for all channels considered here. We have therefore used the same normalisation factor as for the $c_2$ coefficient.

In order to keep the tables as compact as possible, we list only unique form-factor entries. For instance, the abelian coefficients for a given $c_i$ that appear at one-loop level, $\mathcal{F}_{F}$, are identical for all channels ($\gamma\gamma$, $\gamma g$, $gg$) considered here. Similarly at two-loop level, for a given $c_i$, the abelian coefficients, $\mathcal{F}^{(2)}_{FF}$, $\mathcal{F}^{(2)}_{F,h;\text{vac}}$ and $\mathcal{F}^{(2)}_{F,l;\text{vac}}$ are the same for all channels. In contrast to this, we make the observation already done in the $^1S_0$ case \cite{Abreu:2022cco} that, in the light-by-light contribution, the abelian coefficients, $\mathcal{F}^{(2)}_{F,h;\text{lbl}}$ and $\mathcal{F}^{(2)}_{F,l;\text{lbl}}$, are identical for channels where the quarkonium state in the same colour state, either singlet $[1]$ or octet $[8]$, but differ by a factor of two between these $[1]$ and $[8]$ configurations as we will indicate below.

\vspace{0.3cm}
At one-loop level, the coefficients read:
{\small
\begin{center}
\renewcommand{\arraystretch}{1.5}
\begin{longtable}{| m{1.4cm} || m{1cm} | m{2.0cm} | m{2.5cm} | m{4.0cm} | m{1.5cm} |} 
 \hline
 $\gamma \gamma, [1]$ & $\epsilon^{-2}$ & $ \epsilon^{-1}$ & $\epsilon^{0}$ & $\epsilon^{1}$ & $\epsilon^{2}$ \\[5pt]
 \hline\hline
 $\mathcal{F}^{(1)}_F, c_1$ & 0 & $-\frac{3}{2}$ & $-3-4\log{2}$ & $\frac{1}{24}\pi^2-5+4 \log^2{2}-20 \log{2}$ & $-38.66148$ \\[5pt]
 \hline
 $\mathcal{F}^{(1)}_F, c_2$ & 0 & $-\frac{3}{4}$ & $-2+\frac{3}{8}\pi^2-6 \log{2}$ & $\frac{21 }{4}\zeta_3-\frac{3}{2}-\frac{1}{16}\pi^2+6 \log^2{2}-14 \log{2}$ & $-10.87118$ \\[5pt]
 \hline
  $\mathcal{F}^{(1)}_F, c_3$ & 0 & 0 & $-2$ & $\frac{1}{8}\pi ^2 - 2 - 2 \log{2}$ & $-10.99795$ \\[5pt]
 \hline
 \end{longtable}

 \vspace{0.1cm}
 \renewcommand{\arraystretch}{1.5}
 \begin{longtable}{| m{1.4cm} || m{1cm} | m{2.0cm} | m{2.5cm} | m{4.0cm} | m{1.5cm} |}
 \hline
 $gg, [1]$ & $\epsilon^{-2}$ & $ \epsilon^{-1}$ & $\epsilon^{0}$ & $\epsilon^{1}$ & $\epsilon^{2}$ \\[5pt]
 \hline\hline
 $\mathcal{F}^{(1)}_A, c_1$ & 0 & 0 & $\frac{1}{3}-\frac{11 i \pi}{6}-\frac{3}{8}\pi ^2+\frac{8}{3}\log{2}$ & $-\frac{21}{4}\zeta_3+\frac{29}{144}\pi^2+\frac{7}{18}-\frac{245 i \pi }{36}-\frac{8}{3}\log^2{2}+\frac{118}{9}\log{2}+\frac{11}{3} i \pi  \log{2}$ & $12.89474 - i\, 14.02918$ \\[5pt]
 \hline
 $\mathcal{F}^{(1)}_A, c_2$ & $-\frac{1}{2}$ & $\log{2} - \frac{i \pi}{2}$ & $\frac{2}{3}-\frac{11 i \pi }{12}-\frac{7}{48}\pi^2-\log ^2{2}+\frac{7}{3}\log{2}+i \pi \log{2}$ & $-\frac{119}{24}\zeta_3+\frac{i \pi ^3}{8}+\frac{53}{288}\pi ^2+\frac{55}{36}-\frac{197 i \pi }{72}+\frac{2}{3}\log^3{2}-\frac{7}{3}\log^2{2}-i \pi \log^2{2}-\frac{7}{12} \pi^2 \log{2}+\frac{56}{9}\log{2}+\frac{11 i \pi}{6}\log{2}$ & $-6.13509 - i\, 4.07911$ \\[5pt]
 \hline
 $\mathcal{F}^{(1)}_A, c_3$ & $-\frac{1}{2}$ & $\log{2}-\frac{i \pi}{2}$ & $\frac{1}{24}\pi ^2+\frac{1}{6}-\frac{11 i \pi}{12}-\log^2{2}+\frac{5}{6}\log{2}+i \pi  \log{2}$ & $-\frac{7}{3}\zeta_3+\frac{i \pi^3}{8}+\frac{53}{288}\pi^2+\frac{13}{36}-\frac{203 i \pi}{72}+\frac{2}{3}\log^3{2}-\frac{5}{6}\log^2{2}-i \pi  \log^2{2} - \frac{7}{12} \pi^2 \log{2}+\frac{79}{18}\log{2}+\frac{11}{6} i \pi \log{2}$ & $- 2.28306 - i \, 4.54521$ \\[5pt]
 \hline
 \end{longtable}

 \vspace{0.1cm}
 \renewcommand{\arraystretch}{1.5}
 \begin{longtable}{| m{1.4cm} || m{1cm} | m{2.0cm} | m{2.5cm} | m{4.0cm} | m{1.5cm} |}
 \hline
 $\gamma g, [8]$ & $\epsilon^{-2}$ & $ \epsilon^{-1}$ & $\epsilon^{0}$ & $\epsilon^{1}$ & $\epsilon^{2}$ \\[5pt]
 \hline\hline
 $\mathcal{F}^{(1)}_A, c_1$ & 0 & 0 & $\frac{3}{16}\pi^2+\frac{1}{4}+\log{2}$ & $\frac{21}{8}\zeta_3+\frac{29}{96}\pi^2+\frac{1}{8}-\log^2{2}+\frac{7}{2}\log{2}$ & $18.96700$ \\[5pt]
 \hline
 $\mathcal{F}^{(1)}_A, c_2$ & $-\frac{1}{4}$ & $\frac{1}{2}\log{2}-\frac{1}{4}$ & $\frac{5}{96}\pi^2+\frac{3}{8}-\frac{1}{2} \log^2{2}+\frac{3}{2}\log{2}$ & $\frac{49}{48}\zeta_3+\frac{17}{192}\pi^2+\frac{1}{16}+\frac{1}{3}\log^3{2}-\frac{3}{2}\log^2{2}-\frac{1}{24} \pi^2 \log{2}+\frac{7}{4}\log{2}$ & $7.52142$ \\[5pt]
 \hline
 $\mathcal{F}^{(1)}_A, c_3$ & $-\frac{1}{4}$ & $\frac{1}{2}\log{2}-\frac{1}{4}$ & $\frac{7}{48}\pi^2+\frac{3}{8}-\frac{1}{2} \log^2{2}+\frac{3}{4}\log{2}$ & $\frac{7}{3}\zeta_3+\frac{29}{192}\pi^2-\frac{5}{16}+\frac{1}{3}\log^3{2}-\frac{3}{4}\log^2{2}-\frac{1}{24} \pi^2 \log{2}+\log{2}$ & $11.68320$ \\[5pt]
 \hline
 \end{longtable}

 \vspace{0.1cm}
 \renewcommand{\arraystretch}{1.5}
 \begin{longtable}{| m{1.4cm} || m{1cm} | m{2.0cm} | m{2.5cm} | m{4.0cm} | m{1.5cm} |}
 \hline
 $gg, [8]$ & $\epsilon^{-2}$ & $ \epsilon^{-1}$ & $\epsilon^{0}$ & $\epsilon^{1}$ & $\epsilon^{2}$ \\[5pt]
 \hline\hline
 $\mathcal{F}^{(1)}_A, c_1$ & 0 & 0 & $\frac{5}{12}-\frac{11 i \pi }{12}+\frac{7}{3}\log{2}$ & $\frac{29}{72}\pi ^2-\frac{245 i \pi }{72}+\frac{23}{72}-\frac{7}{3} \log^2{2}+\frac{181}{18}\log{2}+\frac{11}{6} i \pi  \log{2}$ & $25.41437 - i\, 7.01459$ \\[5pt]
 \hline
 $\mathcal{F}^{(1)}_A, c_2$ & $-\frac{1}{2}$ & $-\frac{1}{4}-\frac{i \pi}{4}+\log{2}$ & $-\frac{11 i \pi}{24}+\frac{17}{24}-\frac{1}{48}\pi ^2-\log^2{2}+\frac{8}{3}\log{2}+\frac{1}{2} i \pi  \log{2}$ & $-\frac{35}{24}\zeta_3+\frac{i \pi^3}{16}+\frac{13}{72}\pi^2-\frac{197 i \pi}{144}+\frac{119}{144}+\frac{2}{3}\log^3{2}-\frac{8}{3}\log^2{2}-\frac{1}{2} i \pi \log^2{2}-\frac{1}{3} \pi^2 \log{2}+\frac{175}{36}\log{2}+\frac{11}{12} i \pi  \log{2}$ & $4.45388 - i\, 2.03955$ \\[5pt]
 \hline
 $\mathcal{F}^{(1)}_A, c_3$ & $-\frac{1}{2}$ & $-\frac{1}{4}-\frac{i \pi}{4}+\log{2}$ & $\frac{1}{6}\pi ^2-\frac{11 i \pi }{24}+\frac{11}{24}-\log^2{2}+\frac{7}{6}\log{2}+\frac{1}{2} i \pi  \log{2}$ & $\frac{7}{6}\zeta_3+\frac{i \pi^3}{16}+\frac{35}{144}\pi^2-\frac{203 i \pi }{144}-\frac{19}{144}+\frac{2}{3}\log^3{2}-\frac{7}{6}\log^2{2}-\frac{1}{2} i \pi \log^2{2}-\frac{1}{3} \pi^2 \log{2}+\frac{115}{36}\log{2}+\frac{11}{12} i \pi  \log{2}$ & $10.54167 - i\, 2.27260$ \\[5pt]
 \hline
\end{longtable}
\end{center}
}
\vspace{-0.2cm}
\noindent At two-loop level we find the following structures for the bare amplitudes:
{\small
\begin{center}
\renewcommand{\arraystretch}{1.5}
\begin{longtable}{| m{1.5cm} || m{0.5cm} | m{1.7cm} | m{2.6cm} | m{4.5cm} | m{1.6cm} |}
 \hline
  $\gamma \gamma, [1]$ & $\epsilon^{-4}$ & $\epsilon^{-3}$ & $\epsilon^{-2}$ & $\epsilon^{-1}$ & $\epsilon^{0}$ \\[5pt]
 \hline\hline
 $\mathcal{F}^{(2)}_{FF}, c_1$ & 0 & 0 & $-\frac{9}{16}$ & $\frac{3}{80}\pi^2-\frac{179}{32}-14 \log{2}$ & $-33.19740$ \\[5pt]
 \hline
 $\mathcal{F}^{(2)}_{FF}, c_2$ & 0 & 0 & $-\frac{13}{32}$ & $\frac{61}{120}\pi^2+\frac{509}{192}-\frac{27}{2} \log{2}$ & $-10.86064$ \\[5pt]
 \hline
 $\mathcal{F}^{(2)}_{FF}, c_3$ & 0 & 0 & $-\frac{13}{32}$ & $-\frac{353}{96}-\frac{97}{960}\pi^2+\frac{3}{4}\log{2}$ & $-13.78042$ \\[5pt]
 \hline
 $\mathcal{F}^{(2)}_{FA}, c_1$ & 0 & 0 & $-\frac{11}{16}$ & $-\frac{361}{96}-\frac{11}{3} \log{2}$ & $-28.55878$ \\[5pt]
 \hline
 $\mathcal{F}^{(2)}_{FA}, c_2$ & 0 & 0 & $-\frac{11}{32}$ & $\frac{29}{96}\pi^2-\frac{419}{192}-\frac{11}{2} \log{2}$ & $-8.84342$ \\[5pt]
 \hline
 $\mathcal{F}^{(2)}_{FA}, c_3$ & 0 & 0 & 0 & $-\frac{161}{96}-\frac{1}{24}\pi^2$ & $-7.21067$ \\[5pt]
 \hline
 $\mathcal{F}^{(2)}_{Fh;\text{vac}}, c_1$ & 0 & 0 & $\frac{1}{4}$ & $\frac{29}{24}+\frac{4}{3}\log{2}$ & $5.06748$\\
\hline
\end{longtable}

\vspace{0.1cm}
\renewcommand{\arraystretch}{1.5}
\begin{longtable}{| m{1.5cm} || m{0.5cm} | m{1.7cm} | m{2.6cm} | m{4.5cm} | m{1.6cm} |}  
 \hline
  $gg, [1]$ & $\epsilon^{-4}$ & $\epsilon^{-3}$ & $\epsilon^{-2}$ & $\epsilon^{-1}$ & $\epsilon^{0}$ \\[5pt]
 \hline\hline
 $\mathcal{F}^{(2)}_{FA}, c_1$ & 0 & $\frac{15}{16}$ & $\frac{11}{16}+\frac{15 i \pi }{16}+\frac{1}{8}\log{2}$ & $-\frac{71}{96}\pi^2-\frac{217}{96}+\frac{41 i \pi }{16}-\frac{33}{8} \log^2{2}+\frac{73}{12}\log{2}+\frac{1}{8} i \pi  \log{2}$ & $-28.25603 + i\, 25.62622$ \\[5pt]
 \hline
 $\mathcal{F}^{(2)}_{FA}, c_2$ & 0 & 0 & $-\frac{19}{32}+\frac{3 i \pi }{8}-\frac{3}{16}\pi^2+3 \log{2}$ & $-\frac{21}{8}\zeta_3-\frac{3 i \pi ^3}{16}-\frac{53}{192}\pi^2-\frac{911}{192}+\frac{9 i \pi }{4}-9 \log^2{2}+\frac{3}{8} \pi ^2 \log{2}+\frac{93}{16}\log{2}+\frac{9}{4} i \pi  \log{2}$ & $-15.66527 + i\, 9.38879$ \\[5pt]
 \hline
 $\mathcal{F}^{(2)}_{FA}, c_3$ & 0 & $-\frac{15}{64}$ & $\frac{7}{32}+\frac{15}{32}\log{2}$ & $-\frac{197}{96}-\frac{11}{192}\pi^2+\frac{75 i \pi }{64}-\frac{15}{32} \log^2{2}+\frac{7}{8}\log{2}$ & $1.59950 + i\, 6.54061$ \\[5pt]
 \hline
 $\mathcal{F}^{(2)}_{AA}, c_1$ & 0 & 0 & $-\frac{1}{6}+\frac{11 i \pi }{12}+\frac{3}{16}\pi ^2-\frac{4}{3}\log{2}$ & $\frac{21}{8}\zeta_3+\frac{3 i \pi^3}{16}+\frac{1}{9}-\frac{49}{36}\pi^2+\frac{14 i \pi }{9}+4 \log^2{2}-\frac{3}{8} \pi ^2 \log{2}-\frac{34}{9}\log{2}-5 i \pi  \log{2}$ & $-13.29097- i\, 47.78227$ \\[5pt]
 \hline
 $\mathcal{F}^{(2)}_{AA}, c_2$ & $\frac{1}{8}$ & $-\frac{11}{96}+\frac{i \pi }{4}-\frac{1}{2}\log{2}$ & $-\frac{1}{24}\pi ^2-\frac{163}{288}+\frac{11 i \pi }{48}+\log^2{2}-\frac{17}{24}\log{2}-i \pi  \log{2}$ & $\frac{61}{24}\zeta_3+\frac{i \pi^3}{32}-\frac{13 i \pi }{48}-\frac{439}{432}-\frac{311}{576}\pi^2-\frac{4}{3} \log^3{2}+\frac{31}{12}\log^2{2}+2 i \pi  \log^2{2}+\frac{29}{48} \pi^2 \log{2}+\frac{1}{8}\log{2}-\frac{25}{12} i \pi  \log{2}$ & $-5.12709 + i\, 2.87614$ \\
\hline
\end{longtable}
\end{center}
}
\vspace{-0.2cm}
\noindent For the colour-octet states, we note that the light-by-light contributions in the $C_F$ colour factor are the same as for the colour-singlet states by a factor of two,
\begin{equation}
    \mathcal{F}^{(2), [8]}_{Fh;\text{lbl}}=2 \times \mathcal{F}^{(2), [1]}_{Fh;\text{lbl}}, \qquad\qquad \mathcal{F}^{(2), [8]}_{Fl;\text{lbl}}=2 \times \mathcal{F}^{(2), [1]}_{Fl;\text{lbl}}.
\end{equation}
{\small
\begin{center}
\renewcommand{\arraystretch}{1.5}
\begin{longtable}{| m{1.5cm} || m{0.5cm} | m{1.7cm} | m{2.6cm} | m{4.5cm} | m{1.6cm} |}  
 \hline
  $\gamma g, [8]$ & $\epsilon^{-4}$ & $\epsilon^{-3}$ & $\epsilon^{-2}$ & $\epsilon^{-1}$ & $\epsilon^{0}$ \\[5pt]
 \hline\hline
 $\mathcal{F}^{(2)}_{FA}, c_1$ & 0 & $\frac{3}{16}$ & $\frac{5}{8}\log{2}-\frac{3}{8}$ & $-\frac{79}{96}-\frac{19}{240}\pi^2-\frac{21}{8} \log^2{2}+\frac{65}{24}\log{2}$ & $-32.37218$ \\[5pt]
 \hline
 $\mathcal{F}^{(2)}_{FA}, c_2$ & 0 & $-\frac{3}{32}$ & $-\frac{3}{32}\pi^2+\frac{1}{32}+\frac{27}{16}\log{2}$ & $-\frac{21}{16}\zeta_3+\frac{33}{640}\pi^2-\frac{221}{64}-\frac{75}{16} \log^2{2}+\frac{3}{16} \pi^2 \log{2}-\frac{15}{32}\log{2}$ & $-12.71296$ \\[5pt]
 \hline
 $\mathcal{F}^{(2)}_{FA}, c_3$ & 0 & $-\frac{9}{128}$ & $\frac{75}{64}+\frac{9}{64}\log{2}$ & $\frac{1}{12}-\frac{3}{160}\pi^2-\frac{9}{64} \log^2{2}-\frac{49}{32}\log{2}$ & $-3.27014$ \\[5pt]
 \hline
 $\mathcal{F}^{(2)}_{AA}, c_1$ & 0 & 0 & $-\frac{1}{16}-\frac{3}{64}\pi^2-\frac{1}{4}\log{2}$ & $-\frac{21}{32}\zeta_3+\frac{203}{1920}\pi^2+\frac{13}{96}+\frac{3}{4}\log^2{2}+\frac{3}{32} \pi^2 \log{2}-\frac{1}{12}\log{2}$ & $13.17612$ \\[5pt]
 \hline
 $\mathcal{F}^{(2)}_{AA}, c_2$ & $\frac{1}{32}$ & $\frac{1}{192}-\frac{1}{8}\log{2}$ & $-\frac{1}{128}\pi^2-\frac{169}{576}+\frac{1}{4}\log^2{2}-\frac{13}{48}\log{2}$ & $-\frac{55}{192}\zeta_3+\frac{51}{1280}\pi^2-\frac{275}{1728}-\frac{1}{3} \log^3{2}+\frac{19}{24}\log^2{2}+\frac{1}{64} \pi^2 \log{2}+\frac{175}{144}\log{2}$ & $5.22892$ \\[5pt]
 \hline
 $\mathcal{F}^{(2)}_{AA}, c_3$ & $\frac{1}{32}$ & $\frac{1}{192}-\frac{1}{8}\log{2}$ & $-\frac{1}{32}\pi^2-\frac{169}{576}+\frac{1}{4}\log^2{2}-\frac{1}{12}\log{2}$ & $-\frac{59}{96}\zeta_3+\frac{111}{1280}\pi^2-\frac{113}{1728}-\frac{1}{3} \log^3{2}+\frac{11}{48}\log^2{2}+\frac{1}{16} \pi^2 \log{2}+\frac{65}{72}\log{2}$ & $8.63657$ \\[5pt]
 \hline
 $\mathcal{F}^{(2)}_{Ah;\text{vac}}, c_1$ & 0 & 0 & 0 & $-\frac{1}{12}-\frac{1}{16}\pi^2-\frac{1}{3}\log{2}$ & $-2.64817$ \\
\hline
\end{longtable}

\vspace{0.1cm}
\renewcommand{\arraystretch}{1.5}
\begin{longtable}{| m{1.5cm} || m{0.5cm} | m{1.7cm} | m{2.6cm} | m{4.5cm} | m{1.6cm} |}  
 \hline
  $gg, [8]$ & $\epsilon^{-4}$ & $\epsilon^{-3}$ & $\epsilon^{-2}$ & $\epsilon^{-1}$ & $\epsilon^{0}$ \\[5pt]
 \hline\hline
 $\mathcal{F}^{(2)}_{FA}, c_1$ & 0 & $\frac{21}{32}$ & $\frac{5}{16}+\frac{15 i \pi }{32}+\frac{11}{16}\log{2}$ & $-\frac{7}{96}-\frac{431}{960}\pi^2+\frac{41 i \pi }{32}-\frac{75}{16} \log^2{2}+\frac{91}{12}\log{2}+\frac{1}{16} i \pi \log{2}$ & $-32.22081 + i\, 12.81311$ \\[5pt]
 \hline
 $\mathcal{F}^{(2)}_{FA}, c_2$ & 0 & $-\frac{3}{32}$ & $-\frac{3}{32}+\frac{3 i \pi }{16}-\frac{3}{16}\pi^2+\frac{51}{16}\log{2}$ & $-\frac{21}{8}\zeta_3-\frac{3 i \pi ^3}{32}-\frac{303}{64}-\frac{19}{80}\pi^2+\frac{9 i \pi }{8}-\frac{147}{16} \log^2{2}+\frac{3}{8} \pi^2 \log{2}+\frac{83}{16}\log{2}+\frac{9}{8} i \pi  \log{2}$ & $-16.12389 + i\, 4.69439$ \\[5pt]
 \hline
 $\mathcal{F}^{(2)}_{FA}, c_3$ & 0 & $-\frac{3}{16}$ & $\frac{41}{32}+\frac{3}{8}\log{2}$ & $-\frac{5}{48}-\frac{17}{640}\pi^2+\frac{75 i \pi}{128}-\frac{3}{8} \log^2{2}-\frac{35}{32}\log{2}$ & $1.13494 + i\, 3.27031$ \\[5pt]
 \hline
 $\mathcal{F}^{(2)}_{AA}, c_1$ & 0 & 0 & $-\frac{5}{24}+\frac{11 i \pi }{24}-\frac{7}{6}\log{2}$ & $\frac{17}{144}-\frac{539}{1440}\pi^2+\frac{71 i \pi }{72}+\frac{7}{2}\log^2{2}-\frac{55}{18}\log{2}-\frac{29}{12} i \pi  \log{2}$ & $9.78016 - i\, 24.17426$ \\[5pt]
 \hline
 $\mathcal{F}^{(2)}_{AA}, c_2$ & $\frac{1}{8}$ & $\frac{1}{96}+\frac{i \pi }{8}-\frac{1}{2}\log{2}$ & $-\frac{1}{96}\pi^2-\frac{193}{288}+\frac{17 i \pi }{96}+\log^2{2}-\frac{9}{8}\log{2}-\frac{1}{2} i \pi  \log{2}$ & $\frac{35}{48}\zeta_3-\frac{i \pi}{32}-\frac{i \pi^3}{64}-\frac{43}{360}\pi^2-\frac{331}{432}-\frac{4}{3} \log^3{2}+\frac{10}{3}\log^2{2}+i \pi \log^2{2}+\frac{11}{48} \pi^2 \log{2}+\frac{53}{72}\log{2}-\frac{5}{4} i \pi  \log{2}$ & $2.93365 - i\, 3.38398$ \\
\hline
\end{longtable}
\end{center}
}

\section{Renormalised amplitude structure}
\label{sec:renpole}

In this Appendix, we give the coefficients for the renormalised amplitude structure. We proceed in a similar fashion as in the case of the bare amplitude in Appendix~\ref{sec:barepolestructure}. The coefficient for the one-loop amplitudes read as follows:

{\small
\begin{center}
\renewcommand{\arraystretch}{1.5}
\begin{longtable}{| m{1.4cm} || m{1cm} | m{2.0cm} | m{2.5cm} | m{4.0cm} | m{1.5cm} |} 
 \hline
 $\gamma \gamma, [1]$ & $\epsilon^{-2}$ & $ \epsilon^{-1}$ & $\left.\epsilon^{0}\right\vert_{\mu=m_Q}$ & $\left.\epsilon^{1}\right\vert_{\mu=m_Q}$ & $\left.\epsilon^{2}\right\vert_{\mu=m_Q}$ \\[5pt]
 \hline\hline
 $\overline{\mathcal{F}}^{(1)}_F, c_1$ & 0 & 0 & $-1-4 \log{2}$ & $\frac{1}{6}\pi^2 - 1+4 \log^2{2}-20 \log{2}$ & $-29.61758$ \\[5pt]
 \hline
 $\overline{\mathcal{F}}^{(1)}_F, c_2$ & 0 & 0 & $\frac{3}{8}\pi^2 - 1 - 6 \log{2}$ & $\frac{21}{4}\zeta_3 + \frac{1}{2}+6 \log^2{2} - 14 \log{2}$ & $-6.34923$ \\[5pt]
 \hline
 $\overline{\mathcal{F}}^{(1)}_F, c_3$ & 0 & 0 & $-2$ & $\frac{1}{8}\pi^2 - 2 - 2 \log{2}$ & $-10.99795$ \\[5pt]
 \hline
\end{longtable}

\vspace{0.1cm}
\renewcommand{\arraystretch}{1.5}
\begin{longtable}{| m{1.4cm} || m{1cm} | m{2.0cm} | m{2.5cm} | m{4.0cm} | m{1.5cm} |} 
 \hline
 $gg, [1]$ & $\epsilon^{-2}$ & $ \epsilon^{-1}$ & $\left.\epsilon^{0}\right\vert_{\mu=m_Q}$ & $\left.\epsilon^{1}\right\vert_{\mu=m_Q}$ & $\left.\epsilon^{2}\right\vert_{\mu=m_Q}$ \\[5pt]
 \hline\hline
 $\overline{\mathcal{F}}^{(1)}_A, c_1$ & 0 & 0 & $\frac{1}{3}-\frac{11 i \pi }{6}-\frac{3}{8}\pi^2+\frac{8}{3}\log{2}$ & $-\frac{21}{4}\zeta_3+\frac{29}{144}\pi^2+\frac{7}{18}-\frac{245 i \pi }{36}-\frac{8}{3} \log^2{2}+\frac{118}{9}\log{2}+\frac{11}{3} i \pi  \log{2}$ & $12.89474 - i\, 14.02918$\\[5pt]
 \hline
 $\overline{\mathcal{F}}^{(1)}_A, c_2$ & $-\frac{1}{2}$ & $-\frac{1}{2}l_{\mu}-\frac{11}{12}-\frac{i \pi }{2}+\log{2}$ & $\frac{2}{3}-\frac{11 i \pi }{12}-\frac{7}{48}\pi^2 - \log^2{2}+\frac{7}{3}\log{2}+i \pi  \log{2}$ & $-\frac{119}{24}\zeta_3+\frac{i \pi^3}{8}+\frac{53}{288}\pi^2+\frac{55}{36}-\frac{197 i \pi }{72}+\frac{2}{3}\log^3{2}-\frac{7}{3}\log^2{2}-i \pi  \log^2{2}-\frac{7}{12} \pi^2 \log{2}+\frac{56}{9}\log{2}+\frac{11}{6} i \pi  \log{2}$ & $-6.13509 - i\, 4.07911$\\[5pt]
 \hline
 $\overline{\mathcal{F}}^{(1)}_A, c_3$ & $-\frac{1}{2}$ & $-\frac{1}{2}l_{\mu}-\frac{11}{12}-\frac{i \pi }{2}+\log{2}$ & $\frac{1}{24}\pi^2+\frac{1}{6}-\frac{11 i \pi }{12}-\log^2{2}+\frac{5}{6}\log{2}+i \pi  \log{2}$ & $-\frac{7}{3}\zeta_3+\frac{i \pi^3}{8}+\frac{53}{288}\pi^2+\frac{13}{36}-\frac{203 i \pi }{72}+\frac{2}{3}\log^3{2} - \frac{5}{6}\log^2{2} - i \pi  \log^2{2}-\frac{7}{12} \pi^2 \log{2}+\frac{79}{18}\log{2}+\frac{11}{6} i \pi  \log{2}$ & $-2.28306 - i\, 4.54521$\\[5pt]
 \hline
 $\overline{\mathcal{F}}^{(1)}_l, c_1$ & 0 & 0 & 0 & 0 & 0 \\[5pt]
 \hline
 $\overline{\mathcal{F}}^{(1)}_l, c_2$ & 0 & $\frac{1}{3}$ & 0 & 0 & 0 \\[5pt]
 \hline
 $\overline{\mathcal{F}}^{(1)}_l, c_3$ & 0 & $\frac{1}{3}$ & 0 & 0 & 0 \\[5pt]
 \hline
\end{longtable}

\vspace{0.1cm}
\renewcommand{\arraystretch}{1.5}
\begin{longtable}{| m{1.4cm} || m{1cm} | m{2.0cm} | m{2.5cm} | m{4.0cm} | m{1.5cm} |} 
 \hline
 $\gamma g, [8]$ & $\epsilon^{-2}$ & $ \epsilon^{-1}$ & $\left.\epsilon^{0}\right\vert_{\mu=m_Q}$ & $\left.\epsilon^{1}\right\vert_{\mu=m_Q}$ & $\left.\epsilon^{2}\right\vert_{\mu=m_Q}$ \\[5pt]
 \hline\hline
 $\overline{\mathcal{F}}^{(1)}_A, c_1$ & 0 & 0 & $\frac{3}{16}\pi^2 + \frac{1}{4}+\log{2}$ & $\frac{21}{8}\zeta_3 + \frac{29}{96}\pi^2 + \frac{1}{8}-\log^2{2}+\frac{7}{2}\log{2}$ & $18.96700$\\[5pt]
 \hline
 $\overline{\mathcal{F}}^{(1)}_A, c_2$ & $-\frac{1}{4}$ & $-\frac{1}{4}l_{\mu}-\frac{17}{24}+\frac{1}{2}\log{2}$ & $\frac{5}{96}\pi^2 + \frac{3}{8}-\frac{1}{2} \log^2{2}+\frac{3}{2}\log{2}$ & $\frac{49}{48}\zeta_3 + \frac{17}{192}\pi^2 + \frac{1}{16}+\frac{1}{3}\log^3{2} - \frac{3}{2}\log^2{2} - \frac{1}{24} \pi^2 \log{2}+\frac{7}{4}\log{2}$ & $7.52142$ \\[5pt]
 \hline
 $\overline{\mathcal{F}}^{(1)}_A, c_3$ & $-\frac{1}{4}$ & $-\frac{1}{4}l_{\mu} - \frac{17}{24}+\frac{1}{2}\log{2}$ & $\frac{7}{48}\pi^2 + \frac{3}{8}-\frac{1}{2} \log^2{2} + \frac{3}{4}\log{2}$ & $\frac{7}{3}\zeta_3 + \frac{29}{192}\pi^2 - \frac{5}{16}+\frac{1}{3}\log^3{2} - \frac{3}{4}\log^2{2} - \frac{1}{24} \pi^2 \log{2}+\log{2}$ & $11.68320$\\[5pt]
 \hline
 $\overline{\mathcal{F}}^{(1)}_l, c_1$ & 0 & 0 & 0 & 0 & 0 \\[5pt]
 \hline
 $\overline{\mathcal{F}}^{(1)}_l, c_2$ & 0 & $\frac{1}{6}$ & 0 & 0 & 0 \\[5pt]
 \hline
 $\overline{\mathcal{F}}^{(1)}_l, c_3$ & 0 & $\frac{1}{6}$ & 0 & 0 & 0 \\[5pt]
 \hline
\end{longtable}

\vspace{0.1cm}
\begin{longtable}{| m{1.4cm} || m{1cm} | m{2.0cm} | m{2.5cm} | m{4.0cm} | m{1.5cm} |} 
 \hline
 $gg, [8]$ & $\epsilon^{-2}$ & $ \epsilon^{-1}$ & $\left.\epsilon^{0}\right\vert_{\mu=m_Q}$ & $\left.\epsilon^{1}\right\vert_{\mu=m_Q}$ & $\left.\epsilon^{2}\right\vert_{\mu=m_Q}$ \\[5pt]
 \hline\hline
 $\overline{\mathcal{F}}^{(1)}_A, c_1$ & 0 & 0 & $\frac{5}{12}-\frac{11 i \pi }{12}+\frac{7}{3}\log{2}$ & $\frac{29}{72}\pi^2 - \frac{245 i \pi }{72}+\frac{23}{72}-\frac{7}{3} \log^2{2} + \frac{181}{18}\log{2} + \frac{11}{6} i \pi  \log{2}$ & $25.41437 - i\, 7.01459$\\[5pt]
 \hline
 $\overline{\mathcal{F}}^{(1)}_A, c_2$ & $-\frac{1}{2}$ & $-\frac{1}{2}l_{\mu} -\frac{7}{6}-\frac{i \pi }{4}+\log{2}$ & $-\frac{11 i \pi }{24}+\frac{17}{24}-\frac{1}{48}\pi^2 - \log^2{2} + \frac{8}{3}\log{2} + \frac{1}{2} i \pi  \log{2}$ & $-\frac{35}{24}\zeta_3 + \frac{i \pi^3}{16}+\frac{13}{72}\pi^2 - \frac{197 i \pi }{144}+\frac{119}{144}+\frac{2}{3}\log^3{2} - \frac{8}{3}\log^2{2} - \frac{1}{2} i \pi \log^2{2} - \frac{1}{3} \pi^2 \log{2} + \frac{175}{36}\log{2} + \frac{11}{12} i \pi  \log{2}$ & $4.45388 - i\, 2.03955$\\[5pt]
 \hline
 $\overline{\mathcal{F}}^{(1)}_A, c_3$ & $-\frac{1}{2}$ & $-\frac{1}{2}l_{\mu} - \frac{7}{6}-\frac{i \pi }{4}+\log{2}$ & $\frac{1}{6}\pi^2 - \frac{11 i \pi }{24}+\frac{11}{24}-\log^2{2} + \frac{7}{6}\log{2} + \frac{1}{2} i \pi  \log{2}$ & $\frac{7}{6}\zeta_3 + \frac{i \pi^3}{16}+\frac{35}{144}\pi^2 - \frac{203 i \pi }{144}-\frac{19}{144}+\frac{2}{3}\log^3{2} - \frac{7}{6}\log^2{2} - \frac{1}{2} i \pi \log^2{2} - \frac{1}{3} \pi^2 \log{2} + \frac{115}{36}\log{2} + \frac{11}{12} i \pi \log{2}$ & $10.54167 - i\, 2.27260$\\[5pt]
 \hline
 $\overline{\mathcal{F}}^{(1)}_l, c_1$ & 0 & 0 & 0 & 0 & 0 \\[5pt]
 \hline
 $\overline{\mathcal{F}}^{(1)}_l, c_2$ & 0 & $\frac{1}{3}$ & 0 & 0 & 0 \\[5pt]
 \hline
 $\overline{\mathcal{F}}^{(1)}_l, c_3$ & 0 & $\frac{1}{3}$ & 0 & 0 & 0 \\[5pt]
 \hline
\end{longtable}
\end{center}
}
\vspace{-0.2cm}
\noindent We now consider the two-loop renormalised coefficients:
{\small
\begin{center}
\renewcommand{\arraystretch}{1.5}
\begin{longtable}{| m{1.5cm} || m{0.5cm} | m{1.7cm} | m{2.6cm} | m{4.5cm} | m{1.6cm} |}
 \hline
 $\gamma \gamma, [1]$  & $\epsilon^{-4}$ & $\epsilon^{-3}$ & $\epsilon^{-2}$ & $\epsilon^{-1}$ & $\left.\epsilon^{0}\right\vert_{\mu=m_Q}$ \\[5pt]
 \hline\hline
 $\overline{\mathcal{F}}^{(2)}_{FF}, c_1$ & 0 & 0 & 0 & $\frac{9}{40}\pi^2$ & $6.85484$ \\[5pt]
 \hline
 $\overline{\mathcal{F}}^{(2)}_{FF}, c_2$ & 0 & 0 & 0 & $-\frac{13}{240}\pi^2$ & $-7.52047$ \\[5pt]
 \hline
 $\overline{\mathcal{F}}^{(2)}_{FF}, c_3$ & 0 & 0 & 0 & $-\frac{13}{240}\pi^2$ & $-5.93024$ \\[5pt]
 \hline
 $\overline{\mathcal{F}}^{(2)}_{FA}, c_1$ & 0 & 0 & 0 & 0 & $-6.74585$ \\[5pt]
 \hline
 $\overline{\mathcal{F}}^{(2)}_{FA}, c_2$ & 0 & 0 & 0 & $-\frac{1}{24}\pi^2$ & $-3.64930$ \\[5pt]
 \hline
 $\overline{\mathcal{F}}^{(2)}_{FA}, c_3$ & 0 & 0 & 0 & $-\frac{1}{24}\pi^2$ & $-5.78205$ \\[5pt]
 \hline
 $\overline{\mathcal{F}}^{(2)}_{Fh;\text{vac}}, c_1$ & 0 & 0 & 0 & 0 & $-0.13179$ \\
\hline
\end{longtable}

\vspace{0.1cm}
\renewcommand{\arraystretch}{1.5}
\begin{longtable}{| m{1.5cm} || m{0.5cm} | m{1.7cm} | m{2.6cm} | m{4.5cm} | m{1.6cm} |}  
 \hline
 $gg, [1]$  & $\epsilon^{-4}$ & $\epsilon^{-3}$ & $\epsilon^{-2}$ & $\epsilon^{-1}$ & $\left.\epsilon^{0}\right\vert_{\mu=m_Q}$ \\[5pt]
 \hline\hline
 $\overline{\mathcal{F}}^{(2)}_{FA}, c_1$ & 0 & 0 & $\frac{1}{2}+2 \log{2}$ & $l_{\mu}+4 l_{\mu} \log{2}-\frac{1}{12}\pi^2+\frac{17}{12}+\frac{i \pi }{2}-6 \log^2{2}+\frac{38}{3}\log{2}+2 i \pi  \log{2}$ & $5.91232 + i\, 19.34463$ \\[5pt]
 \hline
 $\overline{\mathcal{F}}^{(2)}_{FA}, c_2$ & 0 & 0 & $\frac{1}{2}-\frac{3}{16}\pi^2 + 3 \log{2}$ & $l_{\mu}-\frac{3}{8}\pi^2 l_{\mu}+6 l_{\mu} \log{2}-\frac{21}{8}\zeta_3+\frac{7}{6}-\frac{3 i \pi^3}{16}-\frac{61}{96}\pi^2+\frac{i \pi }{2}-9 \log^2{2}+\frac{3}{8} \pi^2 \log{2}+\frac{27}{2}\log{2}+3 i \pi  \log{2}$ & $-1.63959 + i\, 0.04656$ \\[5pt]
 \hline
 $\overline{\mathcal{F}}^{(2)}_{FA}, c_3$ & 0 & 0 & $1$ & $2 l_{\mu}+\frac{17}{6}+i \pi -\frac{5}{48}\pi^2 - \log{2}$ & $0.08989 + i\, 4.78570$ \\[5pt]
 \hline
 $\overline{\mathcal{F}}^{(2)}_{AA}, c_1$ & 0 & 0 & $-\frac{1}{6}+\frac{11 i \pi }{12}+\frac{3}{16}\pi^2-\frac{4}{3}\log{2}$ & $\frac{3}{8}\pi^2 l_{\mu}-\frac{1}{3}l_{\mu}+\frac{11 i \pi}{6}l_{\mu}-\frac{8}{3} l_{\mu} \log{2}+\frac{21}{8}\zeta_3+\frac{3 i \pi^3}{16}-\frac{1}{2}-\frac{97}{144}\pi^2+\frac{59 i \pi }{12}+4 \log^2{2}-\frac{3}{8} \pi^2 \log{2}-\frac{26}{3}\log{2} - 5 i \pi  \log{2}$ & $-20.39044 - i\, 23.22330$ \\
\hline
\end{longtable}
\end{center}
}
\vspace{-0.2cm}
\noindent Similarly as in the case of the bare amplitude, we can relate the abelian light-by-light contribution between colour-octet and colour-singlet states as follows,
\begin{equation}
    \overline{\mathcal{F}}^{(2), [8]}_{Fh;\text{lbl}}=2 \times \overline{\mathcal{F}}^{(2), [1]}_{Fh;\text{lbl}}, \qquad\qquad \overline{\mathcal{F}}^{(2), [8]}_{Fl;\text{lbl}}=2 \times \overline{\mathcal{F}}^{(2), [1]}_{Fl;\text{lbl}}.
\end{equation}
{\small
\begin{center}
\renewcommand{\arraystretch}{1.5}
\begin{longtable}{| m{1.5cm} || m{0.5cm} | m{1.7cm} | m{2.6cm} | m{4.5cm} | m{1.6cm} |}  
 \hline
 $\gamma g, [8]$  & $\epsilon^{-4}$ & $\epsilon^{-3}$ & $\epsilon^{-2}$ & $\epsilon^{-1}$ & $\left.\epsilon^{0}\right\vert_{\mu=m_Q}$ \\[5pt]
 \hline\hline
 $\overline{\mathcal{F}}^{(2)}_{FA}, c_1$ & 0 & 0 & $\frac{1}{4}+\log{2}$ & $\frac{1}{2}l_{\mu} + 2 l_{\mu} \log{2} - \frac{4}{15}\pi^2 + \frac{23}{24}-3 \log^2{2}+\frac{22}{3}\log{2}$ & $-5.55268$ \\[5pt]
 \hline
 $\overline{\mathcal{F}}^{(2)}_{FA}, c_2$ & 0 & 0 & $\frac{1}{4}-\frac{3}{32}\pi^2 + \frac{3}{2}\log{2}$ & $\frac{1}{2}l_{\mu} - \frac{3}{16}\pi^2 l_{\mu} + 3 l_{\mu} \log{2} - \frac{21}{16}\zeta_3 + \frac{7}{12}-\frac{81}{320}\pi^2 - \frac{9}{2} \log^2{2} + \frac{3}{16} \pi^2 \log{2} + \frac{29}{4}\log{2}$ & $2.65810$ \\[5pt]
 \hline
 $\overline{\mathcal{F}}^{(2)}_{FA}, c_3$ & 0 & 0 & $\frac{1}{2}$ & $l_{\mu}+\frac{23}{12}-\frac{3}{160}\pi^2 - \frac{1}{2}\log{2}$ & $-0.23372$ \\[5pt]
 \hline
 $\overline{\mathcal{F}}^{(2)}_{AA}, c_1$ & 0 & 0 & $-\frac{1}{16}-\frac{3}{64}\pi^2 - \frac{1}{4}\log{2}$ & $-\frac{1}{8}l_{\mu} - \frac{3}{32}\pi^2 l_{\mu} - \frac{1}{2} l_{\mu} \log{2} - \frac{21}{32}\zeta_3 - \frac{5}{24}-\frac{73}{480}\pi^2+\frac{3}{4}\log^2{2}+\frac{3}{32} \pi^2 \log{2} - \frac{35}{24}\log{2}$ & $1.89094$ \\[5pt]
 \hline
 $\overline{\mathcal{F}}^{(2)}_{AA}, c_2$ & $\frac{1}{32}$ & $\frac{1}{16}l_{\mu}+\frac{67}{192}-\frac{1}{8}\log{2}$ & $\frac{1}{16}l_{\mu}^2 + \frac{17}{48}l_{\mu} - \frac{1}{4} l_{\mu} \log{2} - \frac{1}{128}\pi^2 + \frac{421}{1152}+\frac{1}{4}\log^2{2} - \frac{23}{24}\log{2}$ & $\frac{1}{24} l_{\mu}^3 + \frac{35}{192}l_{\mu}^2 - \frac{1}{4} l_{\mu}^2 \log{2} - \frac{1}{64}\pi^2 l_{\mu} - \frac{35}{144}l_{\mu} + \frac{1}{2} l_{\mu} \log^2{2}-\frac{59}{48} l_{\mu} \log{2} -\frac{55}{192}\zeta_3 - \frac{23}{27}-\frac{61}{1920}\pi^2 - \frac{1}{3}\log^3{2} + \frac{71}{48}\log^2{2} + \frac{1}{64} \pi^2 \log{2} - \frac{61}{72}\log{2}$ & $1.81650$ \\
\hline
\end{longtable}

\vspace{0.1cm}
\renewcommand{\arraystretch}{1.5}
\begin{longtable}{| m{1.5cm} || m{0.5cm} | m{1.7cm} | m{2.6cm} | m{4.5cm} | m{1.6cm} |}  
 \hline
 $gg, [8]$  & $\epsilon^{-4}$ & $\epsilon^{-3}$ & $\epsilon^{-2}$ & $\epsilon^{-1}$ & $\left.\epsilon^{0}\right\vert_{\mu=m_Q}$ \\[5pt]
 \hline\hline
 $\overline{\mathcal{F}}^{(2)}_{FA}, c_1$ & 0 & 0 & $\frac{1}{2}+2 \log{2}$ & $l_{\mu} + 4 l_{\mu} \log{2} + \frac{5}{3}-\frac{37}{120}\pi^2 + \frac{i \pi}{4}-6 \log^2{2} + \frac{41}{3}\log{2} + i \pi \log{2}$ & $0.77640 + i\, 9.67232$ \\[5pt]
 \hline
 $\overline{\mathcal{F}}^{(2)}_{FA}, c_2$ & 0 & 0 & $\frac{1}{2}-\frac{3}{16}\pi^2 + 3 \log{2}$ & $l_{\mu} - \frac{3}{8}\pi^2 l_{\mu} + 6 l_{\mu} \log{2} - \frac{21}{8}\zeta_3 + \frac{7}{6}-\frac{11}{20}\pi^2 - \frac{3 i \pi ^3}{32}+\frac{i \pi }{4}-9 \log^2{2} + \frac{3}{8} \pi^2 \log{2} + \frac{3}{2} i \pi  \log{2} + 14 \log{2}$ & $3.66296 + i\, 0.02328$ \\[5pt]
 \hline
 $\overline{\mathcal{F}}^{(2)}_{FA}, c_3$ & 0 & 0 & $1$ & $2 l_{\mu}+\frac{10}{3}+\frac{i \pi }{2}-\frac{1}{20}\pi^2 - \log{2}$ & $2.70225 + i\, 2.39285$ \\[5pt]
 \hline
 $\overline{\mathcal{F}}^{(2)}_{AA}, c_1$ & 0 & 0 & $-\frac{5}{24}+\frac{11 i \pi }{24}-\frac{7}{6}\log{2}$ & $-\frac{5}{12}l_{\mu} + \frac{11 i \pi}{12}l_{\mu} -\frac{7}{3} l_{\mu} \log{2} - \frac{31}{48}-\frac{539}{1440}\pi^2 + \frac{8 i \pi }{3}+\frac{7}{2}\log^2{2} - \frac{22}{3}\log{2} - \frac{29}{12} i \pi \log{2}$ & $-8.81648 - i\, 11.89477$ \\
\hline
\end{longtable}
\end{center}
}

\section{The $^3S_1^{[8]}$ form-factor}
\label{sec:3S18contribution}

In this Appendix, we give a few details on the computation for the $gg \leftrightarrow {^3S_1^{[8]}}$ form-factor that is needed to cancel the infra-red singularity of the $^3P_J$ form-factors. In the following, we make use of the same notation as done in the main text.

According to the Landau-Yang theorem \cite{Landau:1948kw,Yang:1950rg}, the amplitudes for the decay of a massive spin-1 particle into to two massless spin-1 particles vanish. This is indeed the case for the form-factors $\gamma \gamma \leftrightarrow {^3S_1^{[1]}}$ and $gg \leftrightarrow {^3S_1^{[1]}}$. However, when colour-octet states are involved, the amplitude does not necessarily vanish as was already observed in refs.~\cite{Cacciari:2015ela,Beenakker:2015mra,Pleitez:2015cpa,Ma:2014oha}.

In order to investigate this step further, we make use of the axial gauge conditions of the two massless polarisation vectors in eq.~\eqref{eq:polarisationconstraints} and the equivalent condition for the polarisation vector of the massive particle which reads
\begin{equation}
    \varepsilon_{\mathcal{Q}} \cdot \left(k_1 + k_2\right) = 0.
\end{equation}
We can then express the Lorentz structure of the amplitude as follows
\begin{equation}
    \mathcal{A} = \mathcal{A}_{\mu \nu \rho}\; \varepsilon_{1}^{\mu} \varepsilon_{2}^{\nu} \varepsilon_{\mathcal{Q}}^{\rho}  = c_4 \;\frac{1}{m_Q} \big(\left(k_1 - k_2\right) \cdot \varepsilon_{\mathcal{Q}} \big) \left(\varepsilon_1 \cdot \varepsilon_2\right).
\label{eq:tensorstructure3S18}
\end{equation}
In order to project out the form-factor $c_4$ from the tensor $\mathcal{A}_{\mu \nu \rho}$, we follow the same steps as in the main text and define the projection operator as
\begin{equation}
    P_4^{\mu \nu \rho} = \frac{1}{4m_Q}\frac{1}{\left(2-d\right)} \left(g^{\mu \nu} - \frac{1}{2m_Q^2} k_1^{\mu} k_2^{\nu} - \frac{1}{2m_Q^2} k_2^{\mu} k_1^{\nu}\right)\left(k_1^{\rho}-k_2^{\rho}\right),
\end{equation}
which by construction yields
\begin{equation}
    P_4^{\mu \nu \rho} \mathcal{A}_{\mu \nu \rho} = c_4.
\end{equation}

Indeed, the form factors with a colour-singlet final state are symmetric under exchange of the two initial-state particles, hence the amplitude vanishes for these. In the case of the form factors with a colour-octet final-state, we have to take into account also the symmetries coming from the colour algebra. The form-factor $\gamma g \leftrightarrow {^3S_1^{[8]}}$ exhibits only the trivial tensor $\delta^{bc}$, hence vanishes as well. However, the form-factor $g g \leftrightarrow {^3S_1^{[8]}}$ comes with both a symmetric and an anti-symmetric colour tensor, $d^{abc}$ and $f^{abc}$ respectively. In combination with the Lorentz structure in eq.~\eqref{eq:tensorstructure3S18}, we then conclude that only the anti-symmetric colour tensor $f^{abc}$ survives.

Using the explicit representations of the polarisation vectors in Appendix~\ref{sec:appendixpolarisation}, denoting the amplitude configuration as $\mathcal{A}_{J_z,\lambda_1,\lambda_2}$, we find that the only non-vanishing configuration is
\begin{equation}
    \mathcal{A}_{0,1,1}=\mathcal{A}_{0,-1,-1}=\mathcal{A}_4= -2 c_4,
\end{equation}
where the $^3S_1^{[8]}$ state is longitudinal polarised with $J_z=0$.

We can write the amplitude up to one-loop level as
\begin{equation}
    \mathcal{F}_{{^3S_1^{[8]}},c_4}=c_4/\mathcal{N}_{{^3S_1^{[8]}},c_4}=\left(\frac{\alpha_s^B}{\pi}\right) \left[\mathcal{F}_{{^3S_1^{[8]}},c_4}^{(0)}+\left(\frac{\alpha_s^B}{\pi}\right) \mathcal{F}_{{^3S_1^{[8]}},c_4}^{(1)} + \mathcal{O}{\left({\alpha_s^B}\right)^2}\right],
\end{equation}
where the normalisation factor reads
\begin{equation}
    \mathcal{N}_{{^3S_1^{[8]}},c_4} = -i\frac{4\sqrt{2}\,\pi^2}{\sqrt{m_Q}}\, \sqrt{2}\,T_F\, f^{abc}/2.
\end{equation}
It turns out that the tree-level amplitude vanishes, $\mathcal{F}_{{^3S_1^{[8]}},c_4}^{(0)}=0$. In contrast to this, the one-loop amplitude does not vanish. We can use eq.~\eqref{eq:UVrenormalisationtermatordern} and eq.~\eqref{eq:renormalisationCT1expression} to perform UV renormalisation. Here the process-dependent mass counter-term reads
\begin{equation}
    \mathcal{F}_{{^3S_1^{[8]}},c_4}^{(0, 1 \textrm{ mass CT})} = 1.
\end{equation}

The renormalised amplitude can then be written as\footnote{It is understood that the coupling power at tree-level reads $q=1$ for this process, hence global factors related to the leading coupling as for instance $\mu^{2\epsilon}$ can then be factored out.}
\begin{equation}
    \overline{\mathcal{F}}_{{^3S_1^{[8]}},c_4}= \overline{c}_4/\mathcal{N}_{{^3S_1^{[8]}},c_4} =\left(\frac{\alpha_s^{(n_l)}}{\pi}\right)^2 \overline{\mathcal{F}}_{{^3S_1^{[8]}},c_4}^{(1)} + \mathcal{O}{\left({\alpha_s^{(n_l)}}\right)^3},
\end{equation}
where the colour-structure reads
\begin{equation}
    \overline{\mathcal{F}}_{{^3S_1^{[8]}},c_4}^{(1,k)} = C_A\, \overline{\mathcal{F}}_{{^3S_1^{[8]}},c_4; A}^{(1,k)} + C_F\, \overline{\mathcal{F}}_{{^3S_1^{[8]}},c_4; F}^{(1,k)} + T_F n_l\, \overline{\mathcal{F}}_{{^3S_1^{[8]}},c_4; l}^{(1,k)} + T_F n_h\, \overline{\mathcal{F}}_{{^3S_1^{[8]}},c_4; h}^{(1,k)},
    \label{eq:colour3S18}
\end{equation}
where, similarly as in eq.~\eqref{eq:expandamplitudeepsorder}, we have expanded the amplitude in $\epsilon$.

The one-loop amplitude is finite and we give below the coefficients 
up to $\mathcal{O}{\left(\epsilon\right)}$,
\begin{align}
        \overline{\mathcal{F}}_{{^3S_1^{[8]}},c_4; A}^{(1,0)} =& \left(\frac{\pi^2}{8}-\frac{5}{12}-\frac{1}{2}\log{2}\right),
        \\
        \overline{\mathcal{F}}_{{^3S_1^{[8]}},c_4; A}^{(1,1)} =& \left(\frac{7}{4}\zeta_3 + \frac{\pi^2}{12}-\frac{17}{18}+\frac{1}{2}\log^2{2} - \frac{1}{2}\log{2}\right) + l_{\mu}\; \overline{\mathcal{F}}_{{^3S_1^{[8]}},c_4; A}^{(1,0)},
        \\[10pt]
        \overline{\mathcal{F}}_{{^3S_1^{[8]}},c_4; F}^{(1,0)} =& \left(\frac{1}{2}-\frac{\pi^2}{4}+2 \log{2}\right), 
        \\
        \overline{\mathcal{F}}_{{^3S_1^{[8]}},c_4; F}^{(1,1)} =& \left(-\frac{7}{2}\zeta_3-\frac{\pi^2}{12}+1-2 \log^2{2}+4 \log{2}\right) + l_{\mu}\; \overline{\mathcal{F}}_{{^3S_1^{[8]}},c_4; F}^{(1,0)},
        \\[10pt]
        \overline{\mathcal{F}}_{{^3S_1^{[8]}},c_4; l}^{(1,0)} =& -\frac{1}{6},
        \\
        \overline{\mathcal{F}}_{{^3S_1^{[8]}},c_4; l}^{(1,1)} =& \left(-\frac{11}{18}-\frac{i \pi}{6}+\frac{1}{3}\log{2}\right) + l_{\mu}\; \overline{\mathcal{F}}_{{^3S_1^{[8]}},c_4; l}^{(1,0)},
        \\[10pt]
        \overline{\mathcal{F}}_{{^3S_1^{[8]}},c_4; h}^{(1,0)} =& \left(\frac{\pi ^2}{8}-\frac{7}{6}\right),
        \\
        \overline{\mathcal{F}}_{{^3S_1^{[8]}},c_4; h}^{(1,1)} =& \left(\frac{7}{4}\zeta_3 + \frac{\pi^2}{8}-\frac{59}{18}\right) + l_{\mu}\; \overline{\mathcal{F}}_{{^3S_1^{[8]}},c_4; h}^{(1,0)}.
\end{align}

In ref.~\cite{Ma:2014oha}, this one-loop amplitude has been computed up to the finite piece $\mathcal{O}{\left(\epsilon^0\right)}$ only and the result there has been presented by inserting explicitly the colour factors and the number of flavours, corresponding for the charmonium case, $C_A=3$, $C_F=4/3$, $T_F=1/2$, $n_h=1$ and $n_l=3$. In ref.~\cite{Beenakker:2015mra}, the same result has been obtained up $\mathcal{O}{\left(\epsilon^0\right)}$ but they distinguish between the regular contribution and the contributions proportional to $n_h$ and $n_l$. Inserting the corresponding colour factors and flavour numbers, we find that our result is in agreement with both refs.~\cite{Ma:2014oha,Beenakker:2015mra}. In addition to the full colour-structure decomposition in eq.~\eqref{eq:colour3S18}, our result also includes the new $\mathcal{O}{\left(\epsilon\right)}$ term which is needed in the main part of text.

From these coefficients, we can then trivially extract the result for the helicity configuration $\mathcal{A}_4$
\begin{equation}
    \overline{\mathcal{F}}_{{^3S_1^{[8]}},\mathcal{A}_4}= \overline{\mathcal{A}}_4/\mathcal{N}_{{^3S_1^{[8]}},\mathcal{A}_4} =\left(\frac{\alpha_s^{(n_l)}}{\pi}\right)^2 \overline{\mathcal{F}}_{{^3S_1^{[8]}},\mathcal{A}_4}^{(1)} + \mathcal{O}{\left({\alpha_s^{(n_l)}}\right)^3},
\end{equation}
where $\overline{\mathcal{A}}_4=-2\overline{c}_4$ and where we have made use of the same normalisation $\mathcal{N}_{{^3S_1^{[8]}},\mathcal{A}_4}=\mathcal{N}_{{^3S_1^{[8]}},c_4}$. The finite remainder of the one-loop amplitude is
\begin{equation}
\mathcal{F}^{{\rm fin}, (1)}_{^3S_1^{[8]}, \mathcal{A}_4}= \overline{\mathcal{F}}^{(1,0)}_{^3S_1^{[8]}, \mathcal{A}_4}\,,
\end{equation}
and the hard function as defined in eq.~\eqref{eq:definitionhardfunctionA} then reads up to 20 digits,
\begin{align}
    \mathcal{H}_{^3S_1^{[8]}, \mathcal{A}_4}^{(0)} =& 0\,,
    \nonumber \\
    \mathcal{H}_{^3S_1^{[8]}, \mathcal{A}_4}^{(1)} =& 0\,,
    \nonumber \\
    \mathcal{H}_{^3S_1^{[8]}, \mathcal{A}_4}^{(2)} =& 1.7960761912769237093 -0.4467258905111629225\; n_{l}
    \nonumber \\
    & + 0.0277777777777777778\; n_{l}^2\,,
\end{align}
where we set $n_l=3$ for charmonium and $n_l=4$ for bottomonium states.

\section{Scale-dependent coefficients}
\label{sec:appendixscaledependence}

In this Appendix, we have collected the scale-dependent coefficients that are needed to describe the scale dependence of the finite remainder. These were partially derived in our previous paper \cite{Abreu:2022cco}. Here, we also present the new $\overline{\mathcal{D}}^{(i)}_{\mu}$ coefficient due to the inclusion of the $gg \leftrightarrow {^3S_1^{[8]}}$ form factor.

The coefficients associated to $\mu_R$ read
\begin{align}
    \mathcal{C}_{\mu_R}^{(1)}=& \frac{q}{4}\, \beta_0\, l_{\mu_R},
    \\
    \mathcal{D}_{\mu_R}^{(1)}=& \frac{1}{4}\left(1+q\right)\, \beta_0\, l_{\mu_R},
    \\
    \mathcal{C}_{\mu_R}^{(2)}=& \frac{1}{32}q \left(1+q\right)\, \beta_0^2\, l^2_{\mu_R} + \frac{1}{16} q\, \beta_1\, l_{\mu_R},
\end{align}
where
\begin{equation}
    \beta_0 = \frac{11}{3}\,C_A - \frac{4}{3}\, T_F\, n_l,\qquad\qquad \beta_1 = \frac{34}{3}\,C_A^2 - \frac{20}{3}\,C_A\, T_F\, n_l - 4\, C_F\, T_F\, n_l.
\end{equation}

The coefficients associated to the scale $\mu_F$ are more involved
\begin{align}
    \mathcal{C}_{\mu_F}^{(1)}=& -\frac{q\, C_A}{4}\, l_{\mu_F}^2 - \frac{1}{4}\left(\tilde{\mathcal{B}}+q \beta_0\right) l_{\mu_F},
    \\
    \begin{split}
    \mathcal{C}_{\mu_F}^{(2)}=& \frac{1}{32}q^2\, C_A^2\, l_{\mu_F}^4+\frac{1}{16}q\, C_A \left(\tilde{\mathcal{B}}+\frac{1}{3}\left(2 + 3q\right)\beta_0\right)\, l_{\mu_F}^3
    \\
    & + \frac{1}{32}\left(\tilde{\mathcal{B}}^2 + q\left(1+q\right)\beta_0^2 + \left(1 + 2q\right)\beta_0\, \tilde{\mathcal{B}} - 8q\, C_A\, \gamma_{\rm cusp}^{(1)}\right)\, l_{\mu_F}^2
    \\
    & + \left(\gamma_{\rm cusp}^{(1)}\left(q\, C_A \log 2 + \frac{1}{4}i \pi \left(\mathbf{T}_{\mathcal{Q}}^2-2 q\, C_A\right)\right) + \frac{1}{2}\mathbf{T}_{\mathcal{Q}}^2\, \hat{\gamma}^{Q, (1)} + q \gamma^{g, (1)} \right)\, l_{\mu_F},
    \end{split}
\end{align}
where we have defined the quantities
\begin{align}
    \tilde{\mathcal{B}} =& -4q\, C_A \log 2 + \mathbf{T}_{\mathcal{Q}}^2 - i \pi \left(\mathbf{T}_{\mathcal{Q}}^2 - 2q\, C_A\right)\,,\qquad\; \mathbf{T}_{\mathcal{Q}}^2 = \begin{cases}
        0, & \text{if $\mathcal{Q}$ is }[1], \\
        C_A, & \text{if $\mathcal{Q}$ is }[8],
    \end{cases}
    \\
    \gamma^{(1)}_{\text{cusp}} =& \left(\frac{67}{36} - \frac{\pi^2}{12} \right)\,C_A - \frac{5}{9}\, T_F\, n_l\,,
    \\
    \hat{\gamma}^{Q,(1)} =& \left(-\frac{49}{72}+\frac{\pi^2}{24}-\frac{\zeta_3}{4}\right)\,C_A +\frac{5}{18}\,T_F\, n_l\,,
    \\
    \gamma^{g,(1)} =&\left(-\frac{173}{108}+\frac{11\pi^2}{288}+\frac{1}{8}\zeta_3\right)\,C_A^2+\left(\frac{16}{27}-\frac{\pi^2}{72}\right)\,C_A\, T_F\, n_l +\frac{1}{4}\,C_F\, T_F\, n_l\,.
\end{align}

Finally the coefficients for the NRQCD scale $\mu_{\Lambda}$ can be written as
\begin{align}
    \mathcal{C}_{\mu_{\Lambda}}^{(2)} =& \frac{1}{2}\gamma_{\rm Coulomb}\, l_{\mu_{\Lambda}}\,,
    \\
    \overline{\mathcal{D}}^{(1)}_{\mu_{\Lambda}} =& \frac{1}{2}\, \gamma_{{^3S_1^{[8]}}}^{{^3P_J^{[1,8]}}}\, l_{\mu_{\Lambda}}\,.
\end{align}

\section{Analytical expressions}
\label{sec:analyticsFF}

In this Appendix, we have collected the analytical expressions for the two-loop form factor bases defined in Section~\ref{sec:formfactors}. We make them also available in electronic form in ref.~\cite{formFactorPwaveGit}.

These expressions contain functions in the class of multiple polylogarithms and their elliptic generalisation, elliptic multiple polylogarithms, which includes the class of iterated integrals of modular form. We have studied and defined these functions in our companion paper \cite{Abreu:2022vei}.

In addition, the expressions also involve special constants, such as the Catalan constant $C$, the polygamma function $\psi^m{\left(z\right)}$, and $\hpli{\left(0,-,+,-\right)}$ which are defined as follows
\begin{align}
    C=&\sum_{n=0}^{\infty}\frac{\left(-1\right)^n}{\left(2n+1\right)^2}=0.915966\ldots,
    \\[5pt]
    \psi^m{\left(z\right)}=&\frac{d^{m+1}}{dz^{m+1}}\log{\left(\Gamma{\left(z\right)}\right)}\,,
    \\[5pt]
    \begin{split}
    \hpli{\left(0,-,+,-\right)}=&G{\left(0,-1,-1,-1;i\right)}+G{\left(0,-1,-1,1;i\right)}-G{\left(0,-1,1,-1;i\right)}
    \\
    &-G{\left(0,-1,1,1;i\right)}+G{\left(0,1,-1,-1;i\right)}+G{\left(0,1,-1,1;i\right)}
    \\
    &-G{\left(0,1,1,-1;i\right)}-G{\left(0,1,1,1;i\right)}\,,
\end{split}
\end{align}
where above $\Gamma{\left(z\right)}$ is the gamma function.

In the following, where the expression is expressible in terms of functions in the class of multiple polylogarithms, we give them explicitly. However, in the case, where the expressions involve elliptic integrals, as their analytical expressions are rather lengthy, we keep the notation of the master integrals $F_I$ expanded in $\epsilon$ as
\begin{equation}
    F_I = \sum_{k} \epsilon^i\; F_I^{(k)}.
\end{equation}
The definition and the full expressions of these elliptic integrals can be found in the ancillary material of ref.~\cite{Abreu:2022vei} in ref.~\cite{masterIntegralsGit}.

In the following, we give the $a_i^{(2)}$ expressions needed for the regular contributions,
{\small

}

\noindent We now collect the analytical expressions for the bases $b_i^{(2)}$ of the light-by-light contributions
{\small
\begin{align}
    b_{1}^{(2)} =& -\frac{211}{1008}G{\left(0,0,1;3-2 \sqrt{2}\right)}-\frac{61}{252} \Li_3{\left(3-2 \sqrt{2}\right)}-\frac{6833}{210}\zeta_3-\frac{11}{504} \log^3{\left(-1+\sqrt{2}\right)}
    \nonumber \\
    & +\frac{11}{1008} \pi ^2 \log{\left(-1+\sqrt{2}\right)}+\frac{317}{80} \pi ^2 \log{2}-\frac{16721}{8640}\pi ^2+\frac{457}{756}\log{2}+\frac{1243129}{15120}
    \nonumber \\
    &-i \pi \left[\frac{457}{1512}+\frac{13}{84}\pi^2+\frac{11}{336}\log^2{\left(-1+\sqrt{2}\right)}\right] +\sqrt{2} \left[\frac{211}{504} G{\left(0,1;3-2 \sqrt{2}\right)}+\frac{211}{3024}\pi^2 \right.
    \nonumber \\
    &\left. - \frac{211}{504} \log^2{\left(-1+\sqrt{2}\right)}-\frac{211}{504} i \pi \log{\left(-1+\sqrt{2}\right)}\right]+\frac{3}{2} F_{5}^{(0)}+17 F_{14}^{(0)}-\frac{11}{2} F_{24}^{(0)}-40 F_{25}^{(0)}
    \nonumber \\
    &-\frac{201}{20} F_{57}^{(0)}-\frac{1321}{180} F_{64}^{(0)}+\frac{1177}{144} F_{65}^{(0)}\,,
\\[10pt]
    b_{2}^{(2)} =& \frac{31}{30} G\left(0,0,1;3-2 \sqrt{2}\right)+\frac{12}{5} \Li_3{\left(3-2 \sqrt{2}\right)}+\frac{353}{24}\zeta_3+\frac{41}{45} \log^3{\left(-1+\sqrt{2}\right)}
    \nonumber \\
    & -\frac{13}{5} \pi ^2 \log{2} -\frac{41}{90} \pi ^2 \log{\left(-1+\sqrt{2}\right)}-\frac{9}{5} \log{2}+\frac{164}{135}\pi^2-\frac{211}{6}+i \pi \left[\frac{7}{120}\pi^2+\frac{9}{10} \right.
    \nonumber \\
    &\left. +\frac{41}{30} \log ^2{\left(-1+\sqrt{2}\right)}\right] + \sqrt{2} \left[\frac{31}{15} \log ^2{\left(-1+\sqrt{2}\right)}-\frac{31}{90} \pi ^2-\frac{31}{15} G{\left(0,1;3-2 \sqrt{2}\right)} \right.
    \nonumber \\
    & \left. +\frac{31}{15} i \pi \log {\left(-1+\sqrt{2}\right)}\right]-\frac{1}{3} F_{5}^{(0)}-\frac{4}{3} F_{14}^{(0)}+\frac{1}{3} F_{24}^{(0)}+\frac{18}{5} F_{57}^{(0)}+\frac{28}{9} F_{64}^{(0)}-\frac{38}{9} F_{65}^{(0)}\,,
\\[10pt]
    b_{3}^{(2)} =& -\frac{46}{3}\zeta_3+\frac{275}{24}-\frac{1}{2}\pi^2-\frac{1}{3}\log{2}+\frac{7}{3} \pi^2 \log{2} +\frac{i \pi }{6} + \frac{2}{3} F_{5}^{(0)}-\frac{10}{3} F_{14}^{(0)}-\frac{1}{3} F_{24}^{(0)}
    \nonumber \\
    &-4 F_{57}^{(0)}-F_{64}^{(0)}+F_{65}^{(0)}\,,
\\[10pt]
    b_{4}^{(2)} =& \frac{10}{3}\zeta_3-\frac{2}{3} \pi ^2 \log{2}+\frac{43}{108} \pi ^2-\frac{8}{9} \log^2{2}-\frac{146}{27}\log{2}+\frac{7}{54}+i \pi \left[\frac{91}{27}-\frac{4 \pi^2}{9}+\frac{8}{9} \log{2}\right]\,,
\\[10pt]
    b_{5}^{(2)} =& -\frac{35}{12}\zeta_3+\frac{7}{12} \pi^2 \log{2}- \frac{11}{24} \pi^2 + 6 \log{2} + i \pi \left[\frac{7 \pi^2}{18}-\frac{23}{6}\right]\,,
\\[10pt]
    b_{6}^{(2)} =& \frac{17}{24}\zeta_3-\frac{1}{3}\log{2}+\frac{1}{3}+\frac{i \pi }{6}\,,
\\[10pt]
    b_{7}^{(2)} =& -\frac{247}{288} G{\left(0,0,1;3-2 \sqrt{2}\right)}-\frac{121}{72} \Li_3{\left(3-2 \sqrt{2}\right)}+\frac{7411}{480}\zeta_3-\frac{79 }{144} \log^3{\left(-1+\sqrt{2}\right)}
    \nonumber \\
    & +\frac{79}{288} \pi ^2 \log{\left(-1+\sqrt{2}\right)}-\frac{433}{480} \pi ^2 \log{2}+\frac{21}{128} \pi ^2+\frac{1}{9}\log^2{2}+\frac{37}{27}\log{2}-\frac{74227}{1728}
    \nonumber \\
    &-i \pi \left[\frac{79}{96} \log^2{\left(-1+\sqrt{2}\right)}+\frac{1}{9} \log{2}+\frac{37}{54}\right]+\sqrt{2} \left[\frac{247}{144} G{\left(0,1;3-2 \sqrt{2}\right)}+\frac{247}{864}\pi^2\right.
    \nonumber \\
    &\left. -\frac{247}{144} \log^2{\left(-1+\sqrt{2}\right)}-\frac{247}{144} i \pi  \log{\left(-1+\sqrt{2}\right)}\right] -\frac{3}{4} F_{5}^{(0)}-\frac{17}{2} F_{14}^{(0)}+\frac{11}{4} F_{24}^{(0)}
    \nonumber \\
    &+20 F_{25}^{(0)}+\frac{193}{40} F_{57}^{(0)}+\frac{137}{36} F_{64}^{(0)}-\frac{1223}{288} F_{65}^{(0)}\,,
\\[10pt]
    b_{8}^{(2)} =& -\frac{55}{1728}G{\left(0,0,1;3-2 \sqrt{2}\right)} + \frac{215}{432} \Li_3{\left(3-2 \sqrt{2}\right)}-\frac{56771}{2880}\zeta_3 +\frac{305}{864} \log^3{\left(-1+\sqrt{2}\right)}
    \nonumber \\
    & -\frac{305}{1728} \pi ^2 \log{\left(-1+\sqrt{2}\right)}+\frac{5093}{2880} \pi ^2 \log{2}-\frac{4997}{6912} \pi ^2 -\frac{1}{27}\log^2{2} + \frac{145}{144} \log{2} + \frac{8419}{192}
    \nonumber \\
    &+ i \pi \left[\frac{305}{576}  \log^2{\left(-1+\sqrt{2}\right)}+\frac{1}{27}  \log{2}-\frac{145}{288}\right] +\sqrt{2} \left[\frac{55}{864} G{\left(0,1;3-2 \sqrt{2}\right)} +\frac{55}{5184}\pi^2\right.
    \nonumber \\
    & \left. -\frac{55}{864} \log^2{\left(-1+\sqrt{2}\right)}-\frac{55}{864} i \pi  \log{\left(-1+\sqrt{2}\right)}\right]+\frac{1}{6} F_{5}^{(0)}+\frac{2}{3} F_{14}^{(0)}-\frac{1}{3} F_{15}^{(0)}-\frac{1}{6} F_{24}^{(0)}
    \nonumber \\
    &-\frac{1313}{240} F_{57}^{(0)}-\frac{1679}{432} F_{64}^{(0)}+\frac{8827}{1728} F_{65}^{(0)}\,,
\\[10pt]
    b_{9}^{(2)} =& -\frac{131}{216} G{\left(0,0,1;3-2 \sqrt{2}\right)}-\frac{23}{54} \Li_3{\left(3-2 \sqrt{2}\right)}+\frac{13}{108} \log^3{\left(-1+\sqrt{2}\right)}-\frac{17}{36} \pi ^2 \log{2}
    \nonumber \\
    & -\frac{13}{216} \pi ^2 \log{\left(-1+\sqrt{2}\right)}+\frac{55}{18}\zeta_3-\frac{115}{864} \pi^2-\frac{17}{54}\log^2{2}+\frac{7}{2}\log{2}-\frac{1609}{288}
    \nonumber \\
    &+i \pi \left[ \frac{13}{72} \log^2{\left(-1+\sqrt{2}\right)}+ \frac{17}{54} \log{2}-\frac{7}{4} \right] +\sqrt{2} \left[\frac{131}{108} G{\left(0,1;3-2 \sqrt{2}\right)}+\frac{131}{648} \pi^2 \right.
    \nonumber \\
    &\left. -\frac{131}{108} \log^2{\left(-1+\sqrt{2}\right)}-\frac{131}{108} i \pi  \log{\left(-1+\sqrt{2}\right)}\right] -\frac{1}{3} F_{5}^{(0)}+\frac{5}{3} F_{14}^{(0)}-\frac{1}{3} F_{15}^{(0)}+\frac{1}{6} F_{24}^{(0)}
    \nonumber \\
    &+\frac{2}{3} F_{57}^{(0)}+\frac{49}{108} F_{64}^{(0)}-\frac{19}{216} F_{65}^{(0)}\,,
\\[10pt]
    b_{10}^{(2)} =& -\frac{15}{16}\zeta_3+\frac{3}{16} \pi ^2 \log{2}-\frac{5}{27}\pi^2+\frac{2}{9}\log^2{2}+\frac{187}{108}\log{2}-\frac{169}{432}
    \nonumber \\
    &+i \pi\left[\frac{1}{8}\pi^2-\frac{2}{9}\log{2}-\frac{527}{432}\right]\,,
\\[10pt]
    b_{11}^{(2)} =& \frac{115}{48} \zeta_3 -\frac{23}{48} \pi ^2 \log{2} + \frac{943}{2592} \pi^2-\frac{2}{27} \log^2{2}-\frac{509}{108}\log{2}+\frac{31}{72}
    \nonumber \\
    &+i \pi \left[-\frac{23}{72}\pi^2 + \frac{2}{27} \log{2}+\frac{671 }{216}\right]\,,
\\[10pt]
    b_{12}^{(2)} =& -\frac{17}{48} \zeta_3 + \frac{31}{162} \pi^2-\frac{17}{27} \log^2{2}+\frac{53}{54} \log{2}-\frac{5}{18} + i \pi \left[\frac{17}{27} \log{2}-\frac{79}{216}\right]\,.
\end{align}
}

\noindent The bases $c_i^{(2)}$ needed for the vacuum polarisation contribution read
{\small
\begin{align}
    c_1^{(2)} =& -\frac{37}{160}\pi ^2 +\frac{94273}{10080} -\frac{109}{140}F_{64}^{(0)}+\frac{43}{28} F_{65}^{(0)}\,,
\\[10pt]
    c_2^{(2)} =& \frac{8}{15} \zeta_3-\frac{1}{60} \pi ^2 \log{2}+\frac{19}{180}\pi ^2-\frac{264881}{30240} + \frac{1}{5} F_{57}^{(0)}+\frac{1039}{1260} F_{64}^{(0)} - \frac{19}{14} F_{65}^{(0)}\,,
\\[10pt]
    c_3^{(2)} =& \frac{8}{15}\zeta_3 - \frac{1}{60} \pi^2 \log{2}-\frac{5}{48} \pi ^2 -\frac{7729}{864} + \frac{1}{5} F_{57}^{(0)}+\frac{13}{12} F_{64}^{(0)}-2 F_{65}^{(0)}\,,
\\[10pt]
    c_4^{(2)} =& \frac{1}{16}\pi ^2+\frac{43}{36}\,,
\\[10pt]
    c_5^{(2)} =& -\frac{7}{24}\zeta_3 - \frac{1}{16}\pi ^2 -\frac{2}{3} \log^2{2}+\frac{16}{9} \log{2} - \frac{5}{27}\,,
\\[10pt]
    c_6^{(2)} =& -\frac{7}{24}\zeta_3 - \frac{5}{48} \pi ^2 +\frac{61}{108}\,,
\\[10pt]
    c_7^{(2)} =& \frac{21}{160} \Ree{\left[G{\left(0,0,e^{-\frac{i\pi}{3}},-1;1\right)}\right]}+\frac{63}{640} \Ree{\left[G{\left(0,0,e^{-\frac{2 i \pi}{3}},1;1\right)}\right]}+ \frac{7}{48} \zeta_3 \log{2}
    \nonumber \\
    &-\frac{649 }{691200}\pi ^4-\frac{11}{240} \pi ^2 \log^2{2} -\frac{163}{144} G{\left(0,0,1;3-2 \sqrt{2}\right)}-\frac{7}{36} \Li_3{\left(3-2 \sqrt{2}\right)}-\frac{903}{32}\zeta_3
    \nonumber \\
    &+\frac{5}{8} \log^3{\left(-1+\sqrt{2}\right)}-\frac{5}{16} \pi ^2 \log{\left(-1+\sqrt{2}\right)}+\frac{27}{16} \pi ^2 \log{2}-\frac{1}{9}\log^2{2}-\frac{76 }{27}\pi ^2+\frac{313}{72}\log{2}
    \nonumber \\
    &+\frac{674503}{5184}+i \pi \left[\frac{15}{16} \log^2{\left(-1+\sqrt{2}\right)} +\frac{1}{9}  \log{2}- \frac{529}{144}\right]+\sqrt{2} \left[\frac{163}{72} G{\left(0,1;3-2 \sqrt{2}\right)} \right.
    \nonumber \\
    & \left. +\frac{163}{432}\pi ^2-\frac{163}{72} \log^2{\left(-1+\sqrt{2}\right)}-\frac{163}{72} i \pi  \log{\left(-1+\sqrt{2}\right)}\right] + 30 F_{14}^{(0)}+\frac{11}{120} F_{15}^{(0)}+\frac{3}{4} F_{18}^{(0)}
    \nonumber \\
    &-\frac{19}{40} F_{31}^{(0)}-9 F_{57}^{(0)}-\frac{815}{72} F_{64}^{(0)}+\frac{1085}{72} F_{65}^{(0)}\,,
\\[10pt]
    c_8^{(2)} =& \frac{1}{6}\log^4{2}+\frac{17}{40} \pi ^2 \log^2{2}+\frac{1}{27}\log^2{2}-\frac{1}{6} \log^3{\left(2 \sqrt{2}+3\right)} \log{2}+4 \Li_4{\left(\frac{1}{2}\right)}-\frac{8471}{345600} \pi ^4
    \nonumber \\
    &+\frac{10975}{576}\zeta_3-\frac{40043}{576}-\frac{1357}{576} \pi ^2 \log{2}+\log{2} \left[ G{\left(0,0,\frac{\sqrt{2}}{2}+\frac{1}{2};1\right)}+G{\left(0,0,\frac{1}{2}-\frac{\sqrt{2}}{2};1\right)} \right.
    \nonumber \\
    &\left. +G{\left(0,\frac{\sqrt{2}}{2}+\frac{1}{2},1;1\right)} +G{\left(0,\frac{1}{2}-\frac{\sqrt{2}}{2},1;1\right)} -G{\left(\frac{\sqrt{2}}{2}+\frac{1}{2},0,1;1\right)} \right.
    \nonumber \\
    &\left. -G{\left(\frac{1}{2}-\frac{\sqrt{2}}{2},0,1;1\right)} \right] -\frac{1}{6} \pi ^2 \log{\left(2 \sqrt{2}+3\right)} \log{2}-\frac{1}{6} \pi ^2 \log{\left(3-2 \sqrt{2}\right)} \log{2}
    \nonumber \\
    &-\frac{1}{6} \log ^3\left(3-2 \sqrt{2}\right) \log{2}-\frac{7}{24} \zeta_3 \log{2}-\frac{2105}{144}\log{2}-\frac{991}{1728}\pi ^2 \log{\left(-1+\sqrt{2}\right)}
    \nonumber \\
    &-\frac{1}{16} \log ^4\left(3-2 \sqrt{2}\right)+\frac{991}{864} \log^3{\left(-1+\sqrt{2}\right)}+\frac{19229 }{6912}\pi ^2+\frac{2377}{432} \Li_3{\left(3-2 \sqrt{2}\right)}
    \nonumber \\
    &-G{\left(\frac{\sqrt{2}}{2}+\frac{1}{2},0,1,1;1\right)}-G{\left(\frac{1}{2}-\frac{\sqrt{2}}{2},0,1,1;1\right)}+\frac{6535}{1728}G{\left(0,0,1;3-2 \sqrt{2}\right)}
    \nonumber \\
    &+\frac{3}{2} \left[G{\left(0,0,0,\frac{\sqrt{2}}{2}+\frac{1}{2};1\right)}+G{\left(0,0,0,\frac{1}{2}-\frac{\sqrt{2}}{2};1\right)}+ G{\left(0,0,\frac{\sqrt{2}}{2}+\frac{1}{2},1;1\right)}\right.
    \nonumber \\
    &\left.+G{\left(0,0,\frac{1}{2}-\frac{\sqrt{2}}{2},1;1\right)}\right]+\frac{1}{2}\left[ G{\left(0,\frac{\sqrt{2}}{2}+\frac{1}{2},0,1;1\right)}+G{\left(0,\frac{1}{2}-\frac{\sqrt{2}}{2},0,1;1\right)}\right.
    \nonumber \\
    &\left.-G{\left(\frac{\sqrt{2}}{2}+\frac{1}{2},1,0,1;1\right)} - G{\left(\frac{1}{2}-\frac{\sqrt{2}}{2},1,0,1;1\right)}\right] +2 \left[G{\left(0,\frac{1}{2}-\frac{\sqrt{2}}{2},1,1;1\right)}\right.
    \nonumber \\
    &\left.+G{\left(0,\frac{\sqrt{2}}{2}+\frac{1}{2},1,1;1\right)}\right]-\frac{7}{32} \Ree{\left[G{\left(0,0,e^{-\frac{i\pi}{3}},-1;1\right)}\right]}-\frac{21}{128} \Ree{\left[G{\left(0,0,e^{-\frac{2 i \pi}{3}},1;1\right)}\right]}
    \nonumber \\
    &-\frac{7}{160} \Ree{\left[G{\left(0,0,e^{-\frac{i\pi}{3}},-1;1\right)}\right]}-\frac{21}{640} \Ree{\left[G{\left(0,0,e^{-\frac{2 i \pi}{3}},1;1\right)}\right]}-\frac{1}{16} \log^4{\left(2 \sqrt{2}+3\right)}
    \nonumber \\
    &-\frac{1}{2} i \pi  \left[ G{\left(0,0,\frac{\sqrt{2}}{2}+\frac{1}{2};1\right)}+ G{\left(0,0,\frac{1}{2}-\frac{\sqrt{2}}{2};1\right)} + G{\left(0,\frac{\sqrt{2}}{2}+\frac{1}{2},1;1\right)}\right.
    \nonumber \\
    &\left.+G{\left(0,\frac{1}{2}-\frac{\sqrt{2}}{2},1;1\right)} - G{\left(\frac{\sqrt{2}}{2}+\frac{1}{2},0,1;1\right)}-  G{\left(\frac{1}{2}-\frac{\sqrt{2}}{2},0,1;1\right)}\right]
    \nonumber \\
    &-\frac{1}{8} \pi ^2 \log^2{\left(2 \sqrt{2}+3\right)}-\frac{1}{8} \pi ^2 \log ^2\left(3-2 \sqrt{2}\right)+\sqrt{2}\left[-\frac{6535}{864} G{\left(0,1;3-2 \sqrt{2}\right)}\right.
    \nonumber \\
    &\left.+\frac{6535}{864} \log^2{\left(-1+\sqrt{2}\right)}  -\frac{6535}{5184}\pi ^2+\frac{6535}{864} i \pi  \log{\left(-1+\sqrt{2}\right)}\right] +i \pi\left[\frac{2593}{288}-\frac{1}{27} \log{2} \right.
    \nonumber \\
    &\left. +\frac{1}{12} \log^3{\left(2 \sqrt{2}+3\right)}+\frac{1}{12} \log ^3\left(3-2 \sqrt{2}\right)+\frac{1}{12}\pi ^2 \log{\left(2 \sqrt{2}+3\right)}+\frac{1}{12} \pi ^2 \log{\left(3-2 \sqrt{2}\right)}\right.
    \nonumber \\
    &\left.+\frac{991}{576} \log^2{\left(-1+\sqrt{2}\right)}\right] -15 F_{14}^{(0)}-\frac{17}{20} F_{15}^{(0)}-\frac{3}{2} F_{18}^{(0)}+\frac{9}{20} F_{31}^{(0)}+\frac{265}{48} F_{57}^{(0)}+\frac{2627}{432} F_{64}^{(0)}
    \nonumber \\
    &-\frac{16675 }{1728}F_{65}^{(0)}\,,
\\[10pt]
    c_9^{(2)} =& -\frac{7}{32} \Ree{\left[G{\left(0,0,e^{-\frac{i\pi}{3}},-1;1\right)}\right]}-\frac{21}{128} \Ree{\left[G{\left(0,0,e^{-\frac{2 i \pi}{3}},1;1\right)}\right]}-\frac{649 }{691200}\pi ^4
    \nonumber \\
    &+\frac{7}{20} \Ree{\left[G{\left(0,0,e^{-\frac{i\pi}{3}},-1;1\right)}\right]}+\frac{21}{80} \Ree{\left[G{\left(0,0,e^{-\frac{2 i \pi}{3}},1;1\right)}\right]}+\frac{7}{48} \zeta_3 \log{2}
    \nonumber \\
    &-\frac{11}{240} \pi ^2 \log^2{2}+\frac{137}{216} G{\left(0,0,1;3-2 \sqrt{2}\right)}+\frac{10}{27} \Li_3{\left(3-2 \sqrt{2}\right)}-\frac{19}{108}  \log^3{\left(-1+\sqrt{2}\right)}
    \nonumber \\
    &-\frac{2351 }{360}\zeta_3+\frac{233}{360} \pi ^2 \log{2}+\frac{19}{216} \pi ^2 \log{\left(-1+\sqrt{2}\right)}-\frac{97}{864} \pi ^2+\frac{17}{54}\log^2{2} -\frac{89 }{27}\log{2}
    \nonumber \\
    &+\frac{16097}{648}+i \pi \left[\frac{199 }{108}-\frac{19}{72}  \log^2{\left(-1+\sqrt{2}\right)}-\frac{17}{54} \log{2}\right] +\sqrt{2} \left[-\frac{137}{108} G{\left(0,1;3-2 \sqrt{2}\right)} \right.
    \nonumber \\
    &\left.-\frac{137 }{648}\pi ^2 +\frac{137}{108} \log^2{\left(-1+\sqrt{2}\right)} +\frac{137}{108} i \pi  \log{\left(-1+\sqrt{2}\right)}\right] + 12 F_{14}^{(0)}+\frac{17}{40} F_{15}^{(0)}+\frac{3}{4} F_{18}^{(0)}
    \nonumber \\
    &+\frac{1}{40} F_{31}^{(0)}-\frac{83}{30} F_{57}^{(0)}-\frac{125}{54} F_{64}^{(0)}+\frac{409}{216} F_{65}^{(0)}\,,
\\[10pt]
    c_{10}^{(2)} =& \frac{51}{320} \Ree{\left[G{\left(0,0,e^{-\frac{i\pi}{3}},-1;1\right)}\right]}+\frac{153 }{1280}\Ree{\left[G{\left(0,0,e^{-\frac{2 i \pi}{3}},1;1\right)}\right]}-\frac{3119 }{1382400}\pi ^4
    \nonumber \\
    &-\frac{1}{480} \pi ^2 \log^2{2}-\frac{7}{96} \zeta_3 \log{2}+\frac{105 }{16}\zeta_3-\frac{7}{32} \pi ^2 \log{2}+\frac{4759 }{8640}\pi ^2+\frac{7 }{12}\log{2}-\frac{14643169}{362880}
    \nonumber \\
    &-15 F_{14}^{(0)}+\frac{1}{240} F_{15}^{(0)}-\frac{3}{8} F_{18}^{(0)}-\frac{9}{80} F_{31}^{(0)}+\frac{21}{8} F_{57}^{(0)}+\frac{56251 }{15120}F_{64}^{(0)}-\frac{5359 }{1008}F_{65}^{(0)}\,,
\\[10pt]
    c_{11}^{(2)} =& \frac{21}{160} \Ree{\left[G{\left(0,0,e^{-\frac{i\pi}{3}},-1;1\right)}\right]}+\frac{63}{640} \Ree{\left[G{\left(0,0,e^{-\frac{2 i \pi}{3}},1;1\right)}\right]}-\frac{649 }{691200}\pi ^4
    \nonumber \\
    &+\frac{7}{48} \zeta_3 \log{2}-\frac{11}{240}  \pi ^2 \log^2{2}-\frac{487 }{240}\zeta_3+\frac{37}{480} \pi ^2 \log{2}-\frac{25 }{12}\log{2}-\frac{43 }{320}\pi ^2+\frac{18259}{1120}
    \nonumber \\
    &+\frac{15}{2} F_{14}^{(0)}+\frac{11}{120} F_{15}^{(0)}+\frac{3}{4} F_{18}^{(0)}-\frac{9}{40} F_{31}^{(0)}-\frac{37}{40} F_{57}^{(0)}-\frac{881}{630} F_{64}^{(0)}+\frac{655}{336} F_{65}^{(0)}\,,
\\[10pt]
    c_{12}^{(2)} =& \frac{51}{320} \Ree{\left[G{\left(0,0,e^{-\frac{i\pi}{3}},-1;1\right)}\right]}+\frac{153 }{1280}\Ree{\left[G{\left(0,0,e^{-\frac{2 i \pi}{3}},1;1\right)}\right]}-\frac{3119 }{1382400}\pi ^4
    \nonumber \\
    &-\frac{1}{480} \pi ^2 \log^2{2}-\frac{7}{96} \zeta_3 \log{2}+\frac{38}{15} \zeta_3-\frac{19}{240} \pi ^2 \log{2}+\frac{107 }{432}\pi ^2+\frac{3 }{2}\log{2}-\frac{93763}{5184}
    \nonumber \\
    &-6 F_{14}^{(0)}+\frac{1}{240} F_{15}^{(0)}-\frac{3}{8} F_{18}^{(0)}+\frac{11}{80} F_{31}^{(0)}+\frac{19}{20} F_{57}^{(0)}+\frac{337}{216} F_{64}^{(0)}-\frac{167}{72} F_{65}^{(0)}\,,
\\[10pt]
    c_{13}^{(2)} =& \frac{313}{144}\zeta_3-\frac{2}{9} \log^3{2}-\frac{1}{144} \pi ^2 \log{2} +\frac{79}{864}\pi^2+\frac{11}{18} \log^2{2} -\frac{251 }{54} \log{2} +\frac{469}{1296}
    \nonumber \\
    &+i \pi\left[- \frac{5 }{24}\pi ^2 +\frac{1}{3}  \log^2{2}  -\frac{17}{18}  \log{2}+\frac{1457 }{432}\right]\,,
\\[10pt]
    c_{14}^{(2)} =& -\frac{23 }{8}\zeta_3+\frac{29}{48} \pi ^2 \log{2}-\frac{539}{1296} \pi ^2+\frac{13 }{54}\log^2{2}+\frac{307}{54} \log{2}-\frac{65}{72}
    \nonumber \\
    &+i \pi \left[\frac{29}{72}\pi ^2-\frac{2}{27} \log{2}-\frac{863}{216}\right]\,,
\\[10pt]
    c_{15}^{(2)} =& \frac{43}{144} \zeta_3 -\frac{2}{9}\log^3{2} + \frac{2}{9} \pi ^2 \log{2} -\frac{347 }{1296}\pi ^2+\frac{32 }{27}\log^2{2} -\log{2}-\frac{79}{648}
    \nonumber \\
    &+i \pi \left[-\frac{\pi ^2}{18}+\frac{1}{3} \log^2{2}-\frac{32}{27}\log{2}+\frac{29 }{72}\right]\,,
\\[10pt]
    c_{16}^{(2)} =& -\frac{59}{72}\zeta_3-\frac{1}{9} \log^3{2}-\frac{1}{72} \pi ^2 \log{2}-\frac{25}{144} \pi ^2+\frac{19 }{36}\log^2{2}-\frac{83}{216} \log{2}+\frac{49}{648}\,,
\\[10pt]
    c_{17}^{(2)} =& \frac{7 }{96}\zeta_3+\frac{3}{64} \pi ^2+\frac{1}{12}\log^2{2}-\frac{17 }{72}\log{2}-\frac{5}{24}\,,
\\[10pt]
    c_{18}^{(2)} =& -\frac{89}{288}\zeta_3-\frac{1}{9} \log^3{2}- \frac{1}{72} \pi ^2 \log{2}- \frac{5}{288} \pi ^2 +\frac{4}{9} \log^2{2}-\frac{23}{54}\log{2}-\frac{55}{1296}\,.
\end{align}
}

\bibliographystyle{JHEP}
\bibliography{biblio}

\end{document}